\documentclass[useAMS,usenatbib]{mn2e}

\usepackage{pdflscape}
\usepackage{graphicx}

\newcommand{\hiir}{H~{\scshape ii}~region}
\newcommand{\hiirs}{H~{\scshape ii}~regions}
\newcommand{\um}{\,$\umu$m}
\newcommand{\degree}{$^{\circ}$}
\newcommand{\herschel}{{\it Herschel}}
\newcommand{\wise}{{\it WISE}}

\title[First Catalogue of Galactic Bubbles IR Fluxes]{First Extended Catalogue of  Galactic Bubbles InfraRed Fluxes from \wise\, and \herschel\thanks{{\it Herschel} is an ESA space observatory with science instruments provided by European-led Principal Investigator consortia and with important participation from NASA.} Surveys.}
\author[F. Bufano et al.]{F. Bufano$^{1}$\thanks{E-mail:fbufano@oact.inaf.it}, 
P. Leto$^{1}$, D. Carey$^{2}$, G. Umana$^{1}$, C. Buemi$^{1}$, A. Ingallinera$^{1}$,  
\newauthor  A. Bulpitt$^{2}$, F. Cavallaro$^{1,3,4}$, S. Riggi$^{1}$,  C. Trigilio$^{1}$, S. Molinari$^{5}$ \\
$^{1}$INAF-Osservatorio Astrofisico di Catania, Via Santa Sofia 78, I-95123 Catania, Italy\\
$^{2}$School of Computing, University of Leeds, E C Stoner Building, Leeds, LS2 9JT, UK\\
$^{3}$Universit\`a di Catania, Dipartimento di Fisica e Astronomia, Via Santa Sofia, 64, 95123 Catania, Italy\\
$^{4}$CSIRO Astronomy and Space Science, PO Box 76, Epping, NSW 1710, Australia\\
$^{5}$INAF-Istituto di Astrofisica e Planetologia Spaziale, Via Fosso del Cavaliere 100, I-00133 Roma, Italy\\
}
\begin{document}

\date{Accepted ... Received ..; in original form ...}

\pagerange{\pageref{firstpage}--\pageref{lastpage}} \pubyear{2016}

\maketitle

\label{firstpage}

\begin{abstract}
This paper presents the first extended catalogue of far infrared fluxes of Galactic bubbles. 
Fluxes were estimated for 1814 bubbles, defined here as the {\it  golden sample} and were selected from the catalogue produced by \citet{Simpson}. 
The golden sample was comprised of bubbles identified within the \wise\,  dataset (using 12\um \, and 22\um\, images) and  \herschel\,  data (using 70\um, 160\um, 250\um, 350\um\, and 500\um\, wavelength images). 
Flux estimation was achieved initially via classical aperture photometry and then by an alternative image analysis algorithm that used active contours. 
The accuracy of the two methods was tested by comparing the estimated fluxes for a sample of bubbles, made up of 126  \hiirs\,  and 43 Planetary Nebulae, which were identified by \citet{Anderson12}. 
The results of this paper demonstrate that a good agreement between the two was found. 
This is by far the largest and most homogeneous catalogue of infrared fluxes measured for Galactic bubbles and is a step towards the fully automated analysis of astronomical datasets.
\end{abstract}


\section{Introduction}

{\it Bubbles} are one of the most intriguing objects found within recent large-scale infrared (IR) surveys (see e.g. \citealt{Churchwell06}, \citealt{Mizuno10}; \citealt{Wachter10}; \citealt{Simpson}).
The term {\it bubbles} is used to classify the diffuse emissions with a ring, disc or shell-like shape distributed throughout the entire Galactic plane, although they can be the result of different astrophysical phenomena. 
For example, some are related to young  \hiirs, thus to hot massive stars which mold the interstellar medium (ISM), and others to circumstellar envelopes that surround stars at later evolutionary stages, such as Planetary Nebulae (PNe), Luminous Blue Variables (LBVs), Supernova Remnants (SNRs), etc.\\
Bubble studies enable structural and physical properties about these objects to be derived.  For instance, such work allows important information about their central objects, the stellar winds they arise from and the environment in which they expand to be extracted. \\
Churchwell et al. (2006; 2007) have catalogued almost 600 bubbles (typically few arcminutes wide), listing the most prominent
ones  detected in the images from the {\it Spitzer} Galactic Legacy Infrared Mid-Plane Survey Extraordinaire (GLIMPSE; \citealt{GLIMPSE}).
GLIMPSE surveyed  the Galactic plane between sky regions found at  $\mid b \mid \le 1-2$\degree\,  and $\mid l\mid \le$65\degree\, using four different IR wavebands (3.6, 4.5, 5.8 and 8\um). 
Based on the spatial coincidence with known \hiirs , \citet{Churchwell07} claimed that many of the IR bubbles are produced by O and early-B stars. 
The emission observed with the 8\um\, band, in general associated to photo-dissociated regions (PDRs), is mainly due to polycyclic aromatic hydrocarbon (PAH) molecules. These emit via fluorescence at 7.7\um\, and 8.6\um\, \citep{PAH},  when excited by the far-UV photons from the hot central star.
PAH emission at 8\um\, from bubbles associated with H II regions is strong, while e.g. in PNe, it is moderately strong or weak/absent if it comes from C-rich or O-rich PNe, respectively (\citealt{Volk};\citealt{Anderson12}).\\
However, analysis of the images from the {\it Spitzer}/Multiband Imaging Photometer For {\it Spitzer} Inner Galactic Plane (MIPSGAL, \citealt{MIPSGAL}), \citet{Deharveng} noticed  that the emission at 24\um\, of bubbles from \citet{Churchwell06} is frequently observed inside the bubble with a morphology that closely traces the radio continuum emission at 20\,cm from ionized gas.    
They claimed that the emission at this wavelength is dominated by hot thermal dust, containing a contribution from very small grains (probably silicates) that are out of thermal equilibrium.
  \citet{Mizuno10} inspected 24\um\, MIPSGAL images, looking for circularly symmetric and extended emissions. They found a total of 416 bubbles, typically smaller than those
identified by \citet{Churchwell06}  ($\la 1\arcmin$). 
A fraction of the sample ($\sim 16\%$)  was already identified in previous works, and almost the totality of them classified 
as PNe, LBVs or SNRs, leading the authors to conclude, based also on a strong morphological similarity, that their catalogue included primarily evolved stars.\\
 Nowadays, information from existing IR surveys can help to improve knowledge of the bubble structures and their origins in two ways.  
 Firstly, a larger area of the Milky Way has been covered increasing the number of known objects. 
 Secondly bubbles have been observed in different wavebands. 
With this purpose, we analyzed: a) the available data from the Wide-field Infrared Survey Explorer (\wise), which mapped the
entire sky in four IR bands, in particular at 12\um\, and 22\um, resembling the 8\um\, and 24\um\, from {\it Spitzer} GLIMPSE and MIPSGAL
although with a lower resolution; b) the data from the \herschel\, infrared Galactic Plane Survey  (Hi-GAL, \citealt{MolinariHi-Gal}), which covers the entire Galactic plane ($\mid b\mid \le$1\degree) at longer wavelengths than \wise\, tracing e.g. the distribution of the cold dust (see Section \ref{data}). \\
Despite the richness of information available, only few works exploited such IR data for the bubble studies. 
\citet{Anderson12} (hereafter referred to as A12)  analyzed a sample of  bubbles including126  \hiirs\, and 43 known PNe with the
aim of discriminating between the two kind of sources based on their IR colors.
\citet{Paladini12} published a study on 16 known \hiirs\,  in order to understand the mechanisms regulating massive star formation.
Both published IR flux catalogues limited to the studied bubbles, where fluxes were estimated by  ``interactive'' methods. Indeed in these analysis, 
the dimensions of the bubble, thus the radius used for the flux estimation, was visually adjusted and chosen on a case by case basis.\\
In an era of big data, using only this kind of approach would be simply anachronistic when considering the huge flow of information produced by the incoming unbiased surveys 
that will be carried out  e.g. at the Large Synoptic Survey Telescope (LSST), the James Webb Space Telescope (JWST) or the Square Kilometer Array (SKA).\\

In this paper, we present two methods for the automated measurement of bubble fluxes and, as final product, 
a catalogue with the so estimated emitted IR fluxes of a sample of 1814 Galactic bubbles. 
The paper is organized as follows. In Section\,\ref{sample}, we present the selected bubble sample (whose measured fluxes are published in the final catalogue). 
In Section\,\ref{data}, we give a description of the technical characteristics of the employed IR surveys and in Section\,\ref{methods} a description of the methods used for the flux measurements is provided.
In Section\,\ref{catalogue}, the structure of the published catalogue is presented. 
Finally, the results are discussed in Section\,\ref{discussion} and conclusions given in Section\,\ref{conclusions}.



\begin{figure*}
\includegraphics[width=0.275\textwidth]{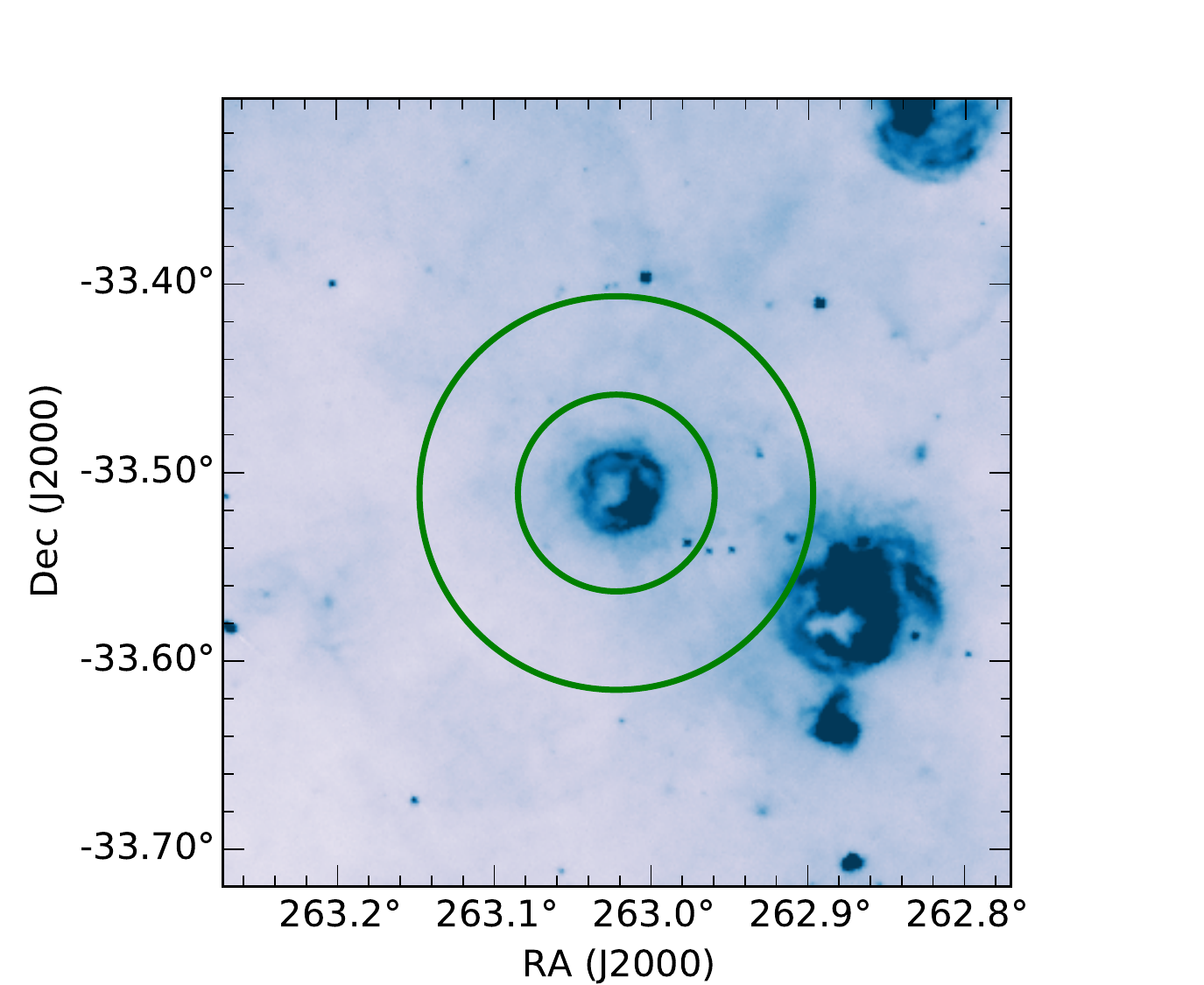}
\hspace{-0.8cm}
\includegraphics[width=0.275\textwidth]{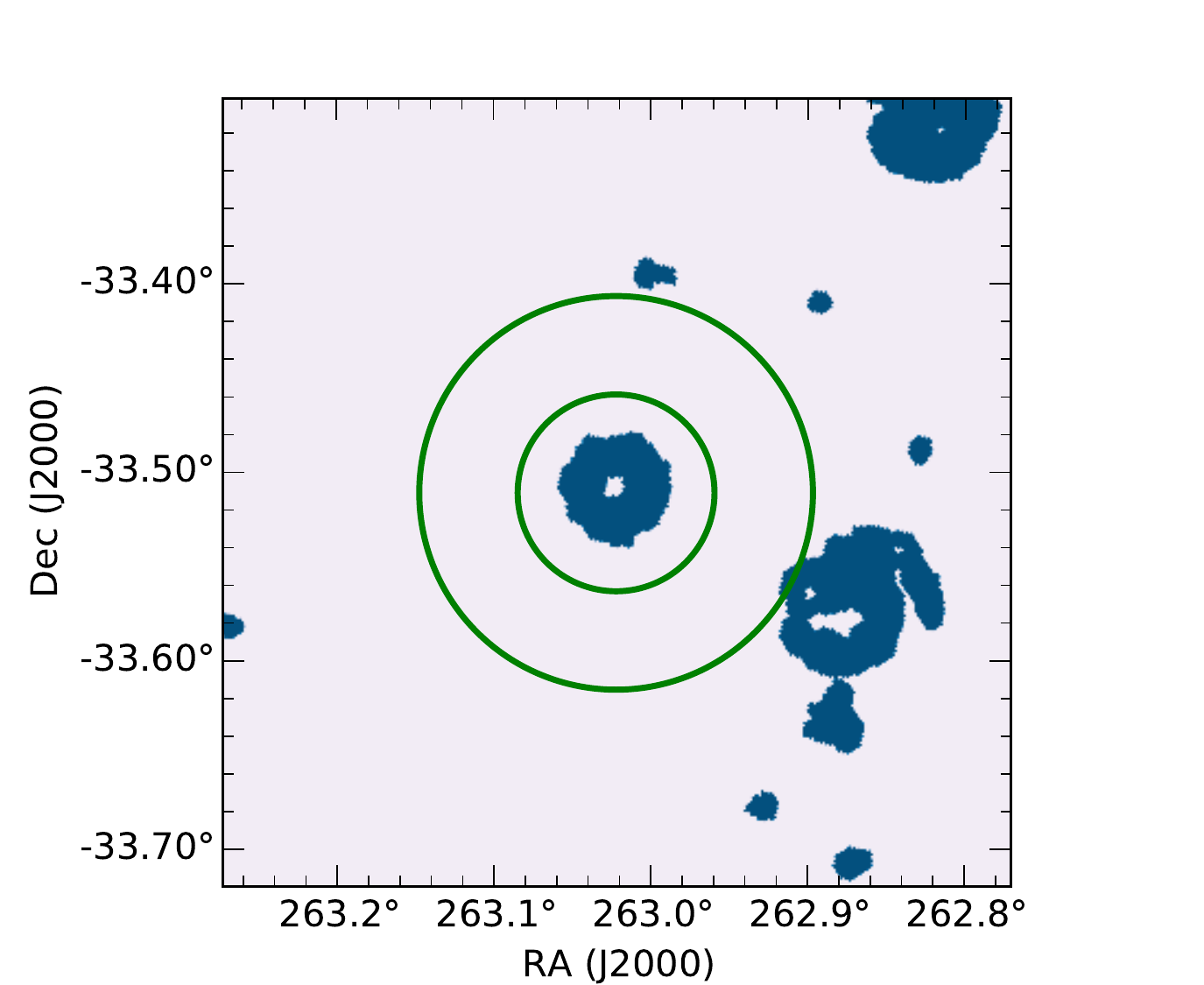}
\hspace{-0.8cm}
\includegraphics[width=0.275\textwidth]{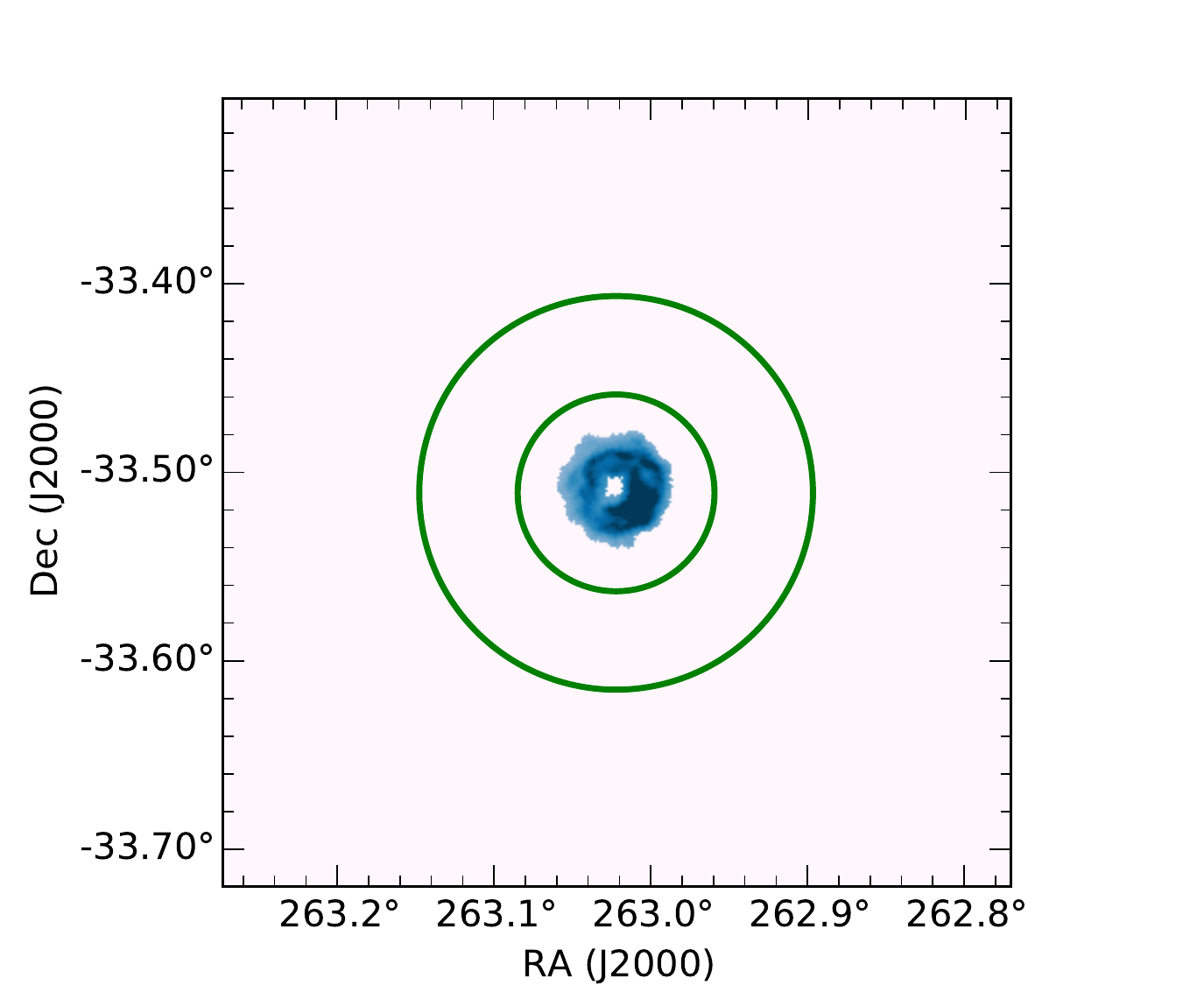}
\hspace{-0.8cm}
\includegraphics[width=0.275\textwidth]{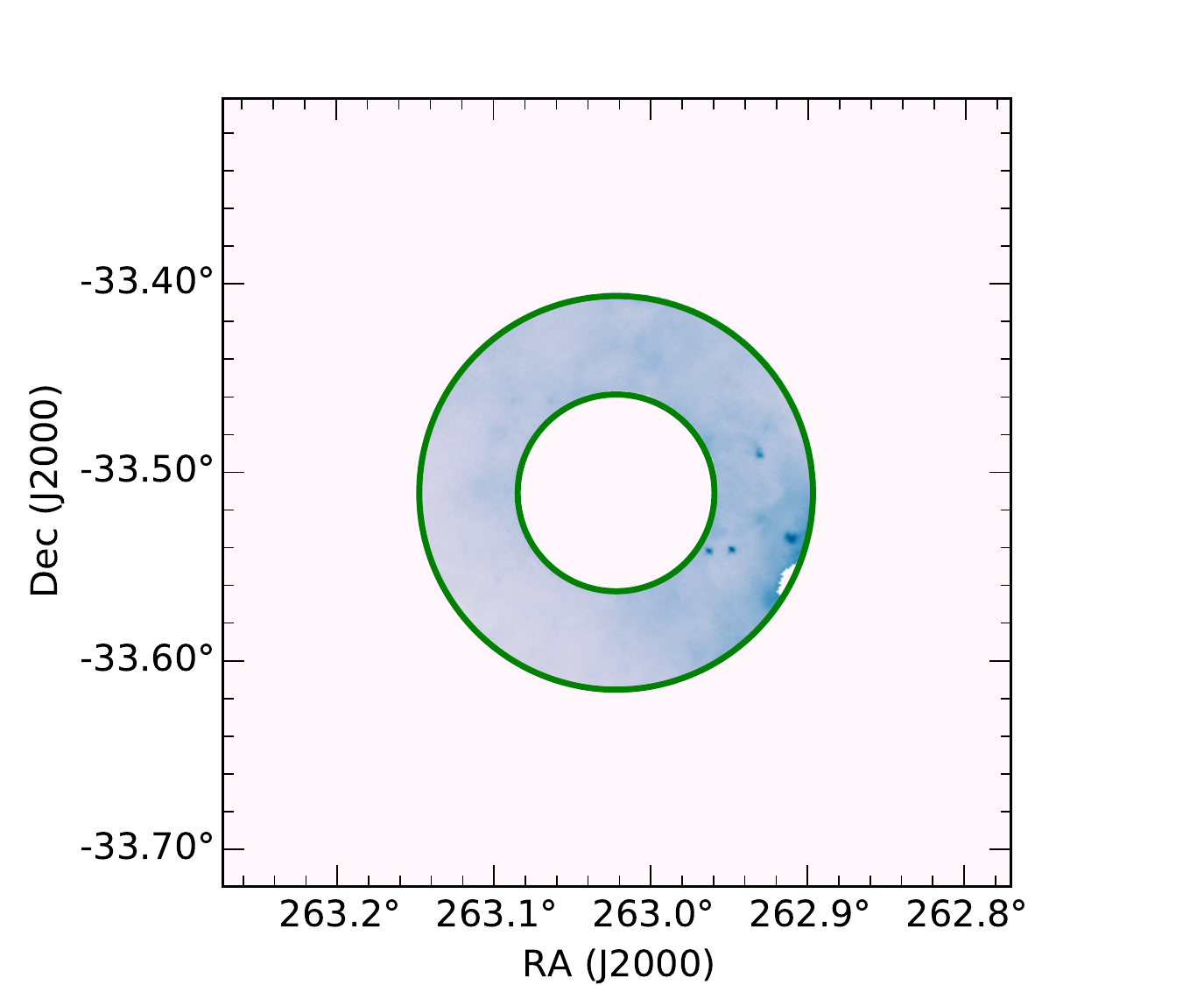}\\

\includegraphics[width=0.275\textwidth]{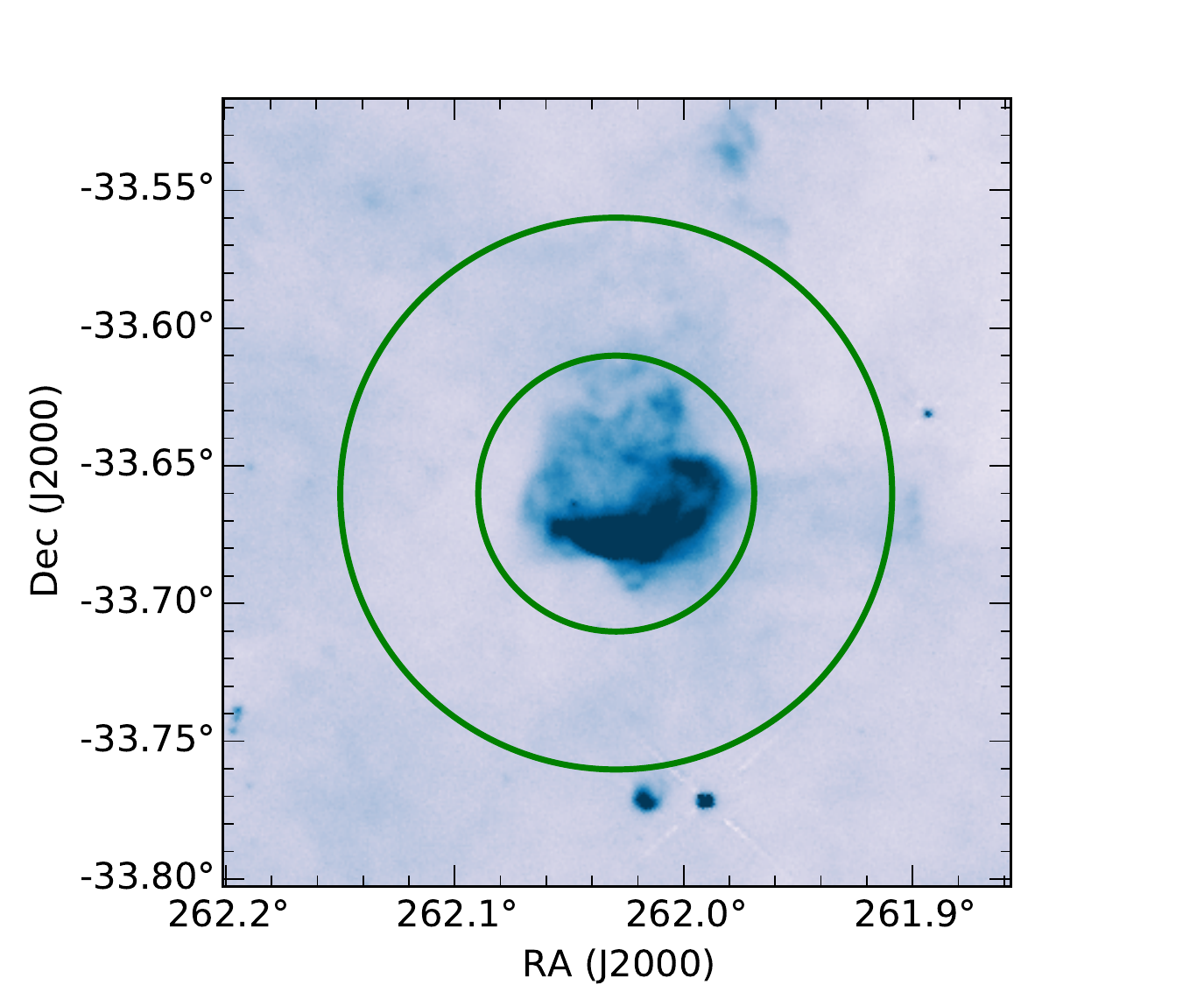}
\hspace{-0.8cm}
\includegraphics[width=0.275\textwidth]{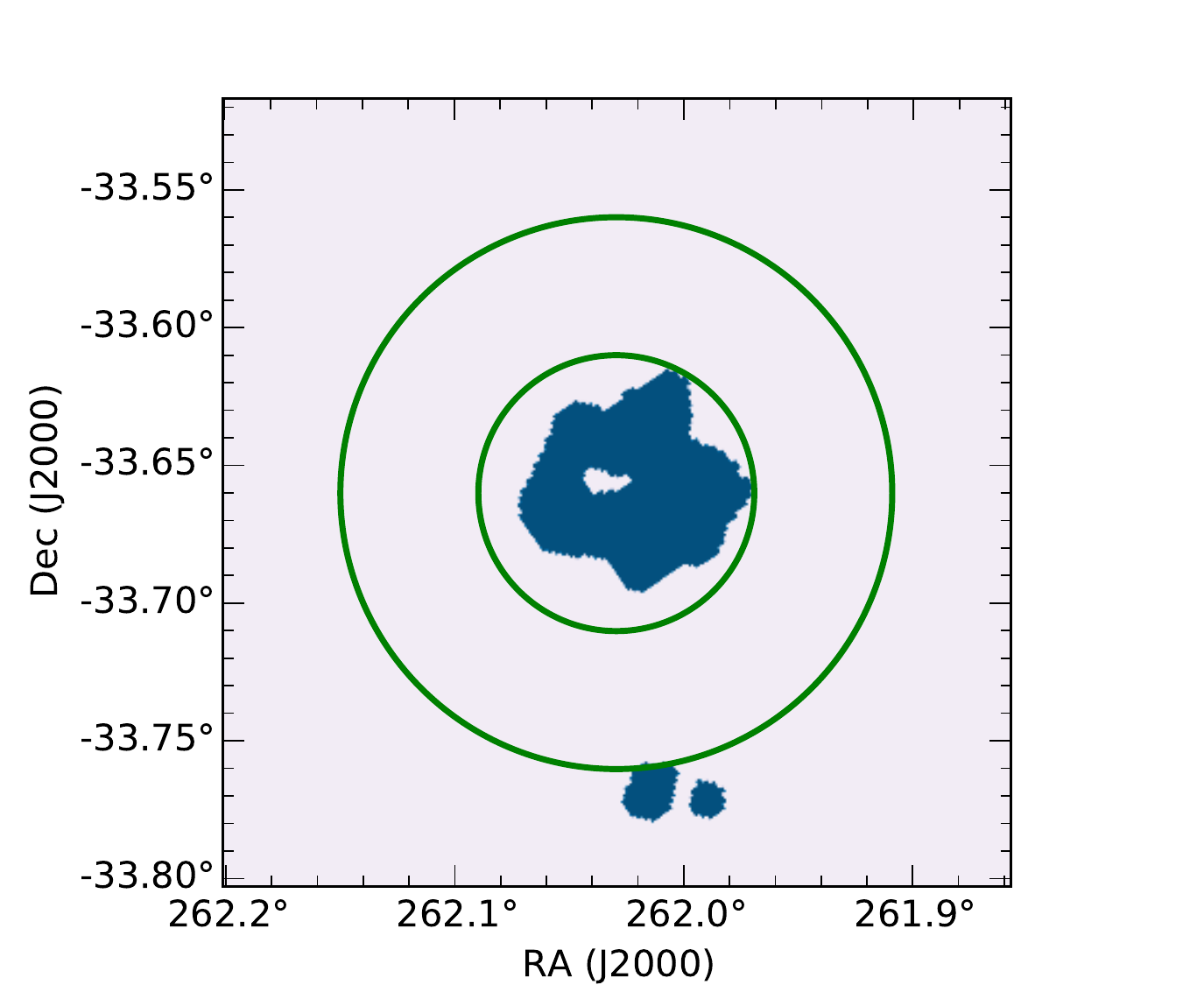}
\hspace{-0.8cm}
\includegraphics[width=0.275\textwidth]{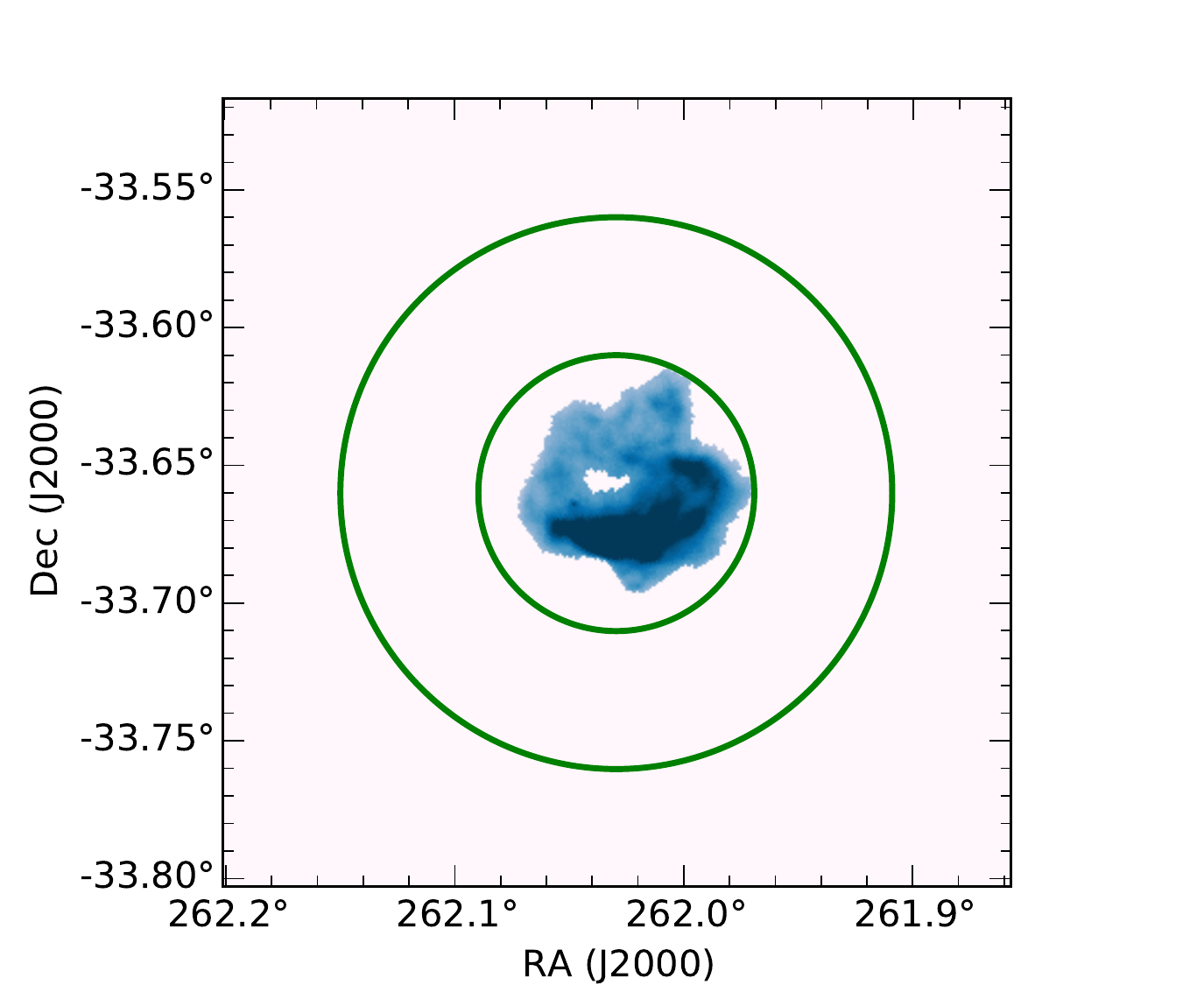}
\hspace{-0.8cm}
\includegraphics[width=0.275\textwidth]{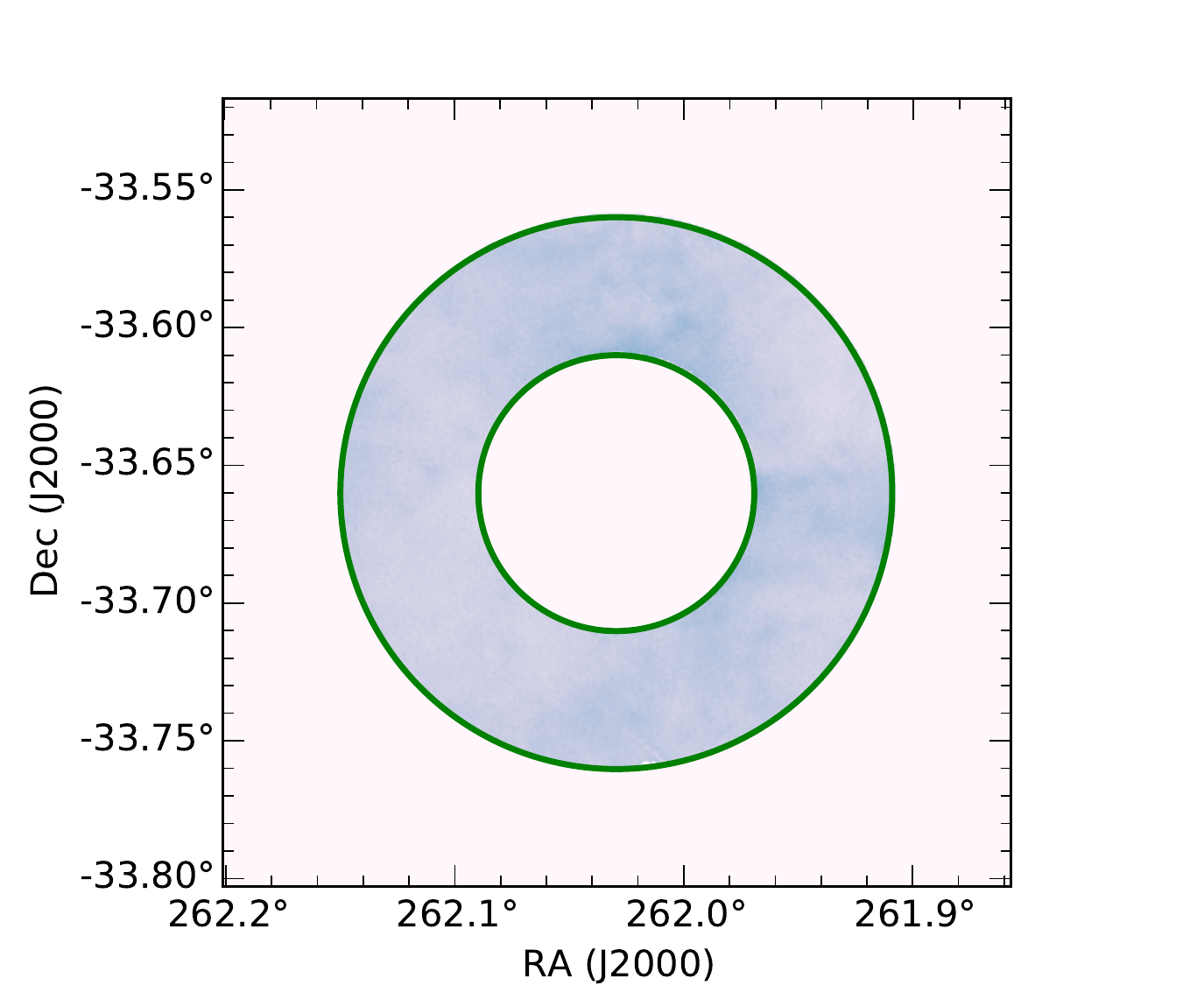}\\

\includegraphics[width=0.275\textwidth]{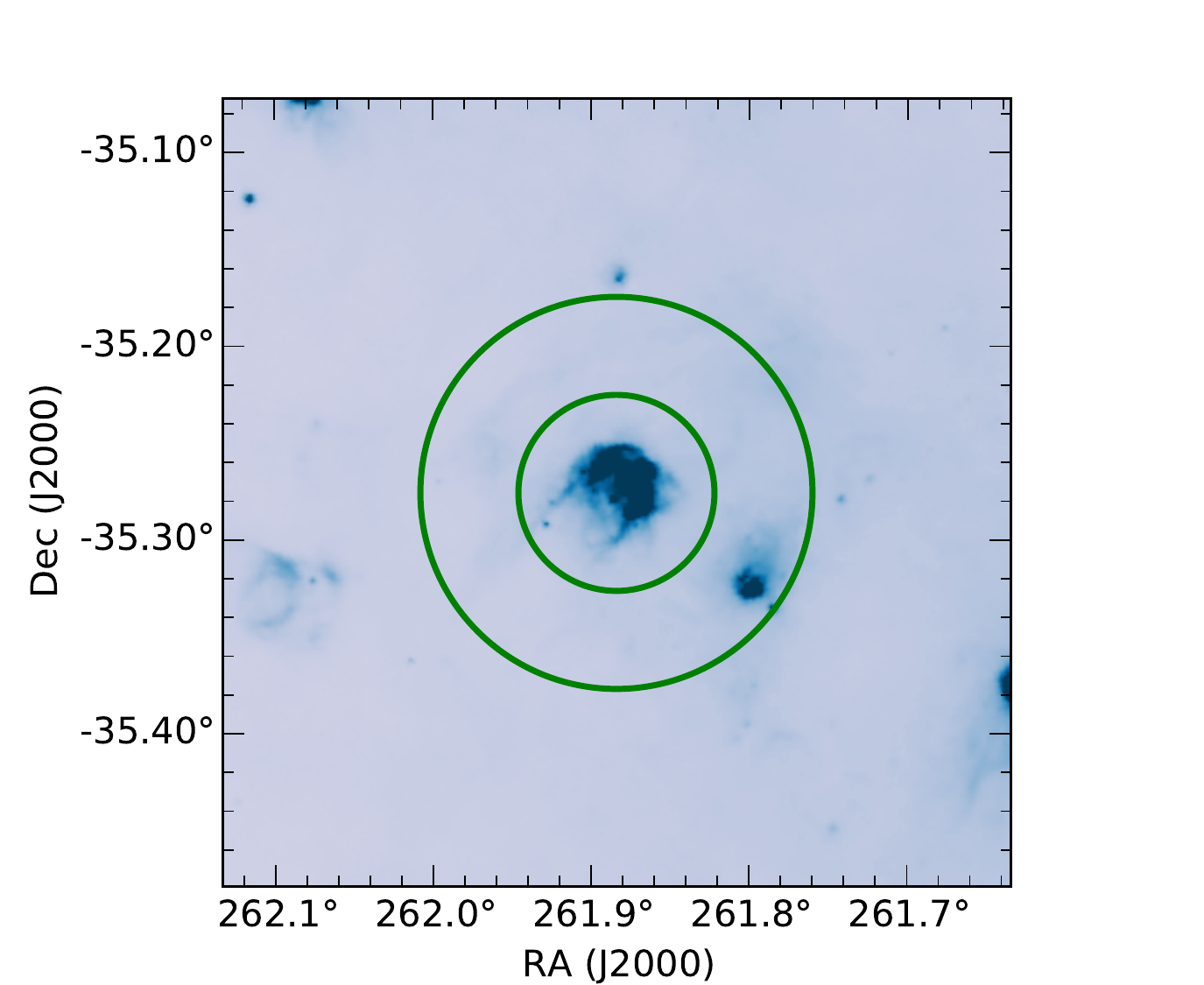}
\hspace{-0.8cm}
\includegraphics[width=0.275\textwidth]{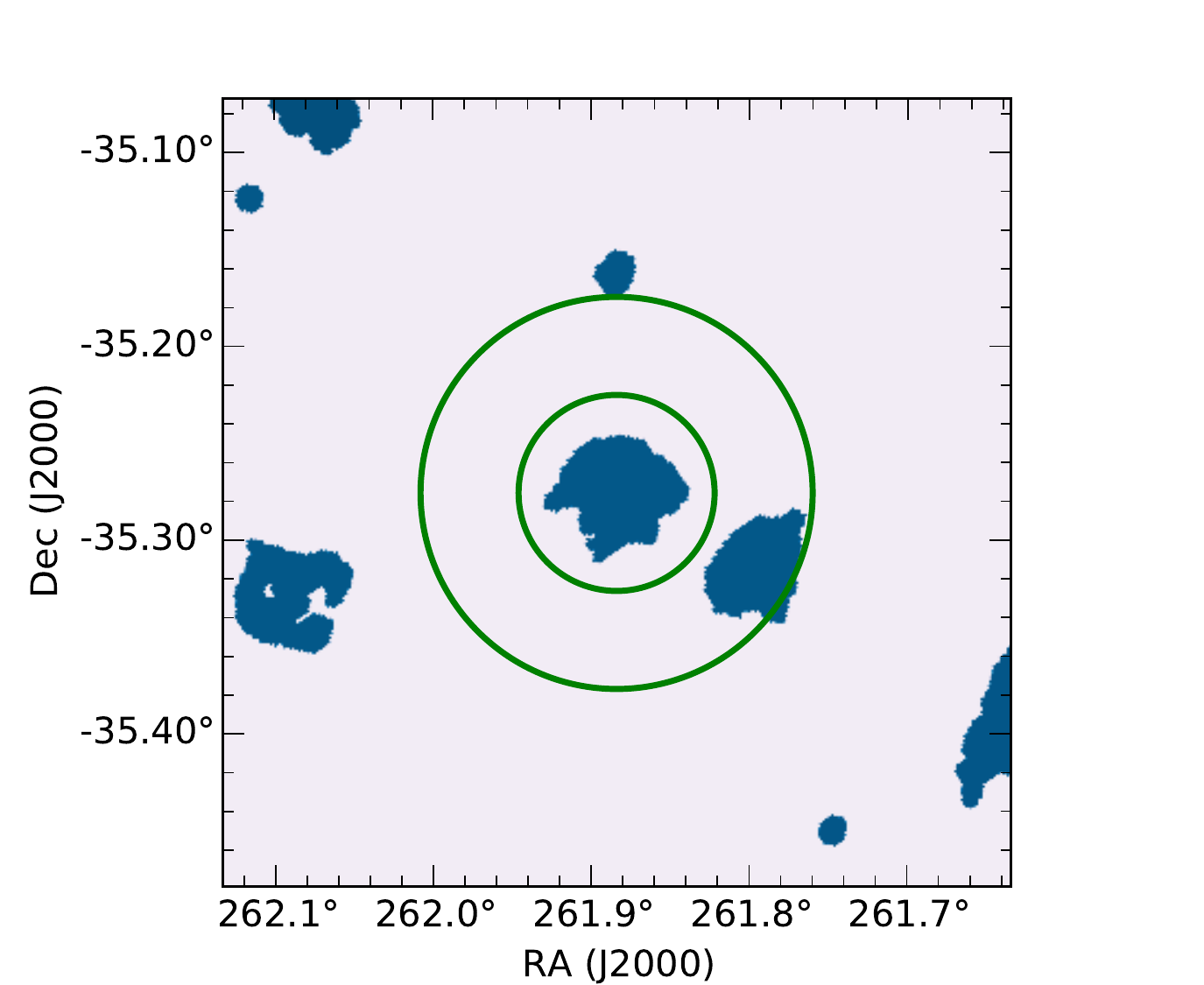}
\hspace{-0.8cm}
\includegraphics[width=0.275\textwidth]{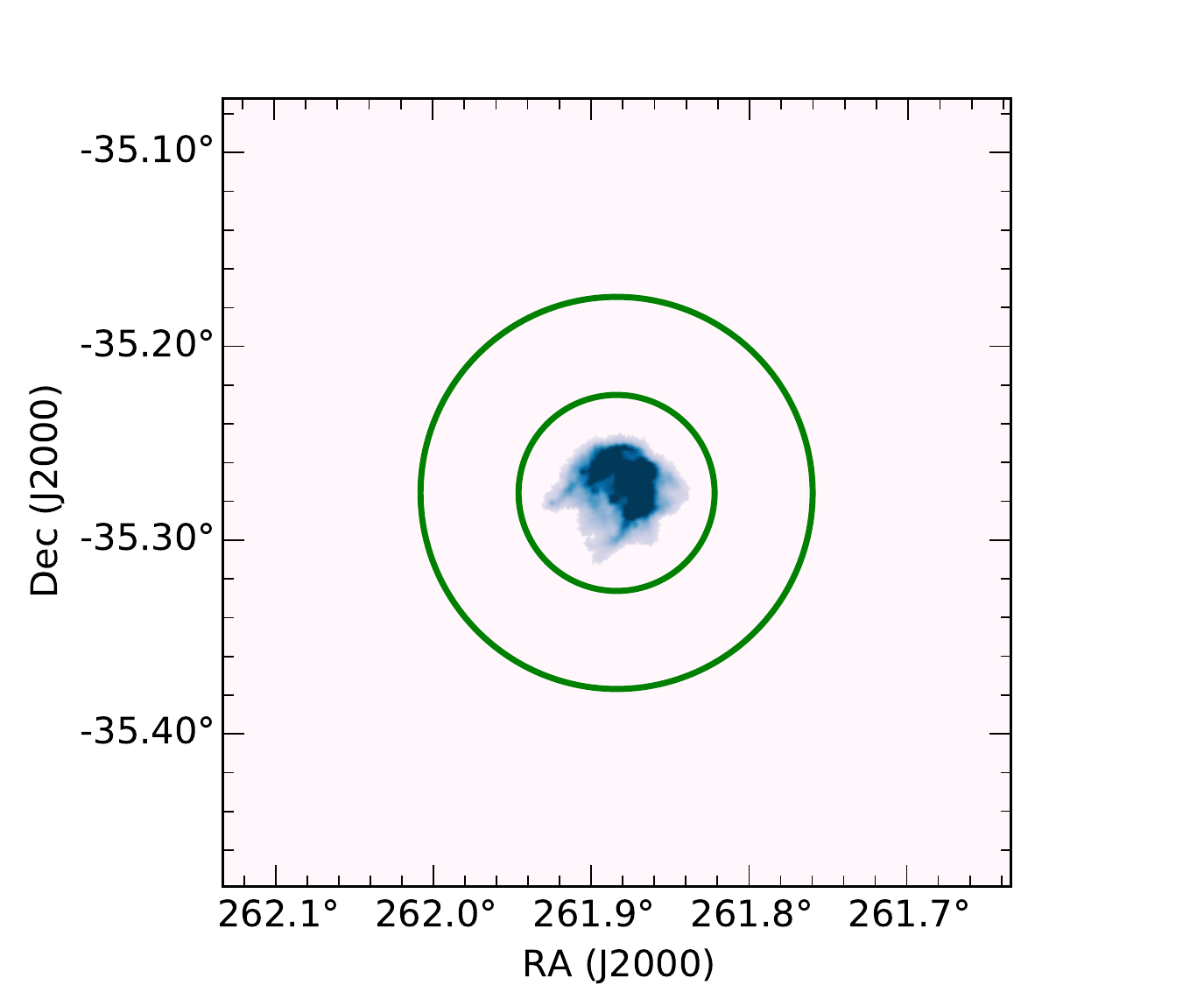}
\hspace{-0.8cm}
\includegraphics[width=0.275\textwidth]{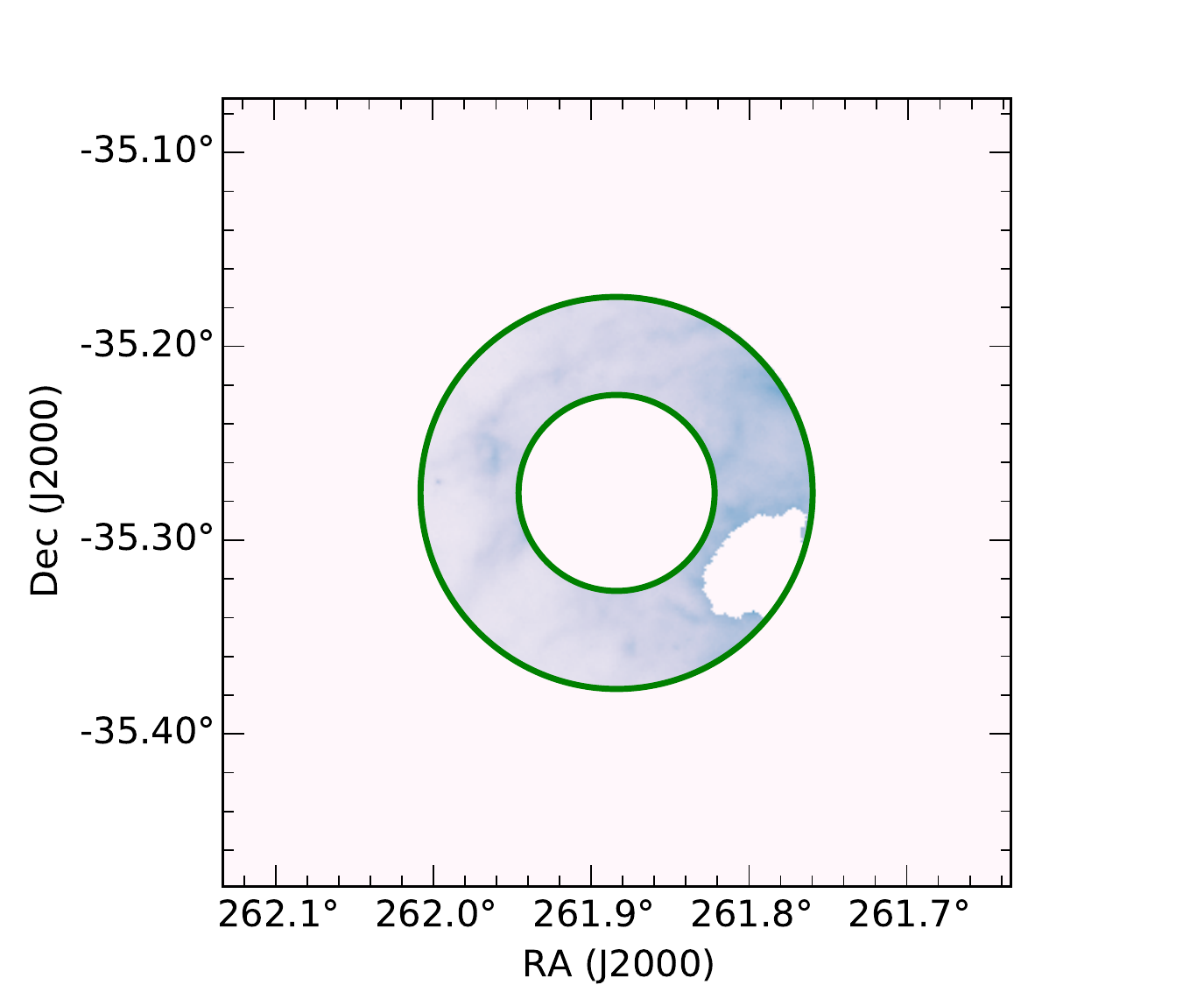}
 \caption{Examples of the application of the two photometric methods on different bubbles, namely 1G354588+000038 ({\it 1st row}), 1G354008+006116 ({\it 2nd row})
 and 1G352598-001860 ({\it 3rd row}). \herschel\, images at 70\um\, have been shown for all the images and bubbles aperture and background region of radius $R_{ph}$ and $R_{bkg}$, respectively, defined with green circles. On the first column,  bubbles fields are shown, on the 2nd column the segmentation mask produced for the specific field and in 3rd and 4th column  the aperture and background region, respectively, obtained using such mask, as described in Sect.\,\ref{methods}. } \label{methods_fig}

\end{figure*}

\section{Bubble Sample Selection} \label{sample}
This paper considered the galactic bubble catalogue produced by \citet{Simpson} as a database of confirmed bubbles. 
The catalogue consists of  5106 bubbles that have been identified by citizen scientists via visual inspection of the GLIMPSE and MIPSGAL infrared images, 
acquired at 8\um\, and 24\um, respectively\footnote{http://www.milkywayproject.org}. 
This data set was made by volunteers marking regions of images where bubbles were located. They drew a circular annulus around bubble features and this was scaled in size and stretched into an elliptical annulus resembling the prominent features of bubbles. The identified bubbles have been further split into two groups: 3744 {\it large}-bubbles, drawn by users as ellipses, and 1362 {\it small}-bubbles, which were too small to be drawn around in detail but can be still identified.
 The catalogue lists  the centroid position and radius for each bubble, averaged over at least five individual usersÕ drawings (see Sect. 3.1 in \citealt{Simpson} for details).
For {\it large}-bubbles, the catalogue also reports  parameters as the inner major and minor axis, outer major diameter, eccentricity and position angle,
 while the effective radius and thickness values are calculated from geometric means of such diameters (as given in Eq.\,1 by \citealt{Simpson}). 
 In particular, since the bubbles were identified on GLIMPSE and MIPSGAL images,
they are distributed exclusively over the inner Galactic plane ($\mid l\mid \le$ 65\degree). \\
As a first step,  we selected only those bubbles located in fields observed by the Hi-GAL survey, obtaining a sample of 4988 bubbles over the original 5106, due to the fact that Hi-GAL  covers the Galactic latitudes $|b|\le$ 1\degree\, at all Galactic longitudes while {\it Spitzer} extends at least up to $|b| =$ 2\degree\, towards the Galactic center region.\\
At the same time, we found that a large number of bubbles were projected over each other
and therefore could contaminate  the final flux estimation. 
Thus we decided to clean the sample and create a {\it golden sample}. Firstly we defined a circular region centered on each bubble centroid  and 
with a radius equal to the outer diameter or to the radius given by \citet{Simpson} in the case of
a {\it large}-  or a {\it small}-bubble, respectively. Then we selected those bubbles whose circular region is not overlapping with that 
of any other bubble.  
We added an extra constraint for the cases where a {\it small} bubble was overlapping a {\it large} one.
 Indeed, having taken for {\it large}-bubbles the radius equal
to the outer diameter  in order to guarantee including the totality of the emission at the different wavelength ranges,
we risk to remove  {\it small}-bubbles at their very border which are not contaminated. 
Therefore if the distance between the two  centroids was larger than the smallest of the two radii (corresponding in most cases to the {\it small}-bubble radius) then both bubbles 
were kept separated and included in the golden sample. 
 In addition this enabled duplications of single bubbles present in both catalogues to be removed. 
 The selected final sample consisted  of a total of 1814 bubbles:   45\% of the $small$-bubbles from \citet{Simpson} were kept, whilst 33\% of the $large$-bubbles. Nonetheless, the $large$-bubbles still represent two thirds of the golden sample.
A catalogue for this  has been produced, listing for each bubble the relative Galactic coordinates (corresponding to their centroids) and their radius ($R_{cat}$),
which is either equal to the bubbles effective radius or to half of its outer diameter for the case of {\it small}- and {\it large-}bubbles respectively.


\section{Data description} \label{data}
The images from which we estimated the fluxes emitted by the bubbles of the golden sample are taken from  the \wise\, and Hi-GAL surveys.

\subsection{\wise\, image database}
\wise\,  \citep{wise} was a mission that mapped the
entire sky in four IR bands, namely 3.4\um , 4.6\um , 12\um\,  and 22\um\, (data used here was from the March 14, 2012 release).

In this work we used the 12\um\, and 22\um\, bands, since they trace similar dust components as that of GLIMPSE 8\um\, and 
MIPSGAL 24\um\, bands, respectively. However, the 12\um\, bandpass  is significantly broader than GLIMPSE's 8.0\um, collecting emissions from PAH features at 11.2\um, 12.7\um\, and 16.4\um\, \citep{PAH}. The PAH features at 7.7\um\, and 8.6\um\, also fall within the bandpass although at diminished sensitivity. 
The spatial resolutions in the two bands are  6\farcs 5, and 12\arcsec\, and the sensitivities are 1 mJy and 6 mJy, respectively. 
The \wise\, image data have units of DN, thus we used a DN- to-Jy conversion factor equal to 1.8326 $\times$ 10$^{-6}$ and 5.2269 $\times$ 10$^{-5}$ for the 12\um\, and 22\um\, bands, respectively (see \wise\, explanatory supplement\footnote{http://wise2.ipac.caltech.edu/docs/release/allsky/expsup/sec2\_ 3f.html}).

\subsection{\herschel\, image database}
The Hi-GAL survey was performed using the Photoconductor Array Camera and Spectrometer (PACS; \citealt{PACS}) and the Spectral and Photometric Imaging Receiver (SPIRE; \citealt{SPIRE}) instruments onboard the \herschel\, Space Observatory \citep{pilbratt}. Hi-GAL maps the Galactic plane (0\degree $\le$ l $\le$ 360\degree, $\mid b\mid \le$ 1\degree) in five wavebands, namely  70\um, 160\um, 250\um, 350\um\,  and
500\um, providing a well-sampled coverage of the frequency range where the spectral energy distribution of cold dust peaks. 
The spatial resolutions of these images are 6\farcs7, 11\arcsec, 18\arcsec, 25\arcsec, and 37\arcsec, respectively. 
Images were reduced using the ROMAGAL data-processing code, for both PACS and SPIRE  data (see \citealt{Traficante} and \citealt{Molinari16}  for details).

\section{Bubble Flux Measurements} \label{methods}
We estimated the flux coming from each of the 1814 bubbles belonging to the golden sample using two different methods:
the first method is a classical aperture photometry, in which we measured the flux within a circular area
centered on the source; the second one uses the same aperture but selects the flux coming from the bubble 
using a segmentation mask, which removes any pixel coming from nearby contaminating sources and from the background. 
Before applying such methods, we prepared our sample images, as described in the following subsection.

\subsection{Dataset Preparation}\label{datasetpreparation}
Using the selected golden sample catalogue, sources were cut out of the  \wise\, and \herschel\, image datasets using a bounding box that was centered on the bubbles centroid and whose width was equal to 10 times $R_{cat}$. Each map (cut out) was projected onto the 
 N-E equatorial direction and scaled to the pixel scale of a reference image using the Montage toolkit\footnote{http://montage.ipac.caltech.edu/}. 
 The image taken as reference was the one with the smallest pixel scale, i.e.  the
\wise\, image at 12\um\, (1.37\arcsec/pix).
This allowed seven images to be produced for each source with an identical pixel grid.


\begin{figure*}
\includegraphics[width=0.485\textwidth]{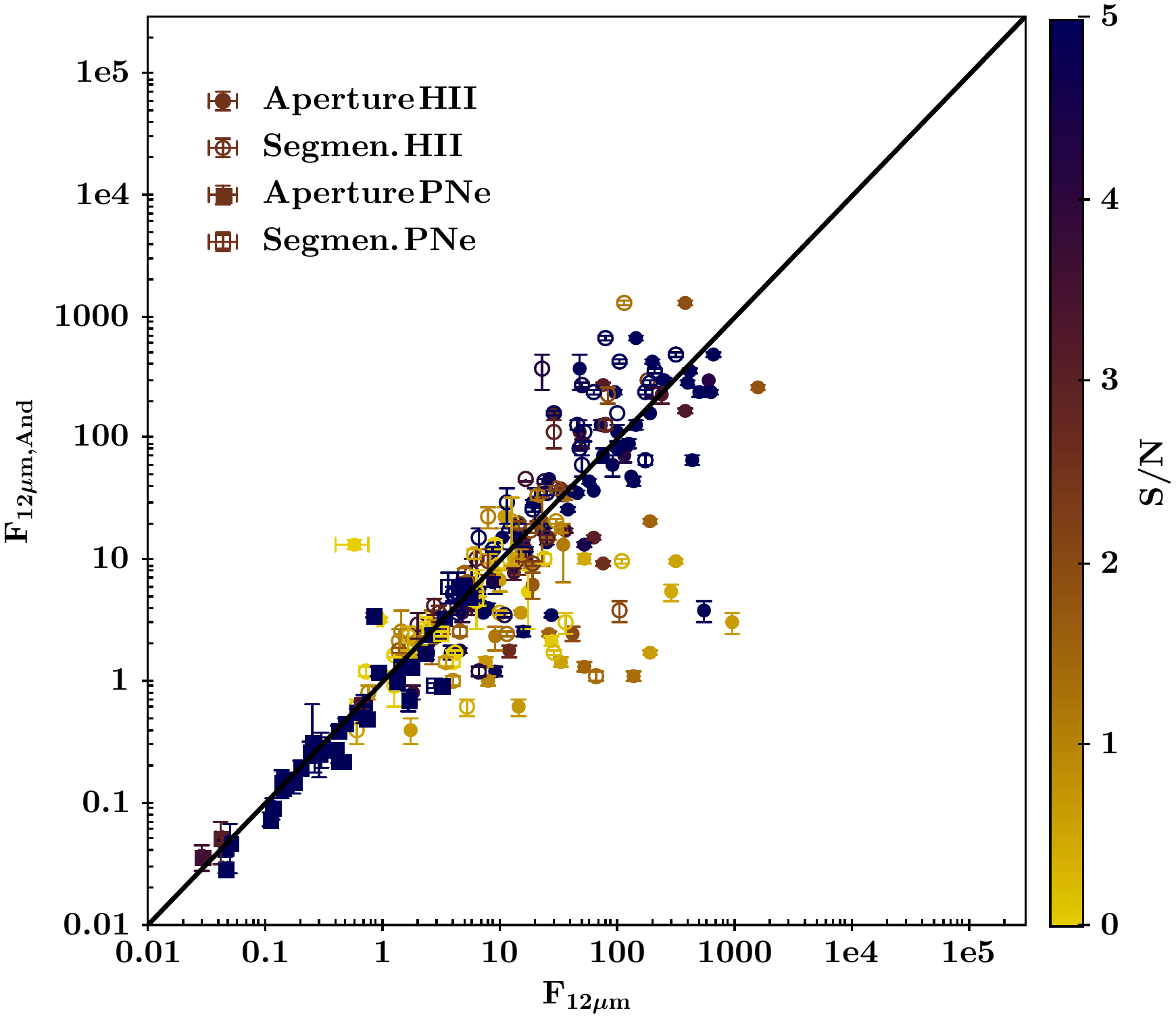}
\includegraphics[width=0.485\textwidth]{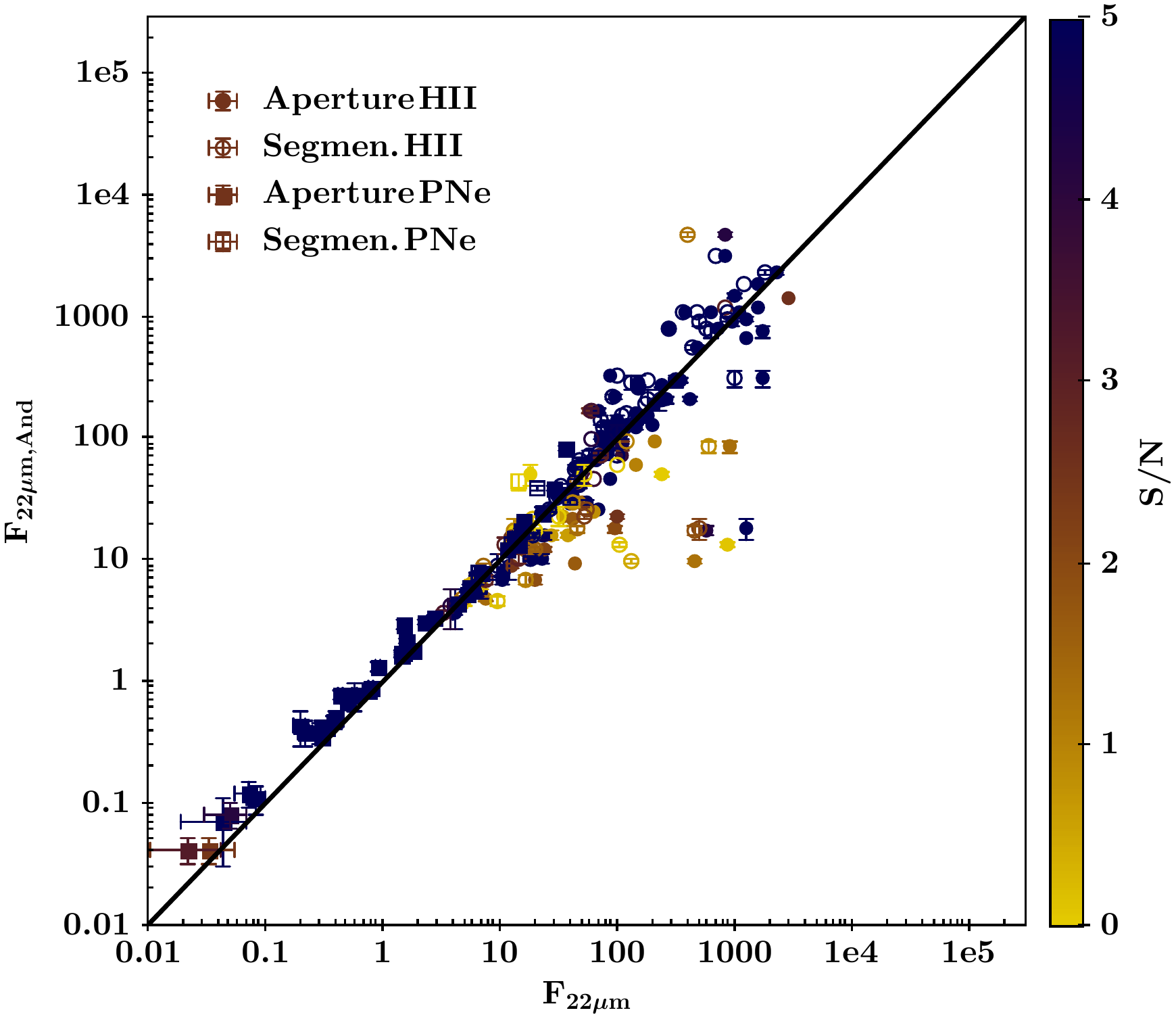}
\includegraphics[width=0.485\textwidth]{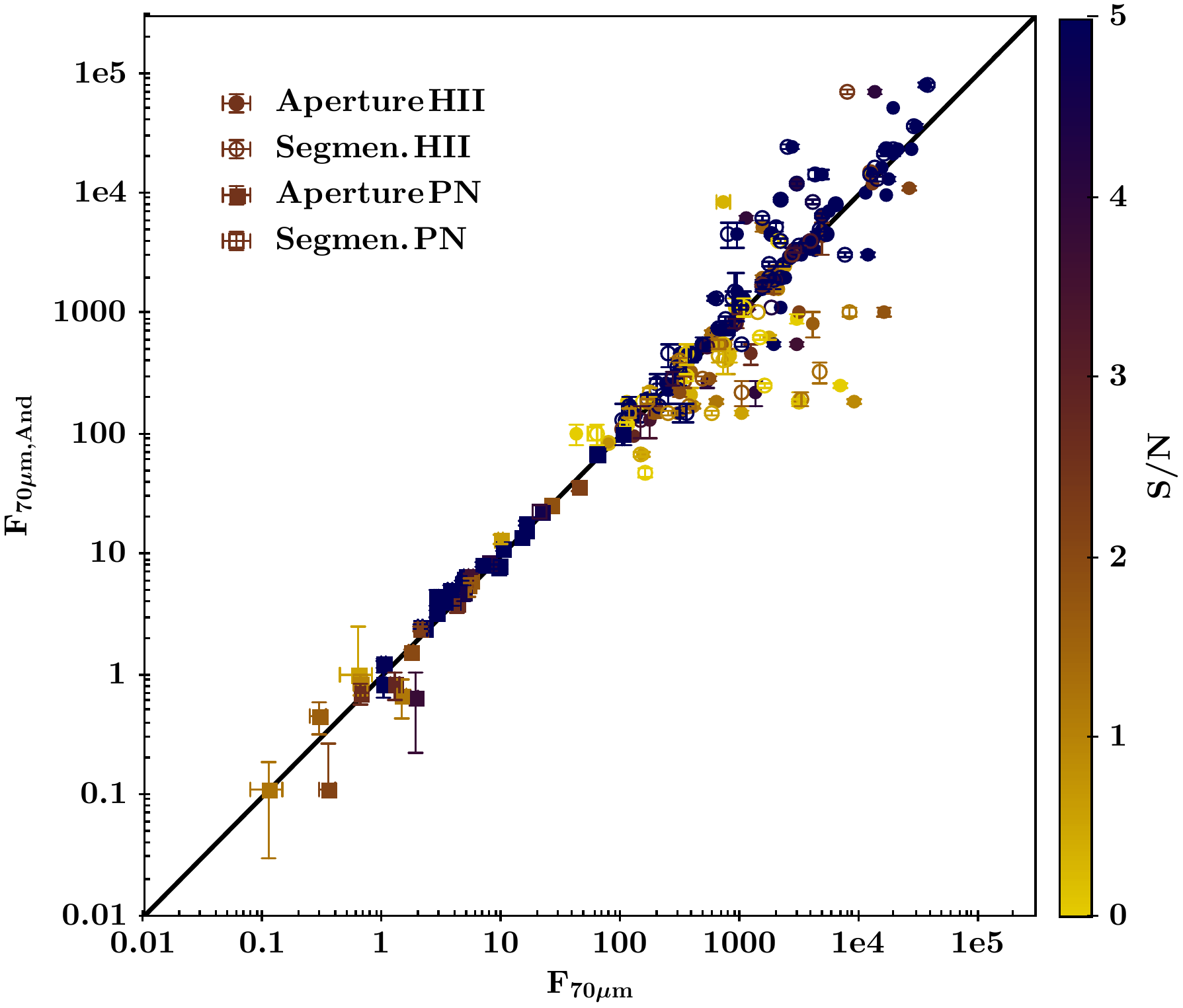}
\includegraphics[width=0.485\textwidth]{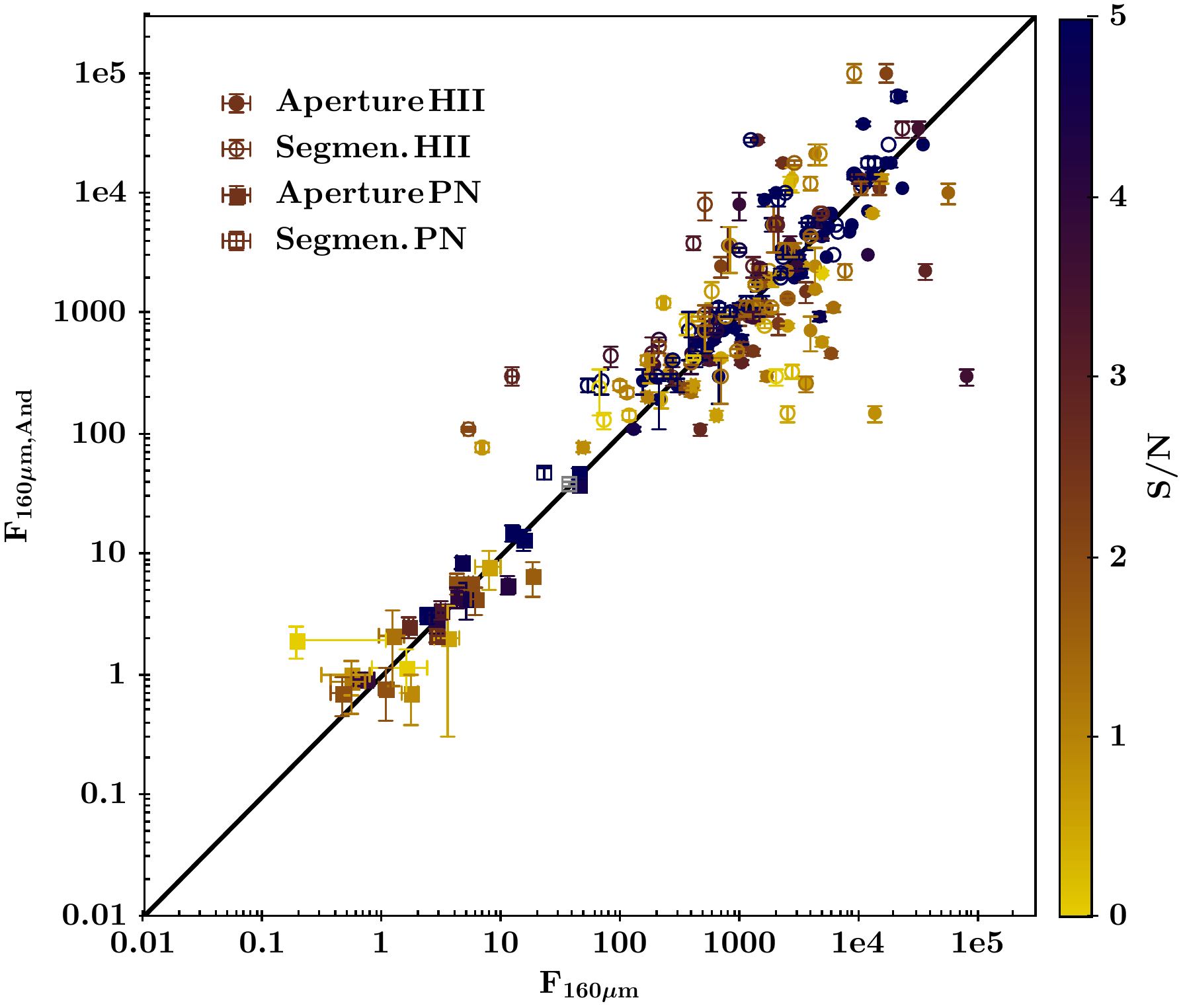}
 \caption{Comparison between photometry obtained with our automated methods and A12 results. Fluxes from \hiirs\, are reported with circles and PNe with squares. For aperture photometry filled dots are used, while dots for the segmentation photometry are empty. Fluxes are given in Jy. The color scale indicates the S/N characteristic of each of our measurements. } \label{Anderson_figure}
\end{figure*}

\begin{figure*}
\includegraphics[width=0.485\textwidth]{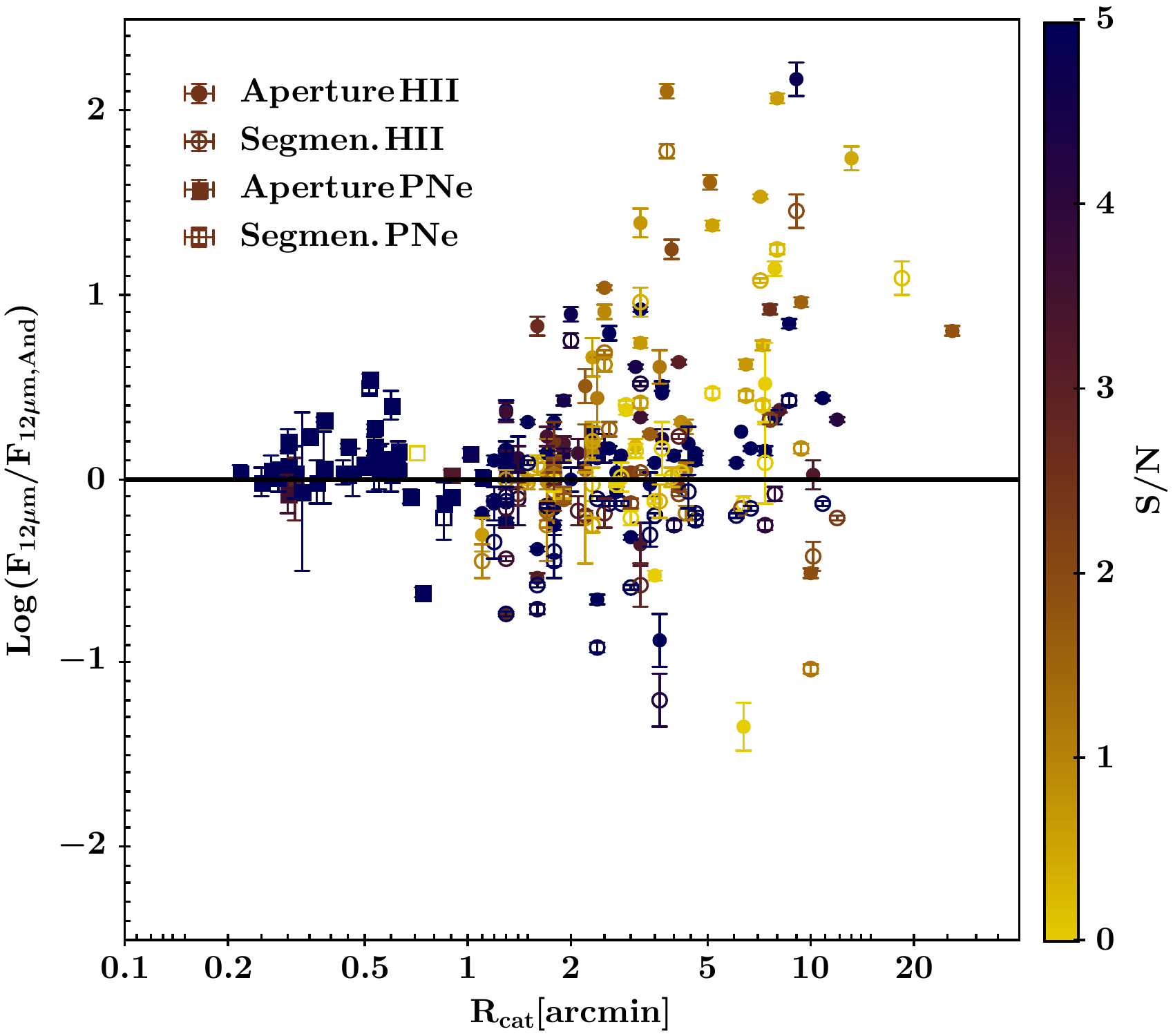}
\includegraphics[width=0.485\textwidth]{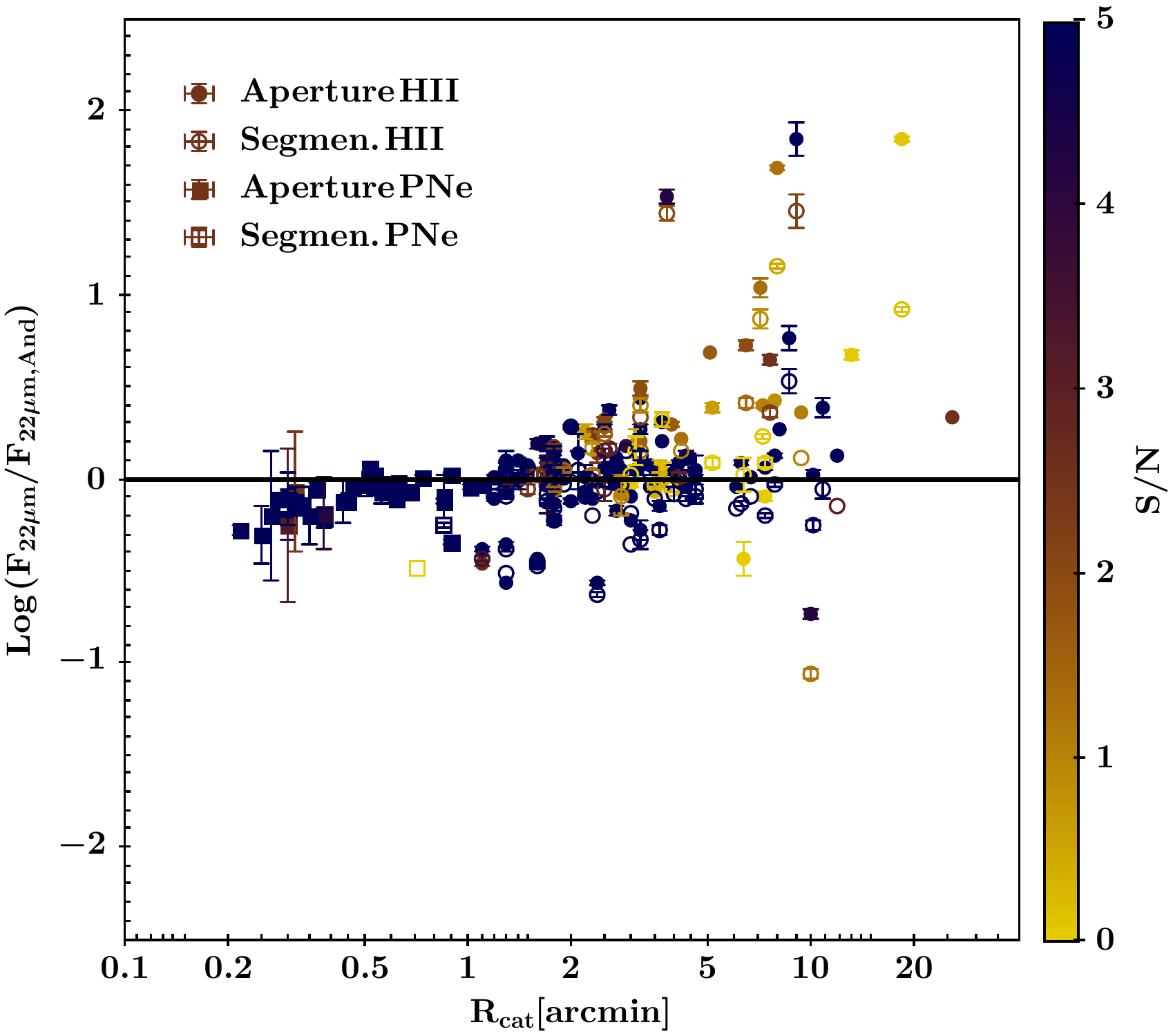}
\includegraphics[width=0.485\textwidth]{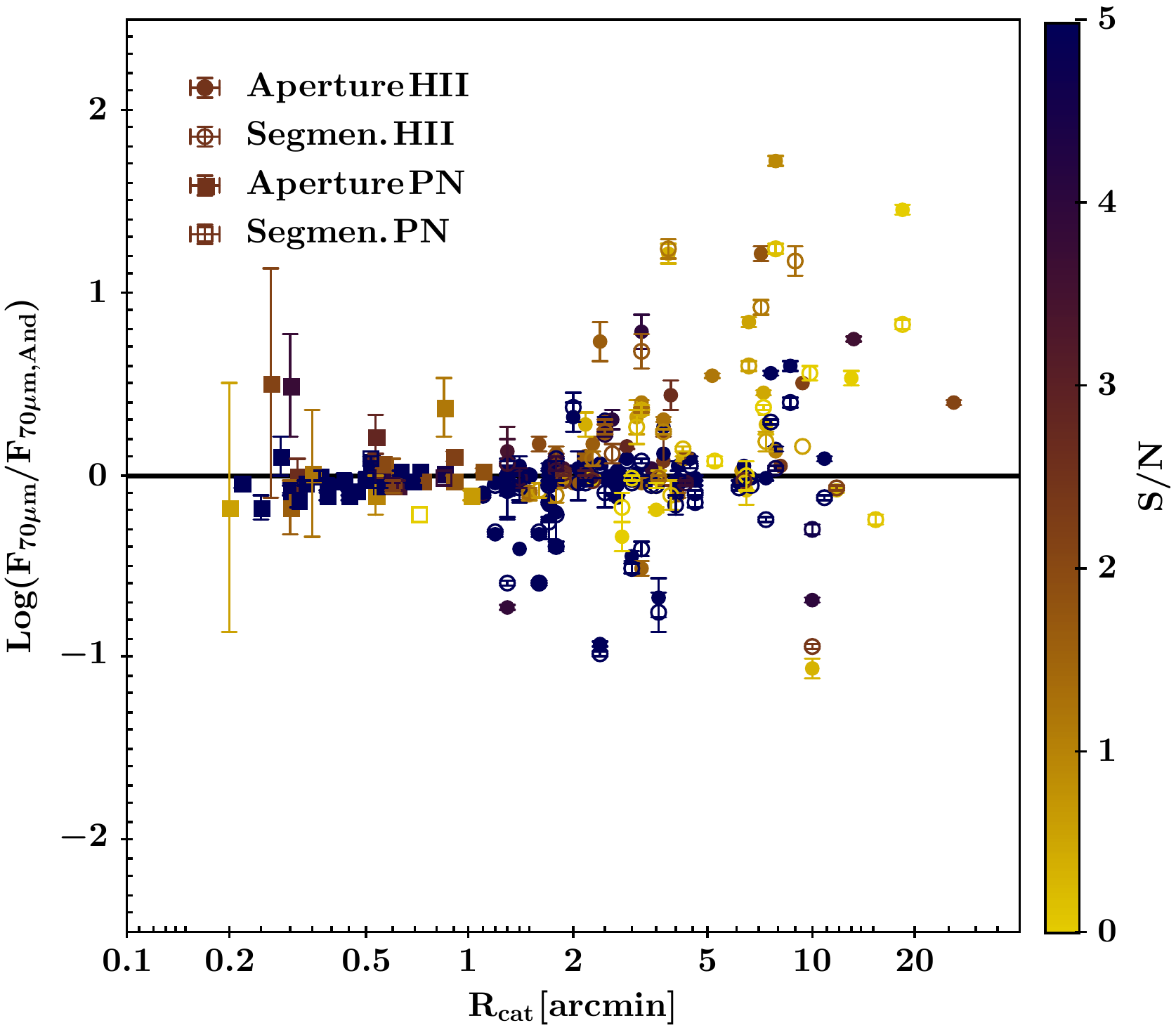}
\includegraphics[width=0.485\textwidth]{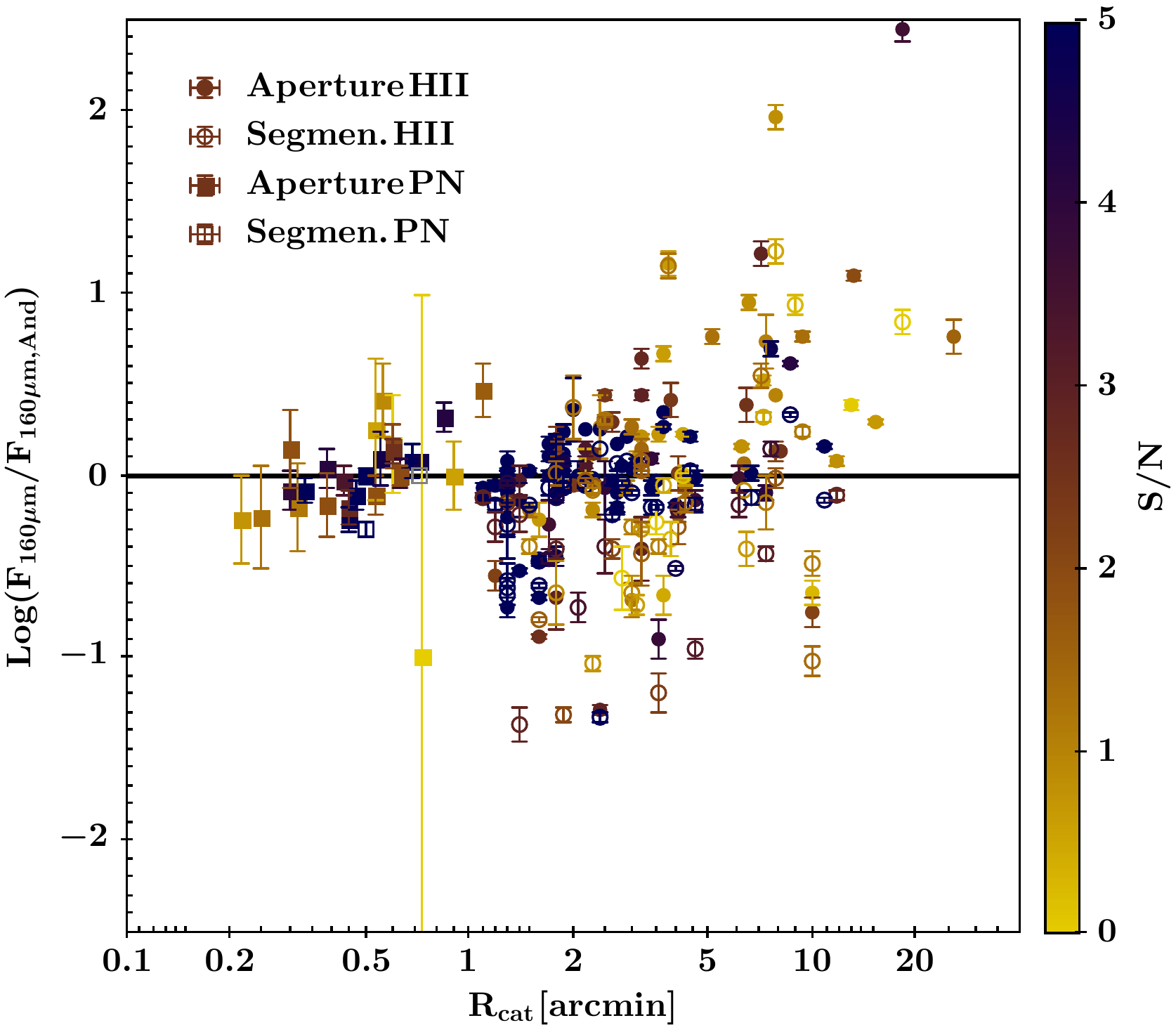}
 \caption{Distribution of flux differences between our methods and A12 measurements as function of the angular extension of the bubble.  Fluxes from \hiirs\, are reported with circles and PNe with squares. For aperture photometry filled dots are used, while dots for the segmentation photometry are empty. Radii are given in arcminutes.The color scale indicates the S/N characteristic of each of our measurements. } \label{Anderson_radius}
\end{figure*}


\subsection{Photometric Methods}\label{photometricmethods}

\subsubsection{ Aperture Photometry} 
We estimated the flux coming from each source in different bands
measuring the flux falling into a circular area centred on the bubble
centroid coordinates given by  \citet{Simpson},
with a radius ({\it R$_{ph}$}) chosen equal to 
\begin{equation}
R_{ph}=\sqrt{(2R_{cat})^2+(FWHM)^2}
\end{equation}
where FWHM  is the beam size for each bandpass, in order to include all the flux coming from the bubble in different bands. 
The local background level has been estimated just outside the aperture, over an annular area between
$R_{ph}$ and $ 2R_{ph}$ and is equal to the sigma-clipped mean  (2-$\sigma$ level).
This was chosen to remove  very bright compact objects or spurious spikes. 
The average background level has then been subtracted from each pixel value within the source aperture,
before computing the total aperture flux of the bubble. 

\noindent{\bf Segmentation Photometry} -- This method made use of a ``segmentation mask'' to select the flux coming from each bubble.  In image processing, segmentation  is the process of partitioning of a digital image into its component parts and that was used here to define bubble regions. 
To enable the segmentation of bubbles,  a localised {\it active contours} algorithm  \citep{Lankton} was used that also incorporated gradient information via {\it Magnetostatic forces} \citep{Xie}. 
In our paper's approach, many of the difficulties associated with the use of localised contours were overcome by adaptively selecting the appropriate kernel sizes that are required by this algorithm (further details in Appendix\,A).
In other words, this paper's active contour algorithm finds bright objects that have large gradients. 
In images where there are many high gradient regions, the contour could grow wildly around the image, at least without human intervention, which is not feasible in this case. 
This was the case for  \herschel\,  images acquired at $\ge$160\um\, as they had high background contamination. 
The same is true for \wise\,  images at 12\um\, as they contained numerous compact field objects. Thus we decided to optimize the method on the 70\um\, images,  since bubble contours at this band
generally include those at shorter wavelengths and, at the same time, trace dust distribution better than longer ones. \\
Thus segmentation masks have been obtained from original \herschel\, 70\um\, images, and consequently they have the same pixel scale.
  Since bubbles images were resampled to the \wise\, 12\um\, pixel scale (as described in section\,\ref{datasetpreparation}),
  we also performed the resampling of the segmentation mask to make them match. 
  Moreover, in order to take into account the instrumental effect on the bubbles contours in
  images at lower resolution than 70\um\, images, 
  we convolved the mask with a Gaussian profile to correct the beam size differences, before applying it to the corresponding image. 
   This smoothed the mask borders and mimicked the instrumental effect,
  assigning a fractional value between 0 and 1 to each pixel, which was finally replaced with 1, to produce the new mask. \\ 
By using the segmentation map, we mask anything falling in the aperture  {\it R$_{ph}$} 
but not expected to be part of the bubble. Any other bright segmented source falling in the background annulus was also removed,
 before estimating the average background level value. As a consequence a shallower clipping level (3-$\sigma$) than in the aperture method was used. 
  The average background level  has been subtracted from each pixel value in the aperture region masked as the bubble, 
  and then summed to estimate the flux of the bubble (``segmentation'' flux).
  Comparing the background average level estimated in this way with that 
of the aperture photometry, we found that they are in agreement within 5\% for 85\% of the bubbles. 
Aperture background level turned out to be higher  than the segmentation one in  around 8\% of the cases, most frequently when the presence of extended emission in the background region increased the background sigma value and thus made the sigma-clipping less effective. 
In other cases (around 7\%), the segmentation did not work correctly in masking bright nearby sources, causing a higher average background level. \\

Examples of the application of the two photometric methods are shown in Figure\,\ref{methods_fig}.
We did not provide flux measurements at a given wavelength for those bubbles that exceed 10\% of saturated pixels within  $R_{ph}$,
since with such high fraction of 'NaN' values the bubble flux estimate would not be reliable. In any case, they represent a very small fraction of the total sample 
(24/1814 for \wise\, images and 4/1814  within  \herschel\, images acquired at 250\um\,  and 500\um\, and 2/1814 for  \herschel\, images taken at 350\um).\\

For both methods, the uncertainty on the flux is calculated as the sum in quadrature of the background and source counts error over the
total number (N) of pixels within {\it R$_{ph}$}  over which the flux was calculated. 
Source counts error is equal to the photon noise in the case of \wise\, images, or to the 
calibration uncertainties for the \herschel\, maps due to the uncertainties in the theoretical models of the
SED of the calibrators and equal to 5\% of the flux for PACS images \citep{Balog} and to 4\% 
for those from SPIRE \citep{Bendo}. Background error is given by the sum in quadrature of the photon noise/calibration error
within the background annulus and the  background standard deviation. \\
Additionally, along with the measured total flux and the relative uncertainty, we provide a value, reported as S/N,
which corresponds to the ratio between the median of the background subtracted pixel values within $R_{ph}$ and the background standard deviation.
Such values can be used as an indicator of the significance of the detection with respect to the background level.   

Finally, we assumed that the contamination from bright compact sources falling in the aperture is negligible,
based on the results of previous work, i.e. A12, and of a statistical analysis conducted on this paper data.
A12 found that as \hiirs  are in general much brighter than any point source within the aperture, the removal
of such point sources has a minimal impact on the derived fluxes.
In the same way, for less extended bubbles, such as PNe, 
the small characteristic angular size makes it unlikely that there is a spatial coincidence with point sources.
Therefore, there is likely to be no consequent contamination of the bubble flux. 
Possible contribution by compact objects  to the bubbles  flux measured in this paper was also checked.
Assuming that compact objects are mainly stars, we estimated their flux contribution at 12\um, considering that
 stellar spectral energy distribution is usually the strongest at this wavelength, relative to others used in this paper.
From the AllWISE Source Catalogue\footnote{http:\/\/irsa.ipac.caltech.edu/cgi-bin/Gator/nph-scan?submit=Select\&projshort=WISE},
we selected the compact objects consistent with a single Point Spread Function and with no saturated pixels, located in the same sky region of the golden sample bubbles. 
We cross-matched the two catalogues and found the number of compact objects included in the aperture radius of each bubble, 
calculating their total flux and contribution to the bubble aperture flux.
We found that for {\it large}-bubbles the fraction of bubbles with a flux contamination higher than 10\% is around 25\% (298/1181),
while for {\it small}-bubbles such fraction is around 15\%. This confirms the marginal contribution of compact objects to the bubbles measured flux already at 12\um,
which is expected to be the most affected band by the contaminants.
It is worth to stress that  3\% of the {\it small}-bubbles has a contamination higher than 50\% most likely due to the coincidence 
of the compact object with the bubble itself. Most importantly, no compact objects are included in the aperture radii of one third of the {\it small}-bubbles.


\begin{figure*}
\includegraphics[width=0.488\textwidth]{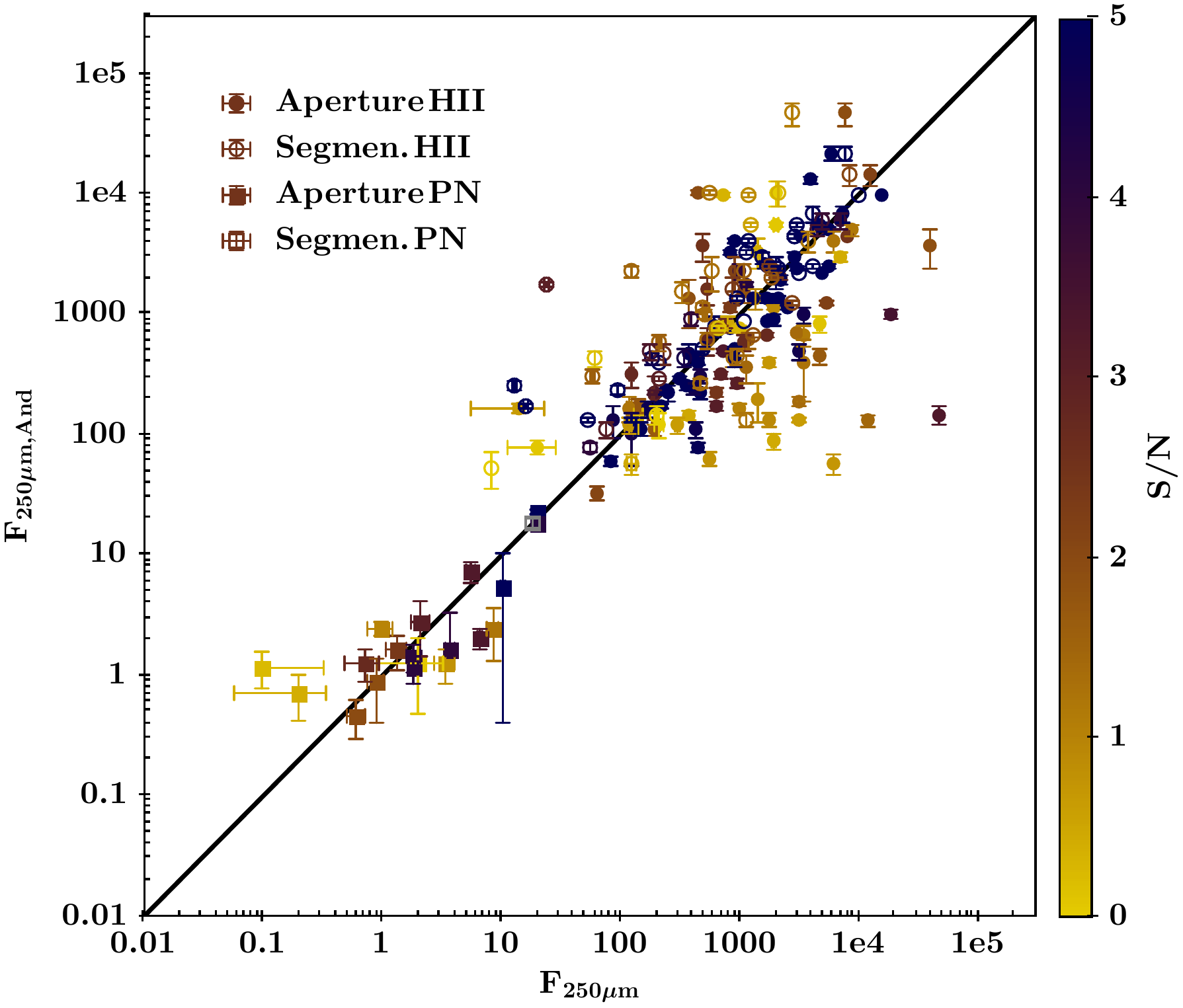}
\includegraphics[width=0.475\textwidth]{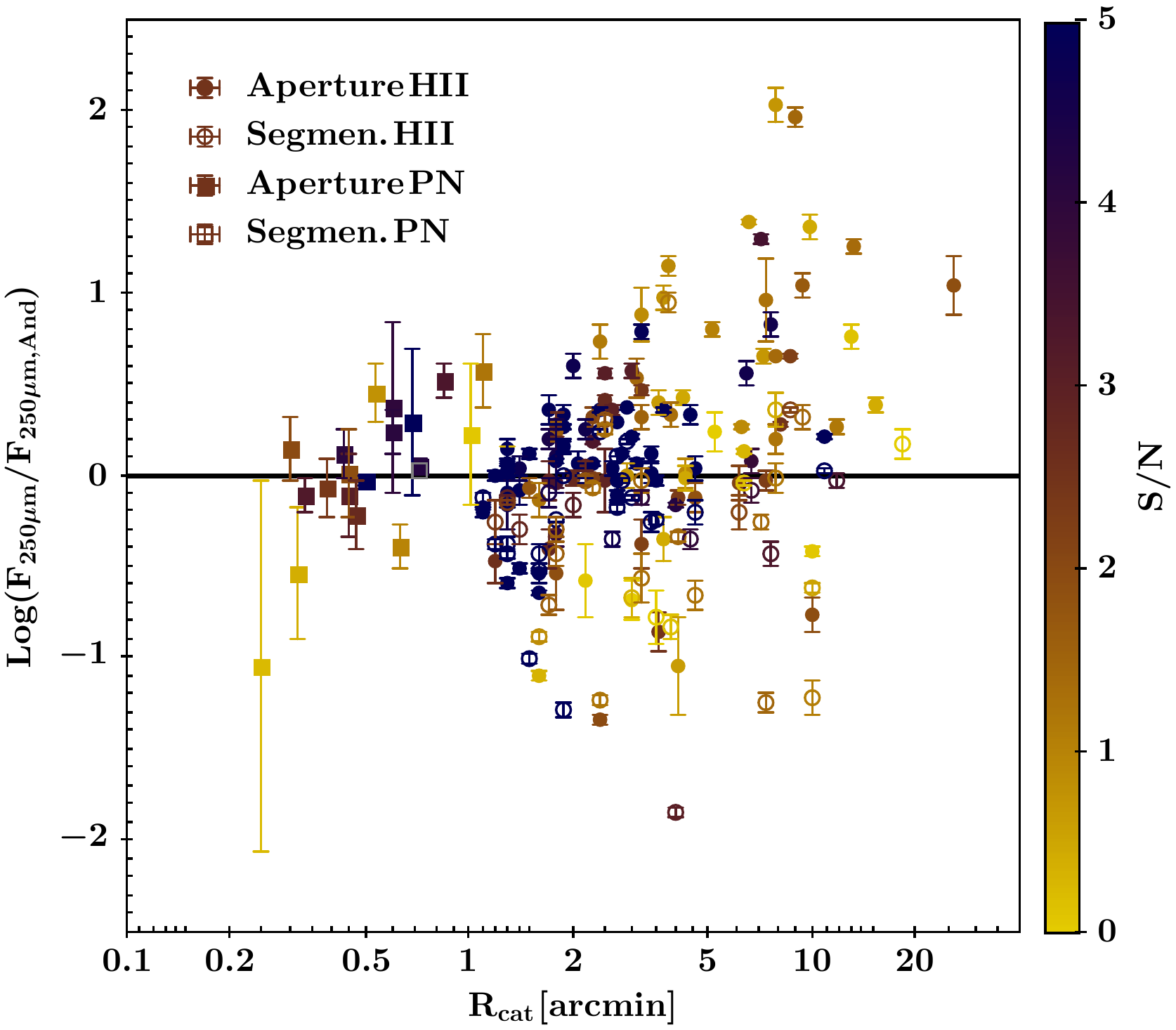}
\includegraphics[width=0.488\textwidth]{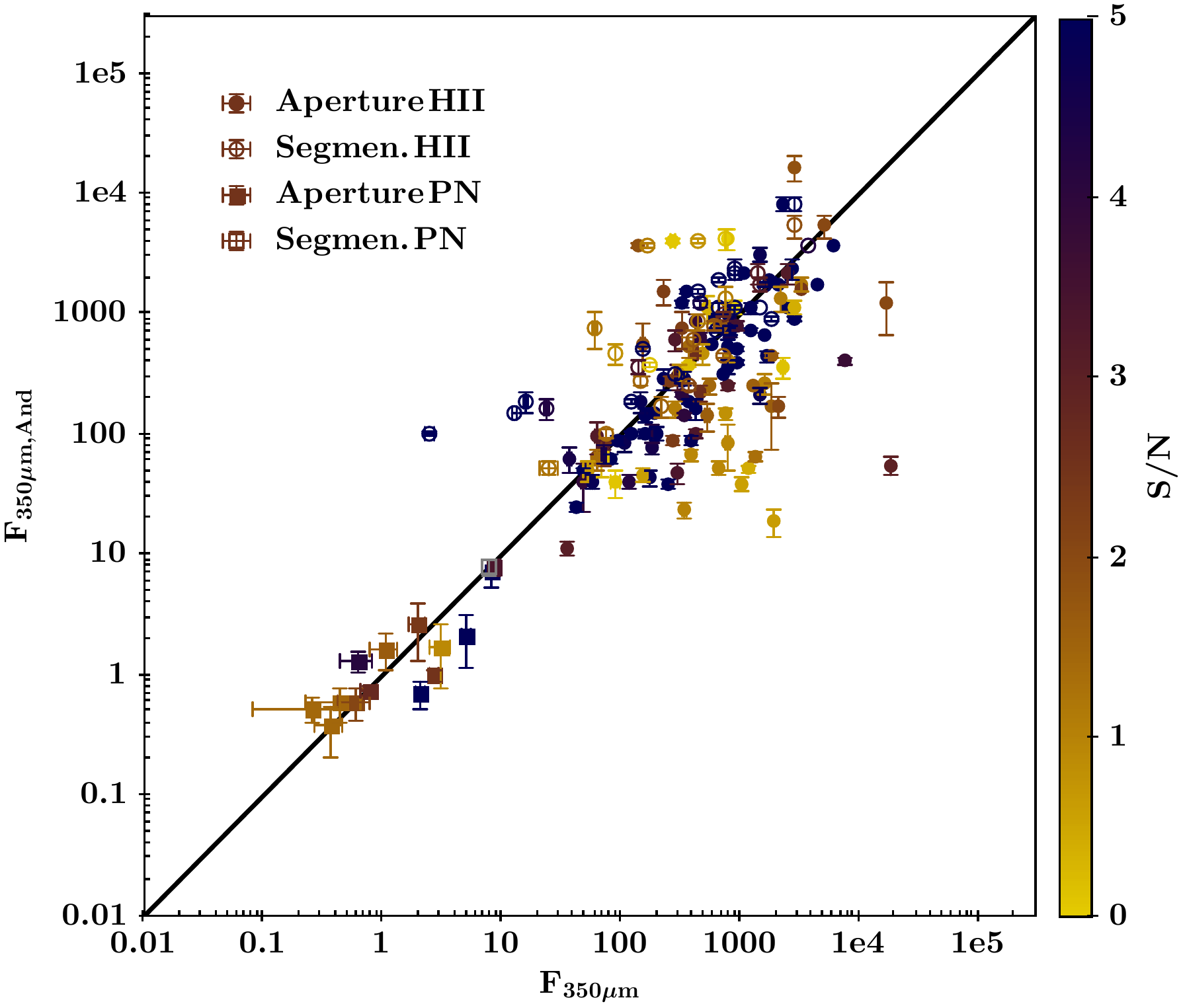}
\includegraphics[width=0.475\textwidth]{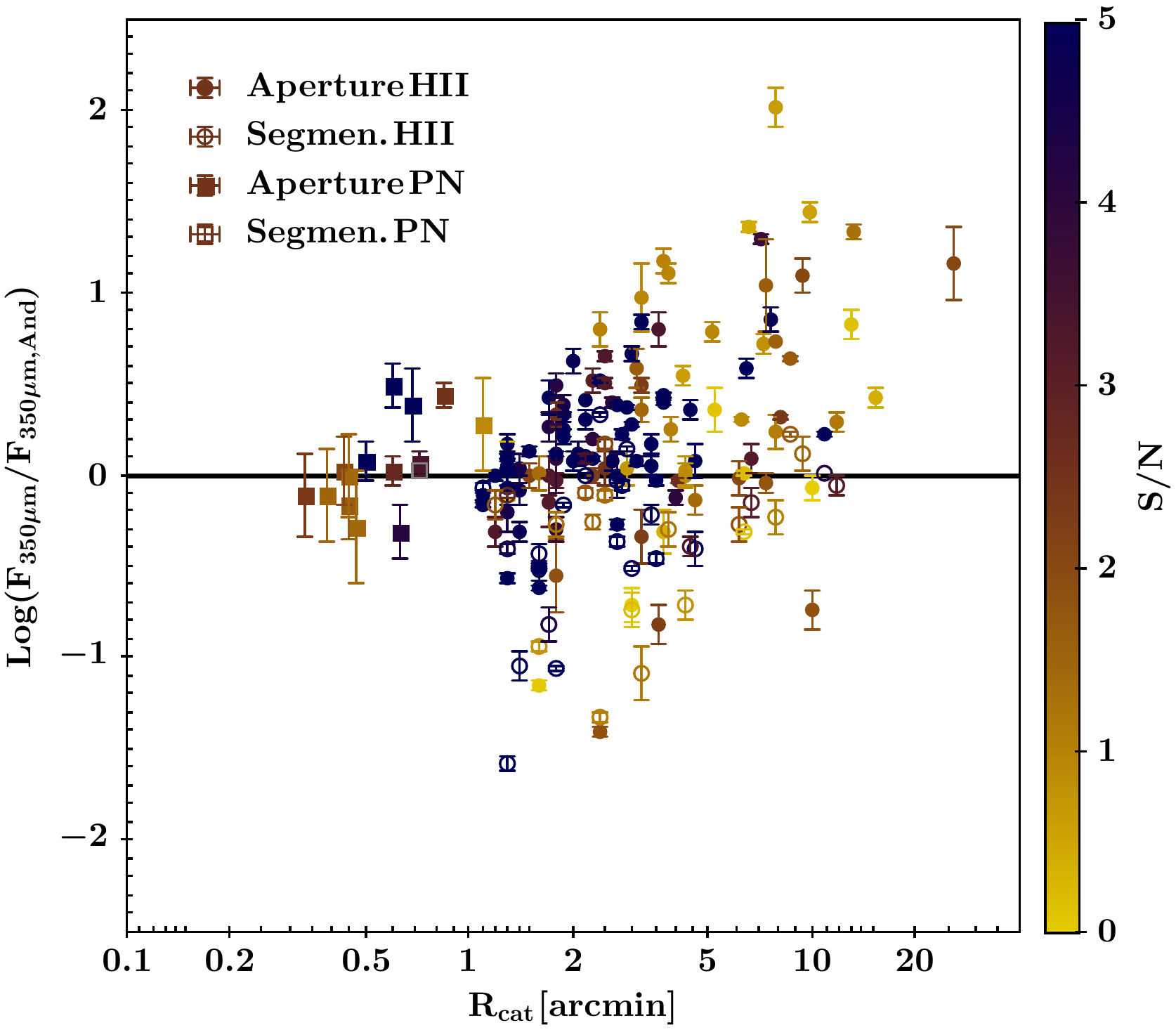}
\includegraphics[width=0.488\textwidth]{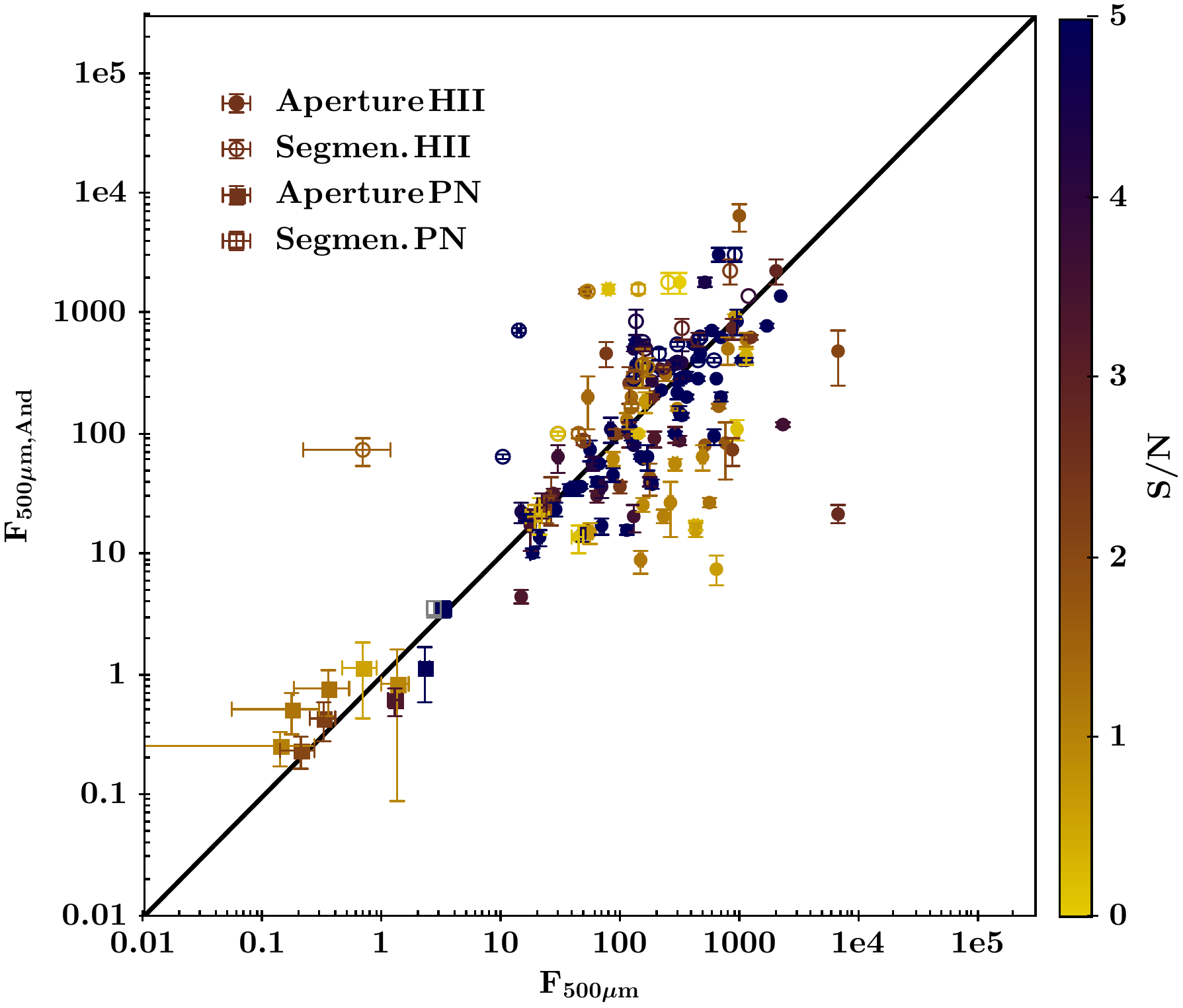}
\includegraphics[width=0.475\textwidth]{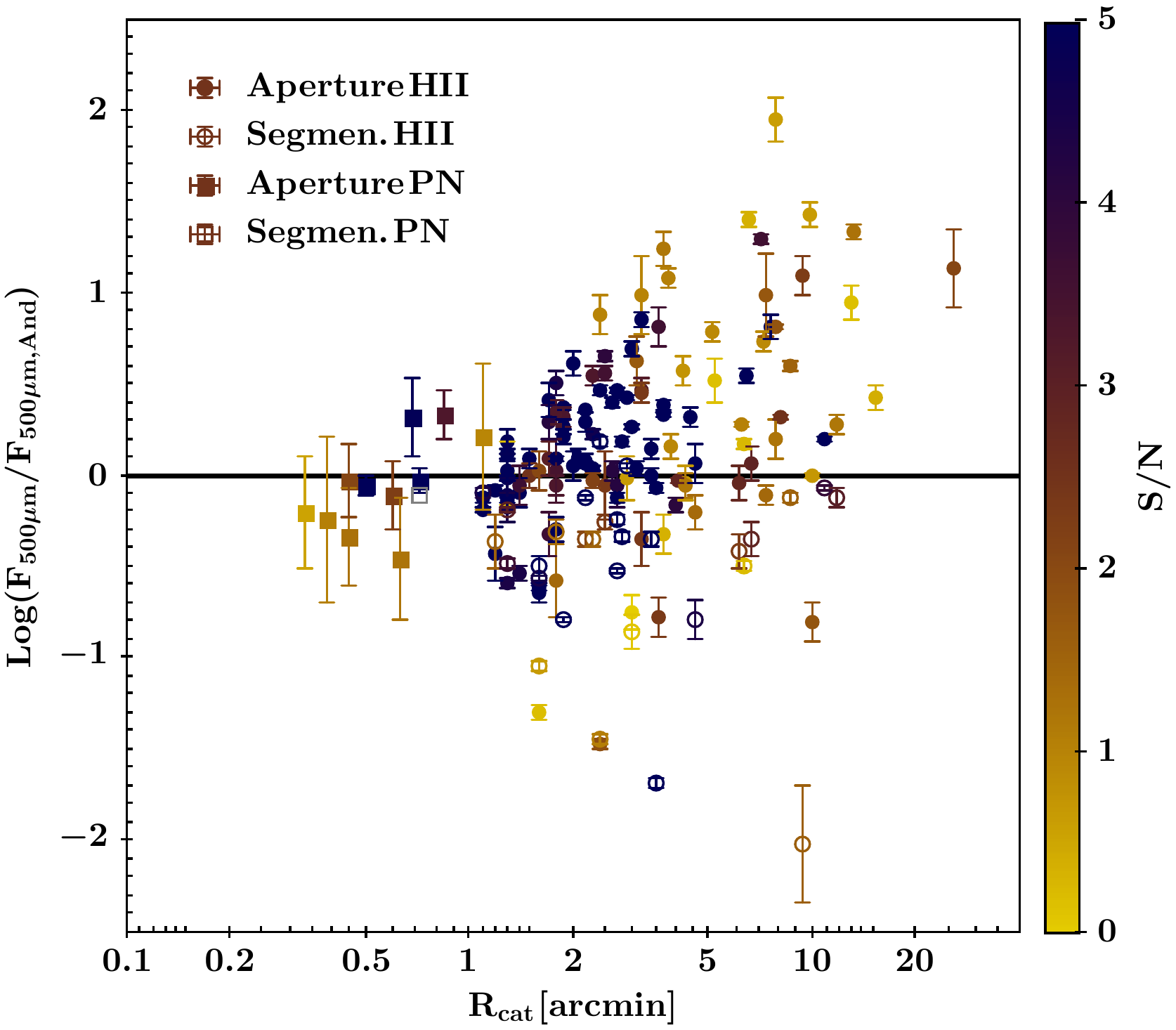}
 \caption{{\it Left Column:} Comparison between photometry obtained with our automated methods and A12 results at 250\um,350\um\,and 500\um. Fluxes from \hiirs\, are reported with circles and PNe with squares. For aperture photometry filled dots are used, while dots for the segmentation photometry are empty. Fluxes are given in Jy. {\it Right Column:} Distribution of flux differences between our methods and A12 measurements as function of the angular extension of the bubble. Radii are given in arcminutes.  In all  the plots, the color scale indicates the S/N characteristic of each of our measurements. } \label{Anderson_red}
\end{figure*}



\begin{figure*}
\includegraphics[width=0.245\textwidth]{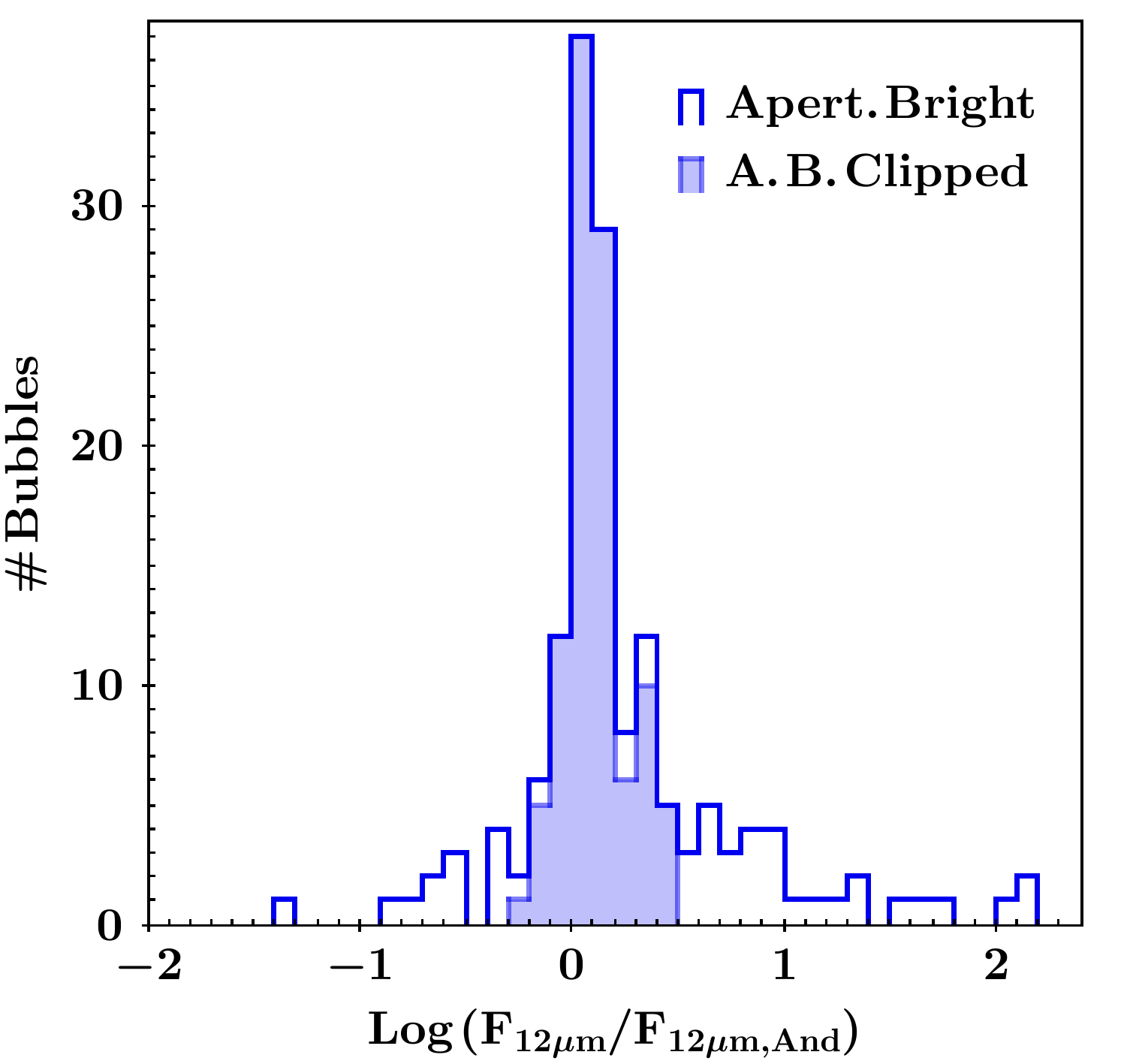}
\includegraphics[width=0.245\textwidth]{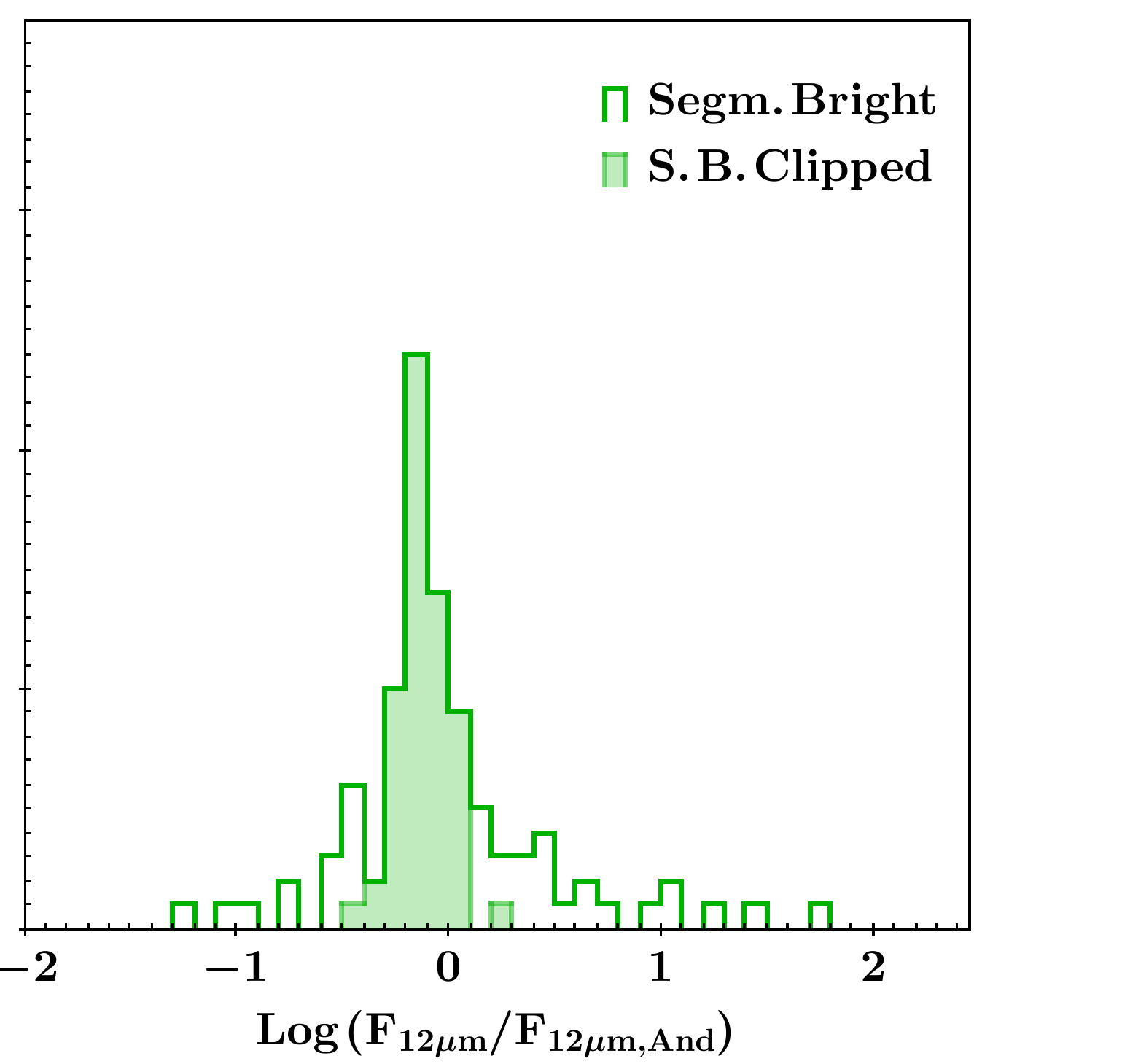}
\includegraphics[width=0.245\textwidth]{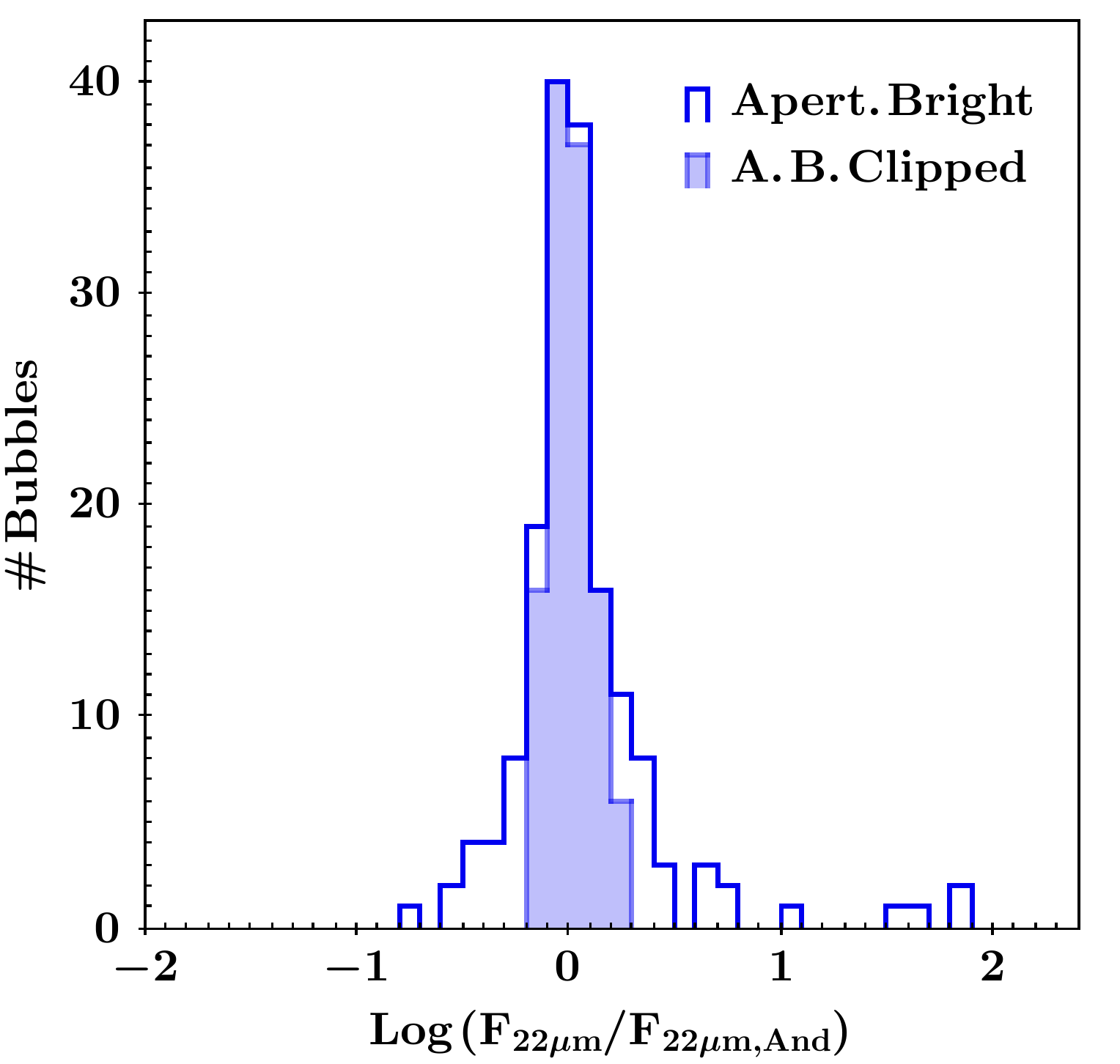}
\includegraphics[width=0.245\textwidth]{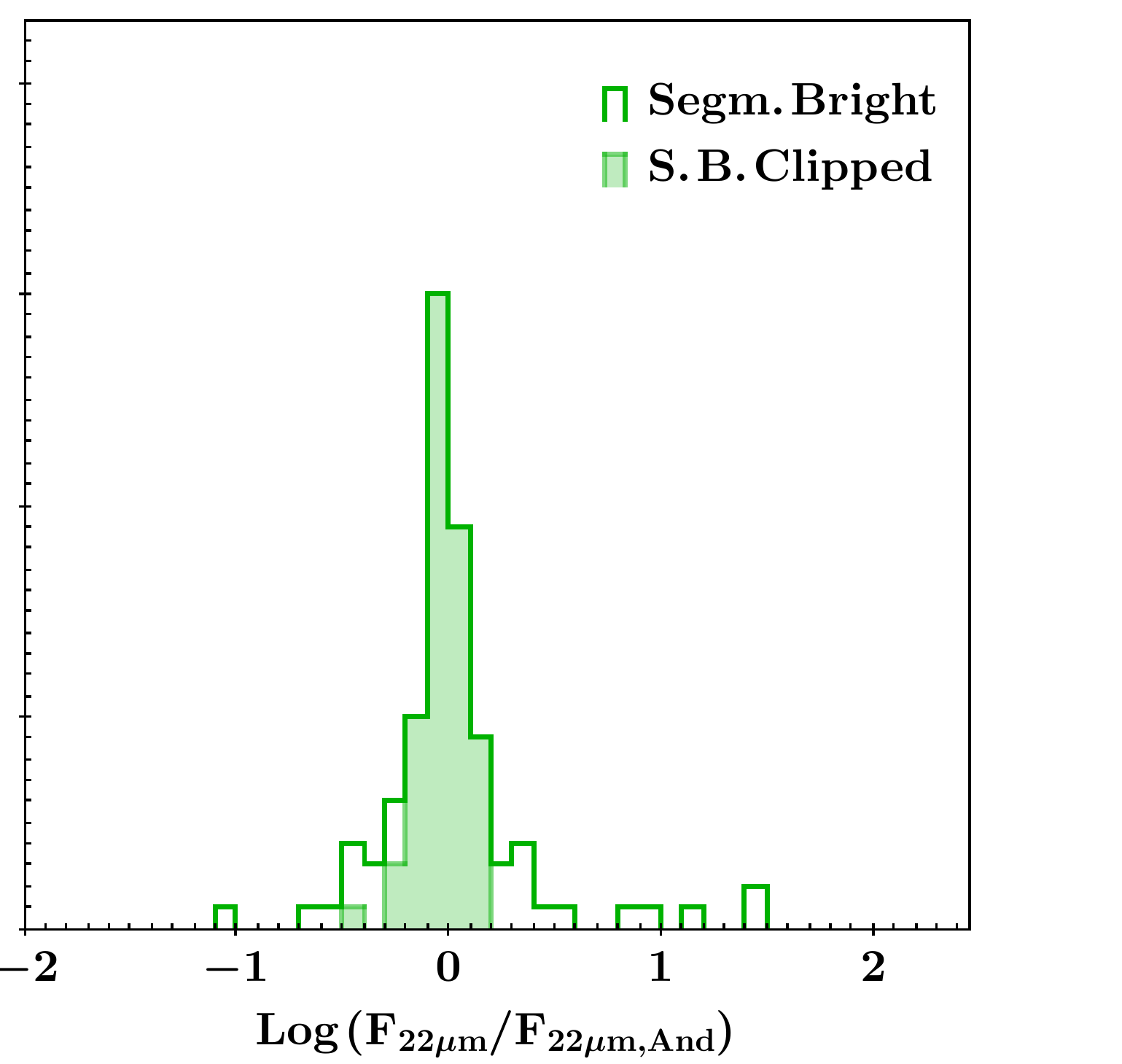}
\includegraphics[width=0.245\textwidth]{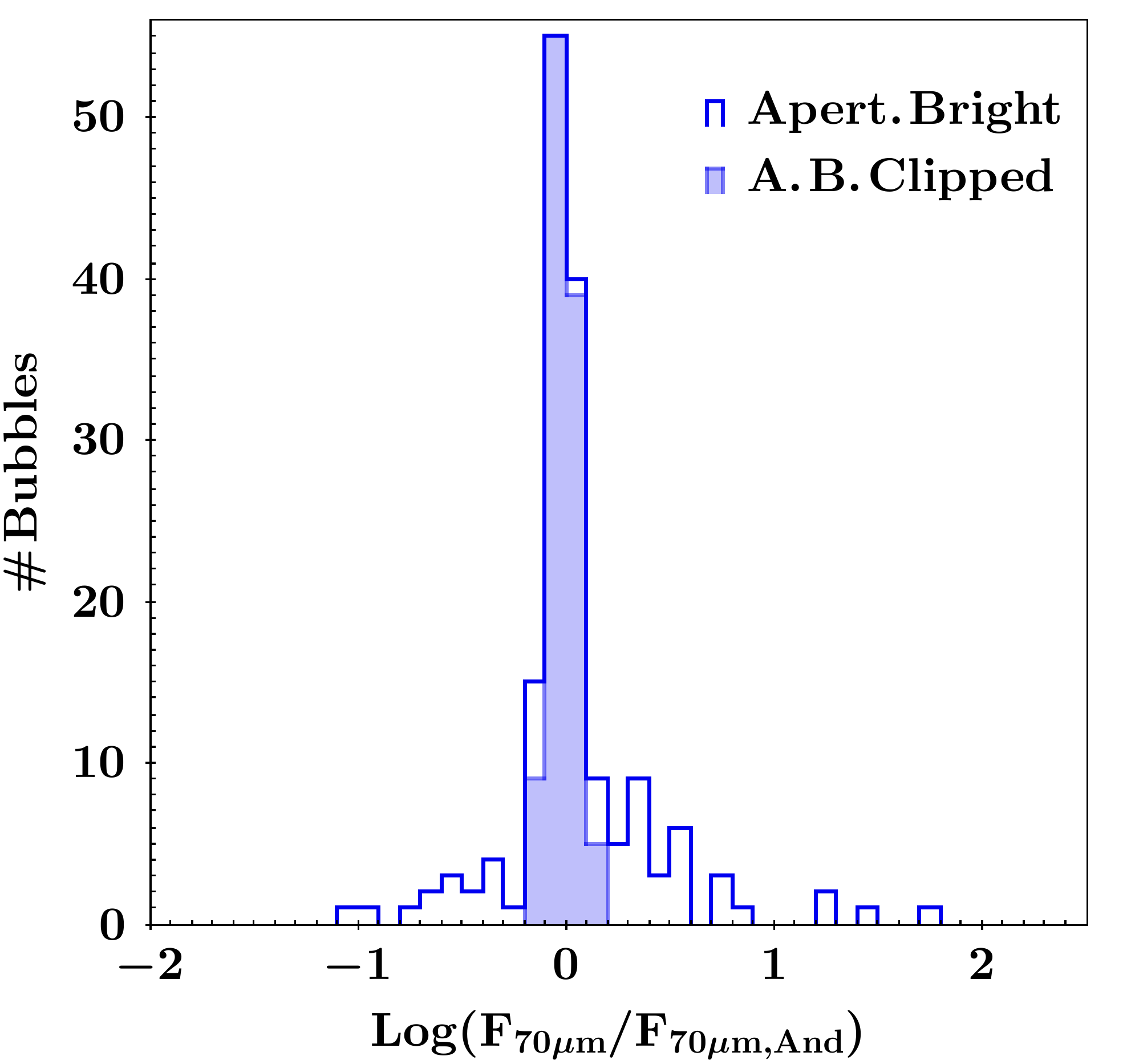}
\includegraphics[width=0.245\textwidth]{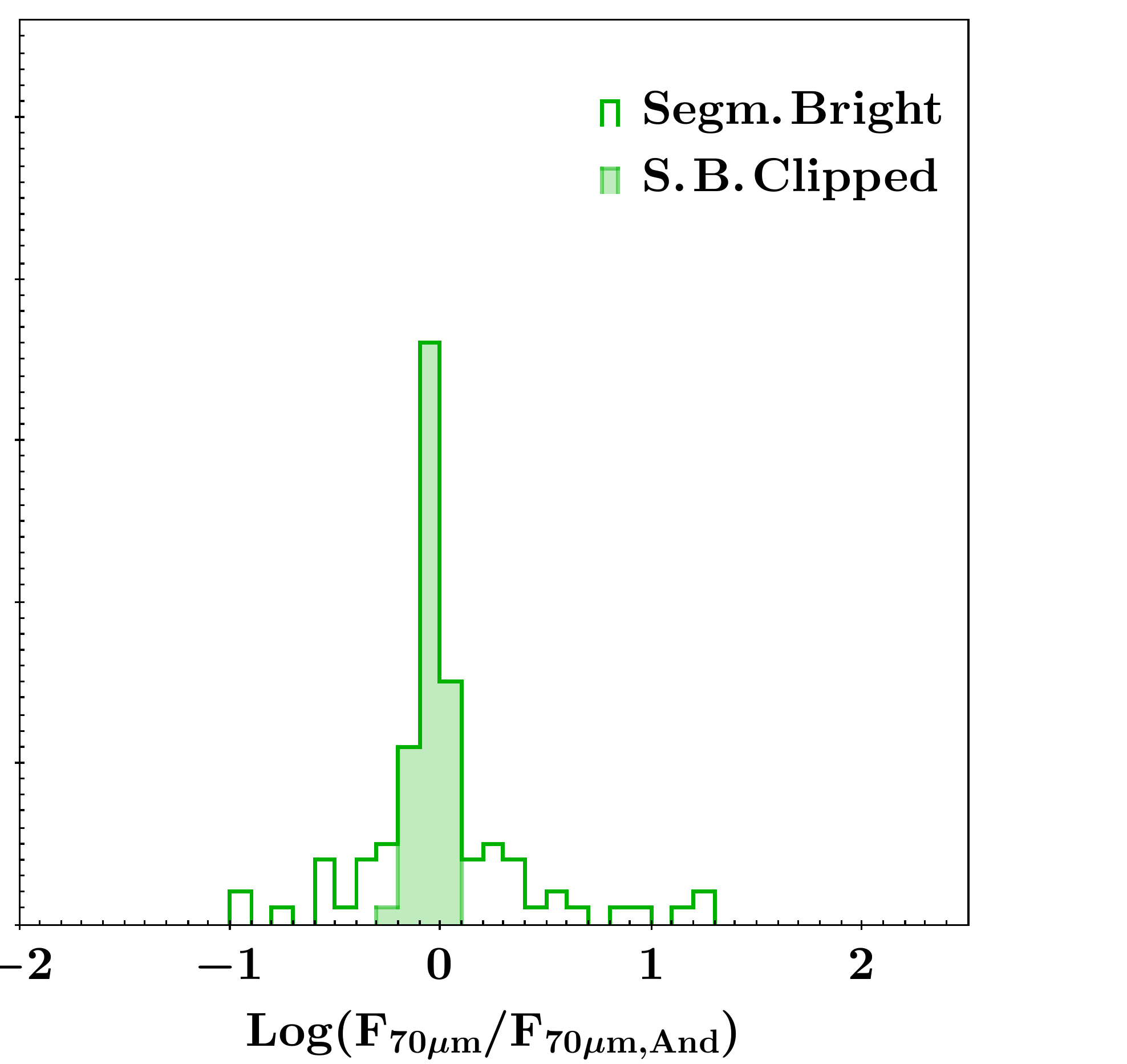}
\includegraphics[width=0.245\textwidth]{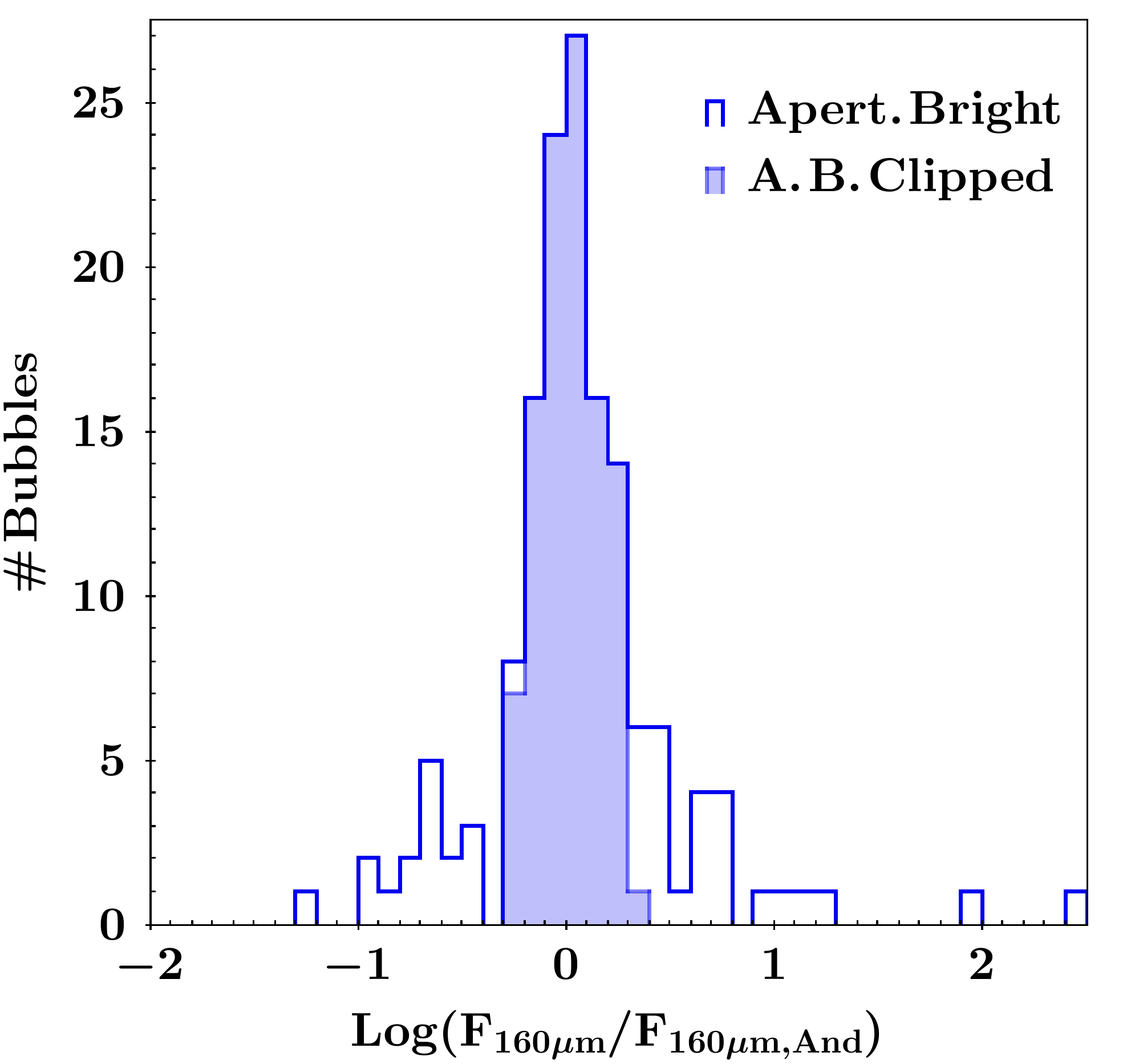}
\includegraphics[width=0.245\textwidth]{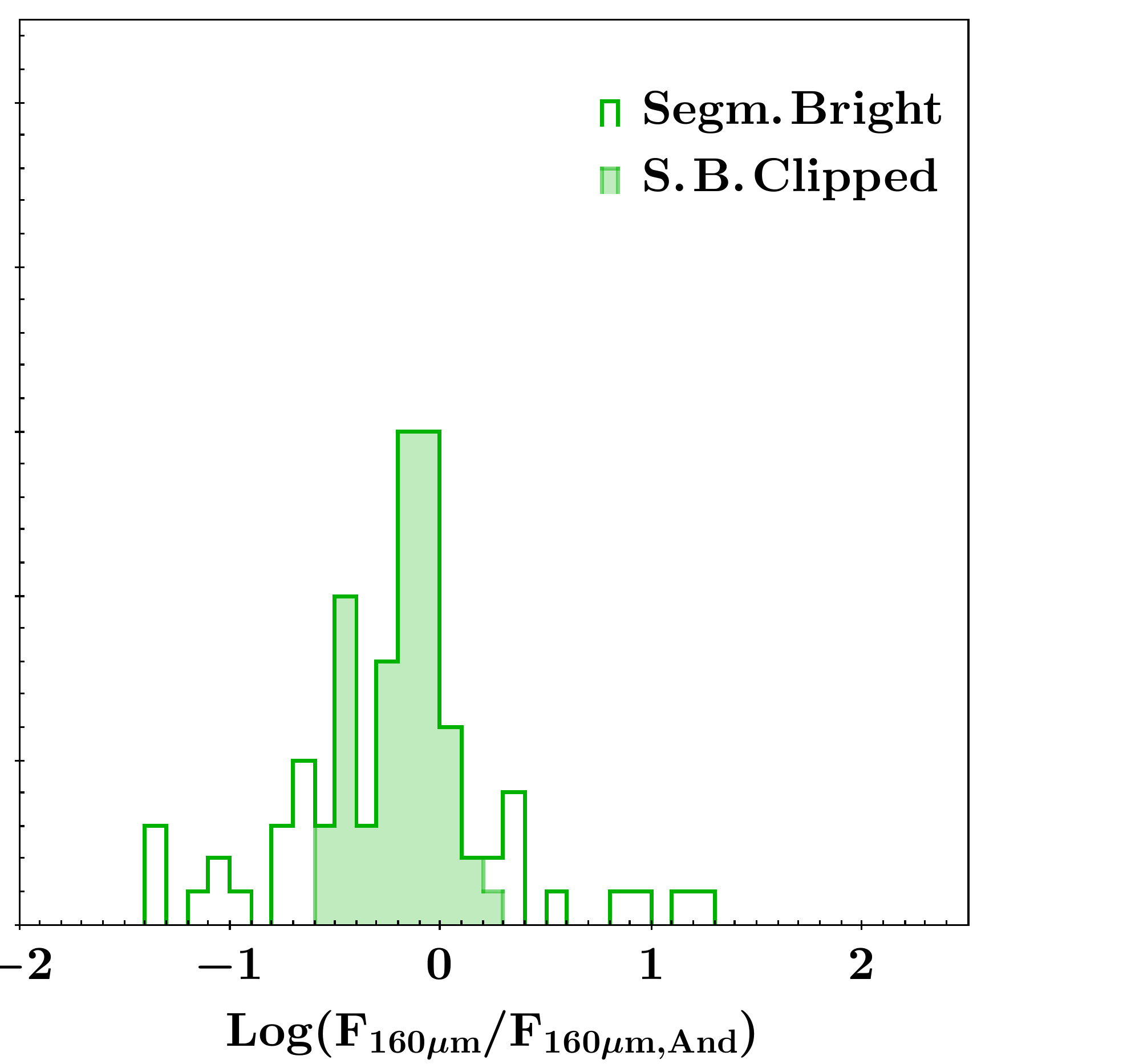}
\includegraphics[width=0.245\textwidth]{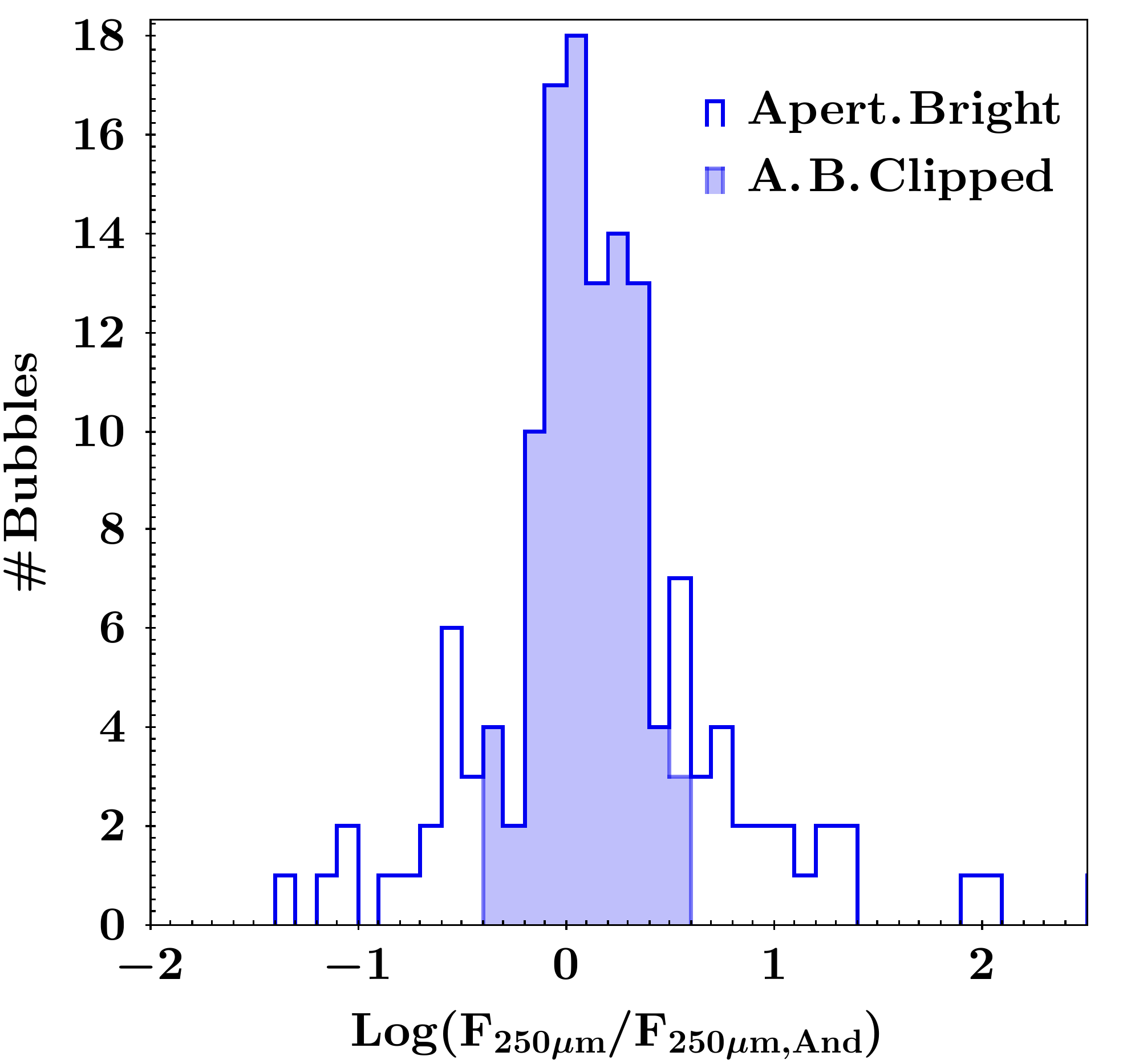}
\includegraphics[width=0.245\textwidth]{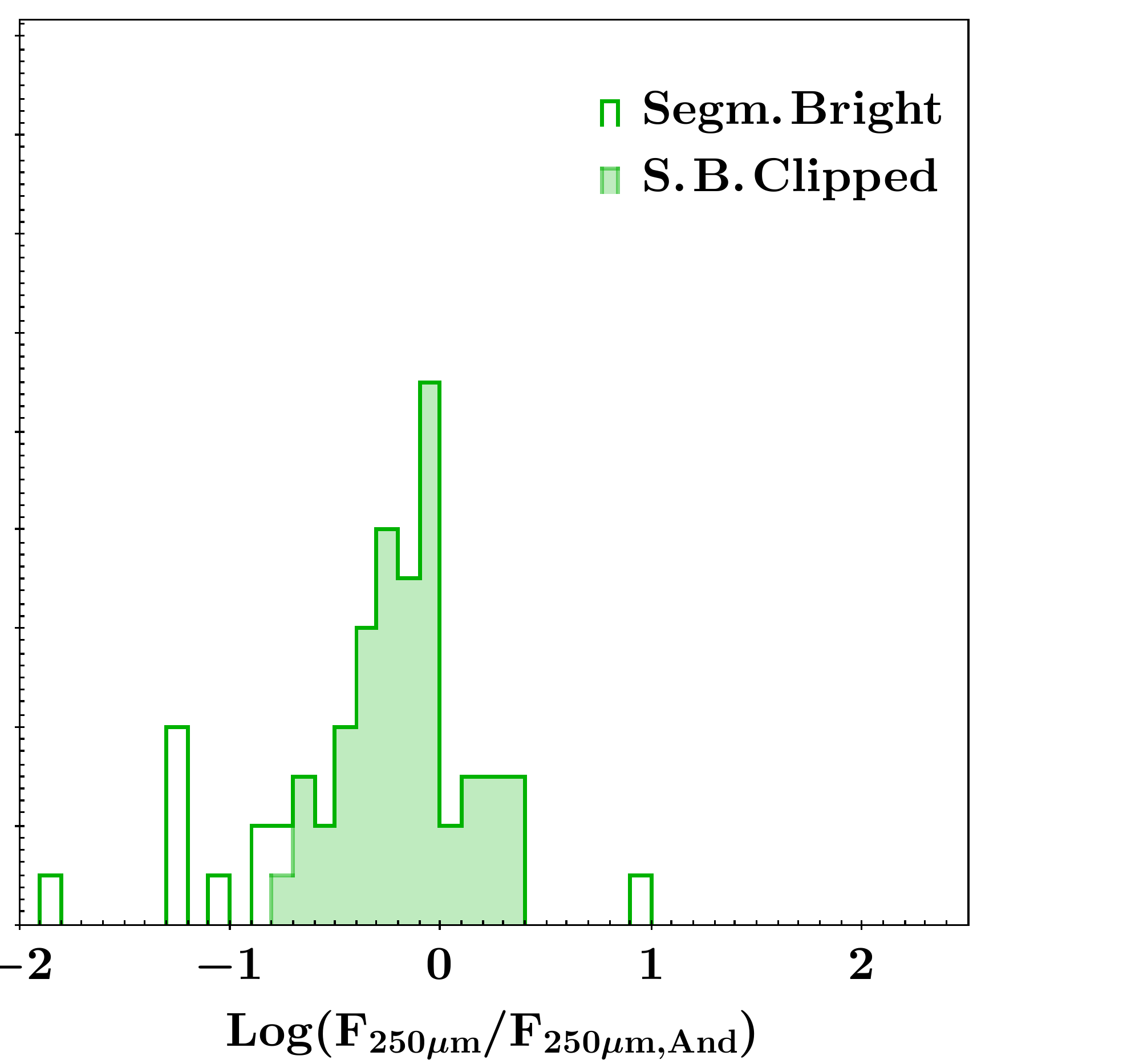}
\includegraphics[width=0.245\textwidth]{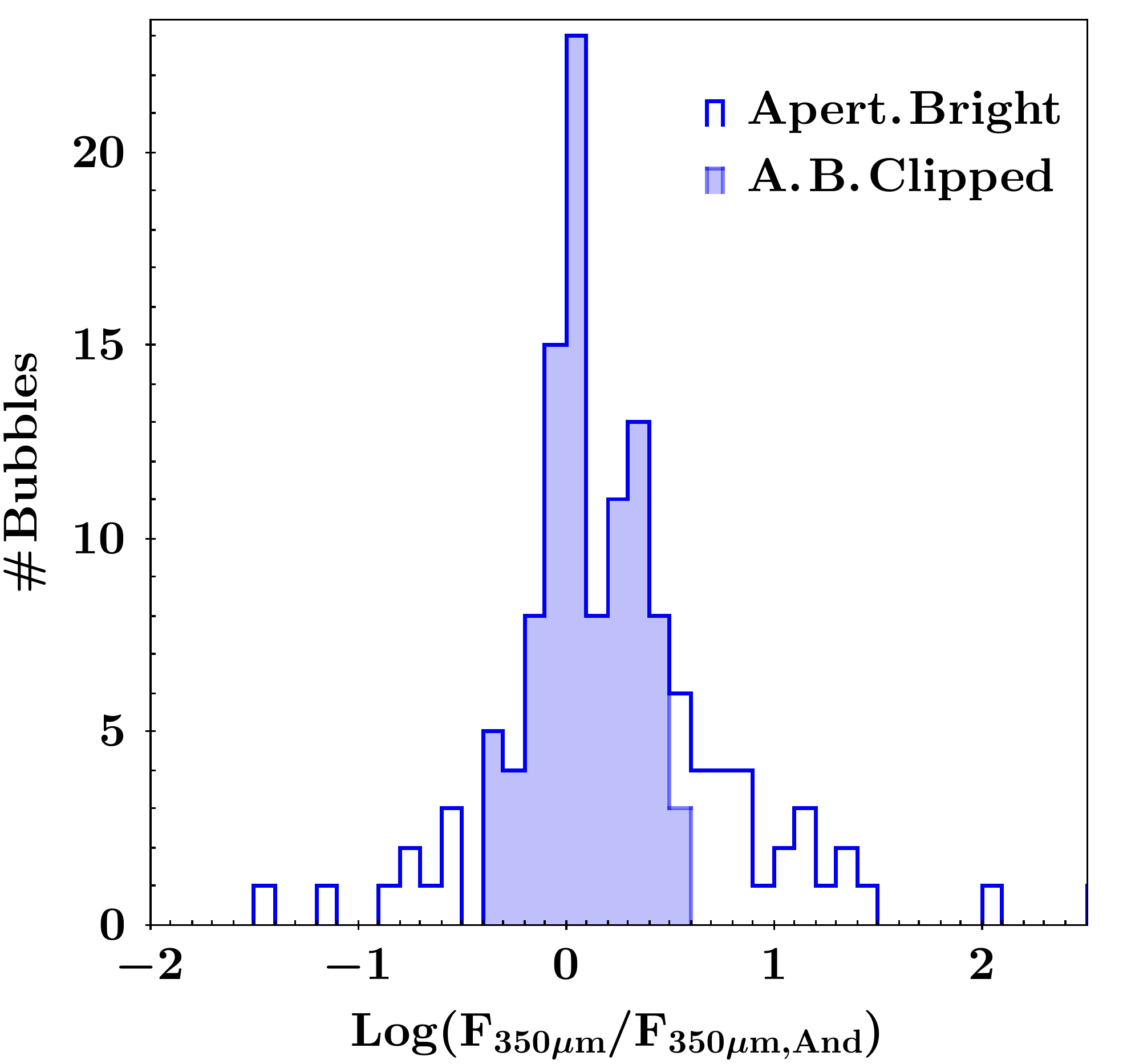}
\includegraphics[width=0.245\textwidth]{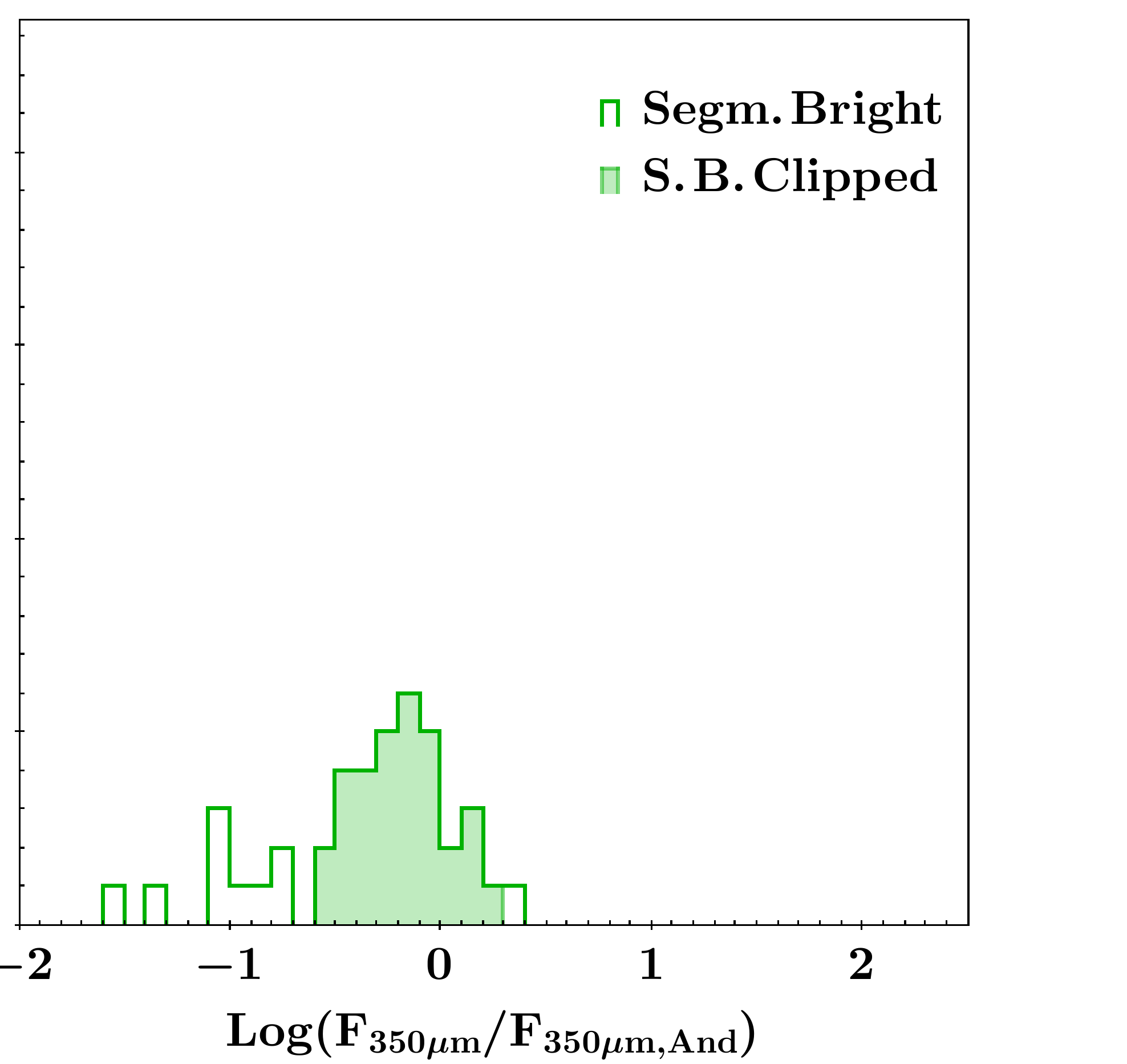}
\includegraphics[width=0.245\textwidth]{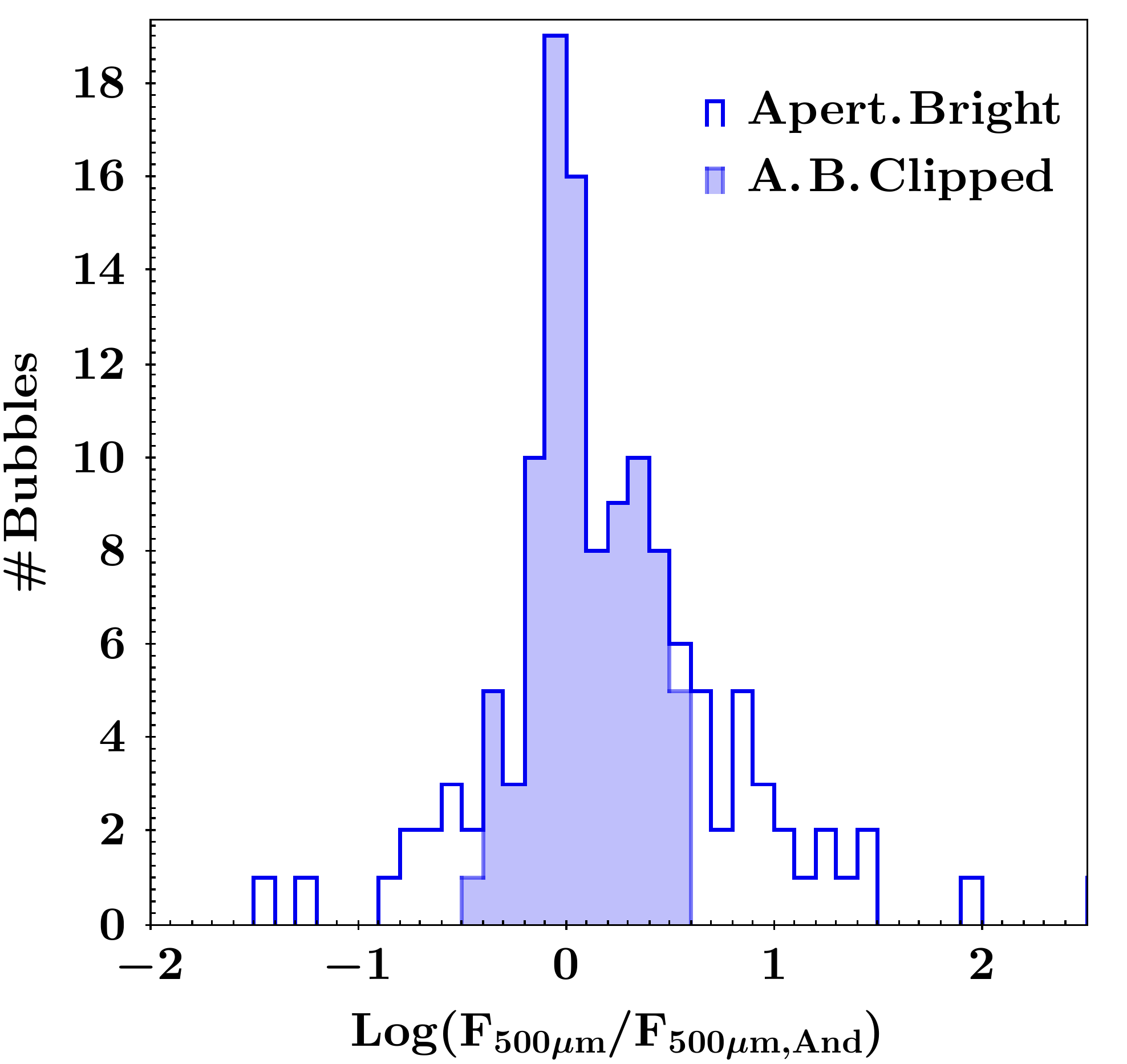}
\includegraphics[width=0.245\textwidth]{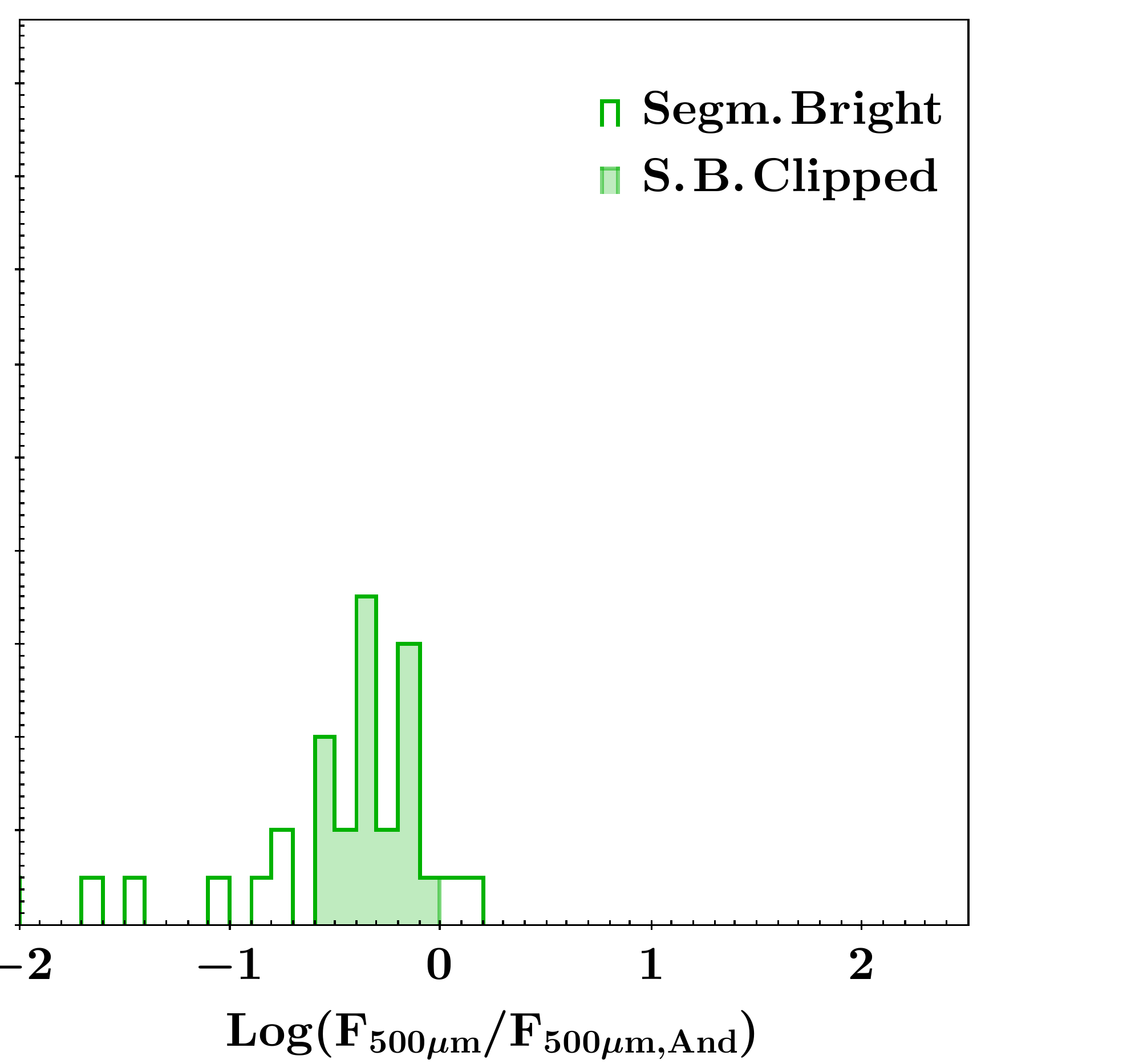}

 \caption{Histograms of the distribution of the difference $\Delta$Log between fluxes from this work and A12. For each photometric method, bubbles samples  (\hiirs\, plus PNe) are formed by the total
 bright bubbles sample or by the limited one (``clipped''), selected after the sigma-clipping (2-$\sigma$ level). These are reported with empty and filled histograms respectively. } \label{Anderson_histo}
\end{figure*}


    \subsection{Comparison with \citet{Anderson12}}\label{and_section}
Recently, A12 analyzed the distribution of FIR emissions from  \hiirs\, and PNe
in order to find a criterion to discriminate between these objects
simply using their IR colours. 
 They collected a sample of  43 PNe and 126 \hiirs. 
 In order to test their diagnostic method, \hiirs\, have been carefully chosen to 
span a wide range of angular sizes ($1\farcm1\le R \le25\farcm9$). In particular, they pay attention at including small size \hiirs, i.e.
young compact \hiirs\, in early evolutionary stages or more evolved \hiirs\, at extreme distances from the observer, since
they can be easily mistaken with typical PNe, having similar dimensions. 
The sample of objects presented, both \hiirs\, and PNe, is numerically limited but it is composed of bright examples of these classes,
thus they are relatively unconfused with nearby sources of emission.
A12 provided the fluxes of these (126+43) bubbles at different IR bandpasses, including 12\um\, and 22\um\, from \wise\, and 70\um, 160\um, 250\um, 350\um\,  and 500\um\, from \herschel.\\
We decided to apply our automated methods to the same object sample and compare our estimated fluxes with those measured by A12, as a quality check of our results.
They measured the fluxes emitted by the \hiirs\, within an aperture of arbitrary size  fixed manually and shaped in a way that
includes all the associated emission at all wavelengths and excludes contaminating compact bright sources in the field.  
The assumption of a unique aperture is considered conservative and safe by the authors, 
since most of their \hiirs\, are bright at IR wavelengths  and have a similar morphology and angular extension at all wavelengths, 
but implies some form of human intervention for deciding the shape.
On the other hand, considering the small angular size of PNe, their photometry is more sensitive to the choice of a unique aperture size for all the wavelengths,
thus they chose to adopt individual apertures at each bandpass.  
A12 published the measured fluxes along with the aperture radius used for each \hiir\, and that at 24\um\, for PNe.\\
Using  radii from A12 catalogue as the dimension ($R_{cat}$) of the bubbles and applying, as previously described, the two methods,  we obtained for each bubble
the aperture and segmentation fluxes.
The comparison between our flux estimates and those from A12 have been plotted in Figure\,\ref{Anderson_figure}:
aperture photometry for the \hiirs\, (circles) and PNe (squares) is reported using filled dots,
while the empty ones refer to the segmentation photometry.  
First in Figure\,\ref{Anderson_figure}, we reported the comparison at 12\um, 22\um, 70\um\, and 160\um, since at these wavelengths the bubble is brighter and
the estimate less sensitive to the background variation, as discussed later in this section.
A general agreement is visible among the flux measurements, especially for the PNe, while a larger scatter is present for the \hiirs. 
The latter could be explained with a stronger contamination by the background to the more extended bubbles. 
This is more evident in Figure\,\ref{Anderson_radius}, in which flux ratios, expressed as 
\begin{equation}\label{eq1}
\Delta Log = Log F_\lambda - Log F_{\lambda, And} = Log ({F_\lambda}/{F_{\lambda, And}})
\end{equation}
are plotted as function of bubble angular extension. We also note that  among extended bubbles,  those with a large difference in flux with respect to A12 have also a very low S/N.\\
 At the same time, we found from a visual inspection of the few bubbles with a large scatter but high S/N (affecting in particular  the \hiirs),
 that the $R_{cat}$ value reported by A12 could largely under- or over-estimate the real dimension of the bubble.\\
 For this reason, in order to estimate an average $ \langle Log(F_{\lambda} / F_{\lambda, And}) \rangle$, that could express the reliability of our methods,
we removed values which differ more than 2-$\sigma$ from the mean. 
Average $ \langle \Delta Log \rangle $ are reported for each bandpass in   Table \ref{And_diff}.
 For the aperture photometry, we also provided the total number of  bubbles from A12 sample detected
 in our images, as well as the number of {\it bright}  bubbles. 
  We labelled bubbles with positive flux as {\it bright}, in contrast with those sources, which are faint  
 over a possibly complex background and have been consequently discarded.
 In a similar way, for segmentation photometry we reported the total number of detected bubbles  successfully masked ({\it segmented}) and 
 the corresponding number of bright bubbles. Finally,  for both methods we gave the sample of bright bubbles used to compute the average 
 after the 2-$\sigma$ clipping ({\it clipped} bubbles). 
 Similarly for  A12 sample, we indicated the number of bright bubbles, i.e. bubbles with a no-null flux measurement by A12, over the total one.
 Results for bandpasses at $\lambda\ge$250\um\, are shown in Figure\,\ref{Anderson_red}.\\
 In Figure\,\ref{Anderson_histo}, we reported the histograms of  $\Delta$Log for each bandpass:
 histograms are shown for both methods and refer to the distribution of the bright bubbles  sample and 
 of the more limited one, selected after the clipping ({\it clipped sample}).
From Figure\,\ref{Anderson_histo}, we can notice that there is in general  a very good agreement between the two flux estimates.
In particular, the difference between our aperture photometry and A12,  
is less than 5\%  at 22\um, 70\um\, and 160\um, and increases to about $\sim$25\% at longer wavelengths, 
where we expect the emission from the bubble to be less intense relative to the background.
The $ \langle \Delta Log \rangle$ at 12\um\, is also $\sim$28\%,  even though we used the old DN-to-Jy conversion factor
(2.9045$\times$10$^{-6}$ from the Explanatory Supplement of the Preliminary data release products),
as  A12 did, to make results consistent.\\
Moving to the comparison with the results obtained with the segmentation method,
we found that the segmentation method has a good agreement at shorter wavelengths ($\langle 1-  F_{\lambda} / F_{\lambda, And}\rangle< $10\%), and a larger difference ($\sim$25-35\%) at longer ones.
In particular, we can notice that it tends to provide on average lower fluxes ($ \langle \Delta Log \rangle < 0$): 
this is likely due to the fact that the segmentation helps in better masking the flux falling in the aperture, allowing  only that coming from the sources to be selected
and removing the contaminating flux from the background.
The origin of such differences could also be in the method, which could generally be too strong and likely remove  the pixels of the more external parts of the bubble.
On the other hand, as already discussed in Sect.\,\ref{photometricmethods}, in this work we chose to obtain the masks from 70\um\, images,\
knowing that 70\um\, emission contours generally include those at shorter wavelengths. 
 Moreover, we can assume that bubble shape at $\lambda> $70\um\, does not change significantly, since emissions at such wavelengths originate 
from the same  component of the bubble, namely the cold dust.
As a consequence, the  estimated discrepancies $\langle \Delta Log \rangle$ can be more likely arisen from background contamination, 
which largely increases at redder bandpasses.\\
Finally, it is worth noticing that, when we considered the results on PNe of the segmentation method, we found
that the active contours failed in finding most of the bubbles, possibly because of their small angular size and/or of a possible faint emission
of such objects at 70\um. We discuss about the reason of the segmentation method failure in finding bubbles contours in Section\,\ref{discussion}.

\begin{table*}
 \caption{Average difference between fluxes from this work and A12, where $\Delta$Log has been calculated as in Eq.\,\ref{eq1}. The number of bubbles composing the ``detected'', ``bright'' and ``clipped'' sample (see Sect.\,\ref{and_section}) are given } \label{And_diff}
 \begin{minipage}{160mm}        
\begin{tabular}{@{}rcccccc}
     \hline
     \hline
      &\multicolumn{5}{c}{Aperture Photometry}\\ 
      \hline 
      Band &$\langle \Delta Log \rangle$  & \multicolumn{2}{c}{\hiir} & & \multicolumn{2}{c}{PN} \\
  \cline{3-4}
   \cline{6-7}
&&Clip./Bright/Detect.&Bright/Detect.(A12)&  &Clip./Bright/Detect.&Bright/Detect.(A12)\\
    \hline
12\um\footnote{Similarly to A12, for \wise\, 12\um\, it has been used a different DN-to-Jy conversion factor, equal to 2.9045$\times$10$^{-6}$ taken from the Explanatory Supplement of the preliminary data release products. }  &      0.11 $\pm$ 0.14 & 73/115/126    &      126/126  &&  32/39/43  &     40/43    \\
 22\um   &  0.01 $\pm$ 0.10   & 84/123/126   &     126/126 &&   31/41/43  &     42/43     \\
 70\um	 & -0.01 $\pm$ 0.06   & 75/122/126   &     126/126 &&   33/43/43  &     43/43     \\
160\um   &  0.02 $\pm$ 0.14    & 81/120/126   &     126/126 &&   24/28/43  &     31/43    \\ 
250\um   &  0.10 $\pm$ 0.20   & 81/121/124   &     126/126 &&   17/19/43  &     21/43     \\
350\um   &  0.10 $\pm$ 0.22   & 84/120/126   &     126/126 &&   14/14/43  &     16/43     \\
500\um   &  0.09 $\pm$ 0.23   & 83/120/126   &     126/126 &&   11/11/43   &    12/43    \\ 
     \hline
     \hline
      &\multicolumn{5}{c}{Segmentation Photometry}\\ 
      \hline 
            Band &$\langle \Delta Log \rangle$  & \multicolumn{2}{c}{\hiir} & & \multicolumn{2}{c}{PN} \\
  \cline{3-4}
   \cline{6-7}
&&Clip./Bright/Segmen.&Bright/Detect.(A12)&  &Clip./Bright/Segmen.&Bright/Detect.(A12)\\
  12\um$^{a}$& -0.12 $\pm$ 0.11  & 57/94/94    &     126/126  &&  4/4/4  &     40/43    \\
 22\um & -0.02 $\pm$ 0.11    & 68/94/94  &       126/126 	&&  4/4/4   &    42/43\\
 70\um & -0.04 $\pm$ 0.07    & 59/97/97  &       126/126 & &  4/4/4   &      43/43\\
160\um & -0.18 $\pm$ 0.18  & 61/86/95  &       126/126 & &  2/2/4   &       31/43\\
250\um & -0.16 $\pm$ 0.26  & 52/62/98  &      126/126 & &   1/1/4   &       21/43\\
350\um & -0.18 $\pm$ 0.20  & 31/41/97  &       126/126  &&  1/1/4   &       16/43\\
500\um & -0.31 $\pm$ 0.15   &21/30/98  &       126/126  &&  1/1/4   &       12/43\\
\hline
\hline
 \end{tabular}
 \end{minipage}
 \end{table*}



\section{Catalogue Format}    \label{catalogue}

 As an example of the final catalogue, we reported in Appendix\,\ref{photometrytables} tables with the fluxes measured with the two methods for
 a collection of bubbles taken from the golden sample: \wise\, 12\um\, and 22\um\, values are given in Table \ref{wiseap} and Table \ref{wiseseg}, respectively  and \herschel\, 70\um, 160\um, 250\um, 350\um\,  and 500\um\, values in Table \ref{hgap} and Table \ref{hgseg}.\\
 All the Tables give the source name followed by the Galactic longitude and latitude and the angular size ($R_{cat}$) taken from \citet{Simpson} and used, as previously described (Sect.\,\ref{sample}), in our work.
 Total flux (F$_{\lambda}$) is given in Jy, as well as the associated uncertainty. \\
Missing flux estimates are indicated with a ``--'' if the survey image only partially covers
the bubble or in case it has a high fraction of saturated/NaN pixels ($>$10\%), since in both cases the measurement would not be reliable.
 In the \wise\, images, there are 24 missing bubbles, all of them discarded for a high fraction of saturated pixels, while for \herschel\, there are a maximum of 7 
 (for 250\um\, and 500\um\, images) with about half of them not completely covered.  
The number of the remaining bubbles, with ``positive'' detection is reported in Table\,\ref{bubbles} ({\it Detected} bubbles) for each bandpass together
with their percentage with respect to the whole golden sample catalogue (1814 bubbles).\\
Flux estimates are indicated with a `` *** " when the emission of the bubble is too faint, i.e. the source average flux per pixel is lower than the estimated
average background level.
Thus in Table\,\ref{bubbles}, we reported also the total number of  {\it bright} bubbles (all the bubbles with a flux estimate listed in the catalogue)
and their percentage with respect to the number of detected bubbles.
In the case of the segmentation photometry, a measurement of the flux could also be missing (indicated with a ``--'' ) when the active contours method fails
in finding the bubble,  i.e. no segmentation mask corresponding to the bubble is produced. Also in this case we report the number of  {\it segmented} bubbles 
with the relative fraction over the detected sample, and the number of {\it bright} bubbles  along with the fraction respect the segmented one.
Table\,\ref{bubbles} thus gives a global view of what is available in the entire flux catalogue.\\

 \begin{table*}
 \caption{Bubbles sample listed in the IR flux catalogue.} \label{bubbles}
 \centering
 \begin{minipage}{160mm}        
 \begin{tabular}{@{}lccccccccc}   
      \hline
     & \multicolumn{4}{c}{Aperture Photometry} &&\multicolumn{4}{c}{Segmentation Photometry} \\  
 \cline{2-5}
 \cline{7-10} 
Bandpass & Detected&\%\footnote{Percentage of Detected bubbles over the 1814 from the golden sample.}&Bright&\%\footnote{Percentage of Bright bubbles over the Detected ones}&&Segmented&\%\footnote{Percentage of Segmented bubbles over the Detected ones.}&Bright&\%\footnote{Percentage of Bright bubbles over the Segmented ones}\\
\hline
12\um&   1790  &     98.6\%&1704&95.2\%&  &1018  &    56.9\%&1014&99.6\%\\
22\um  &  1791  &    98.7\%&1726&96.4\%& & 1019   &    56.9\%&1017&99.8\%\\
70\um  &  1814  &      100.\%&1763&97.2\%&& 1024  &    56.5\%&1024&100.\%\\
160\um  &  1814  &   100.\%&1763&97.2\%& &1024  &      56.4\%&663&64.7\%\\
250\um  & 1807  &   99.6\%&1675&92.7\%& &1022  &      56.6\%&351&34.3\%\\
350\um  &   1811  &   99.8\%&1655&91.4\%&    &1022 &      56.4\%&208&20.3\%\\
500\um    & 1807  &    99.6\%&1635 & 90.5\%& &1022 &      56.6\%&143&14.0\%\\
   \hline
 \end{tabular}
 \end{minipage}
 \end{table*}

\begin{figure*}
\includegraphics[width=0.47\textwidth]{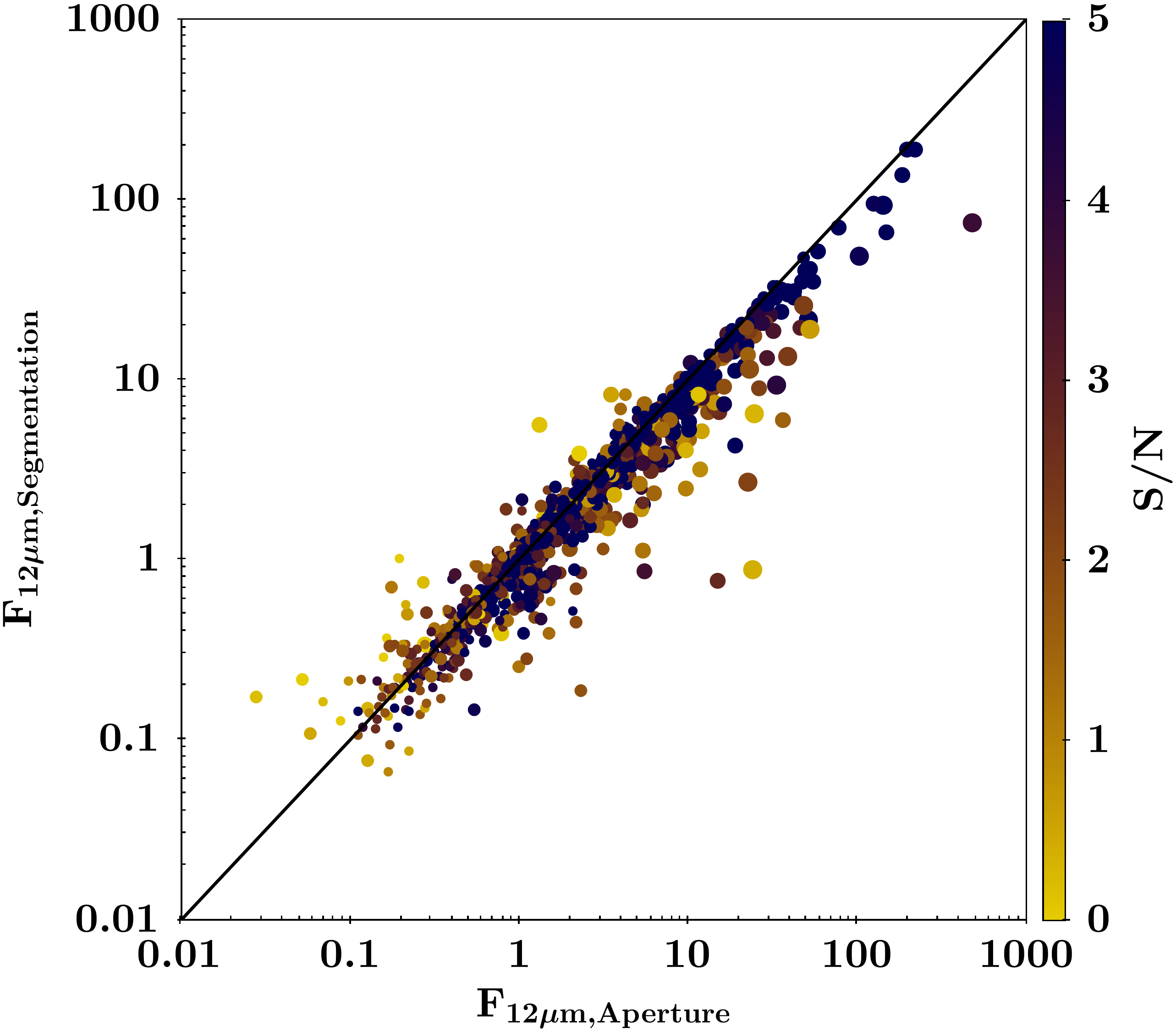}
\includegraphics[width=0.47\textwidth]{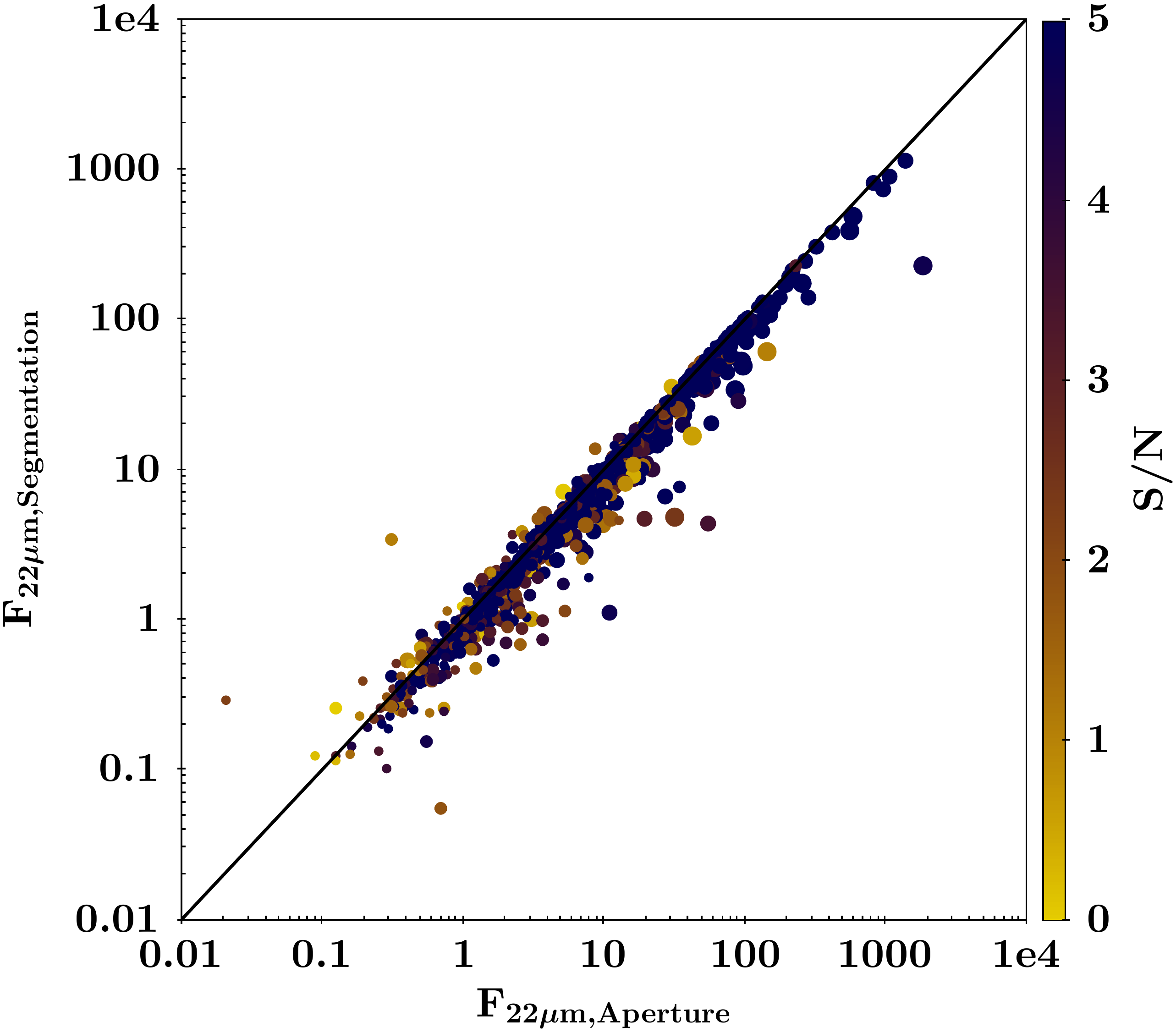}
\includegraphics[width=0.47\textwidth]{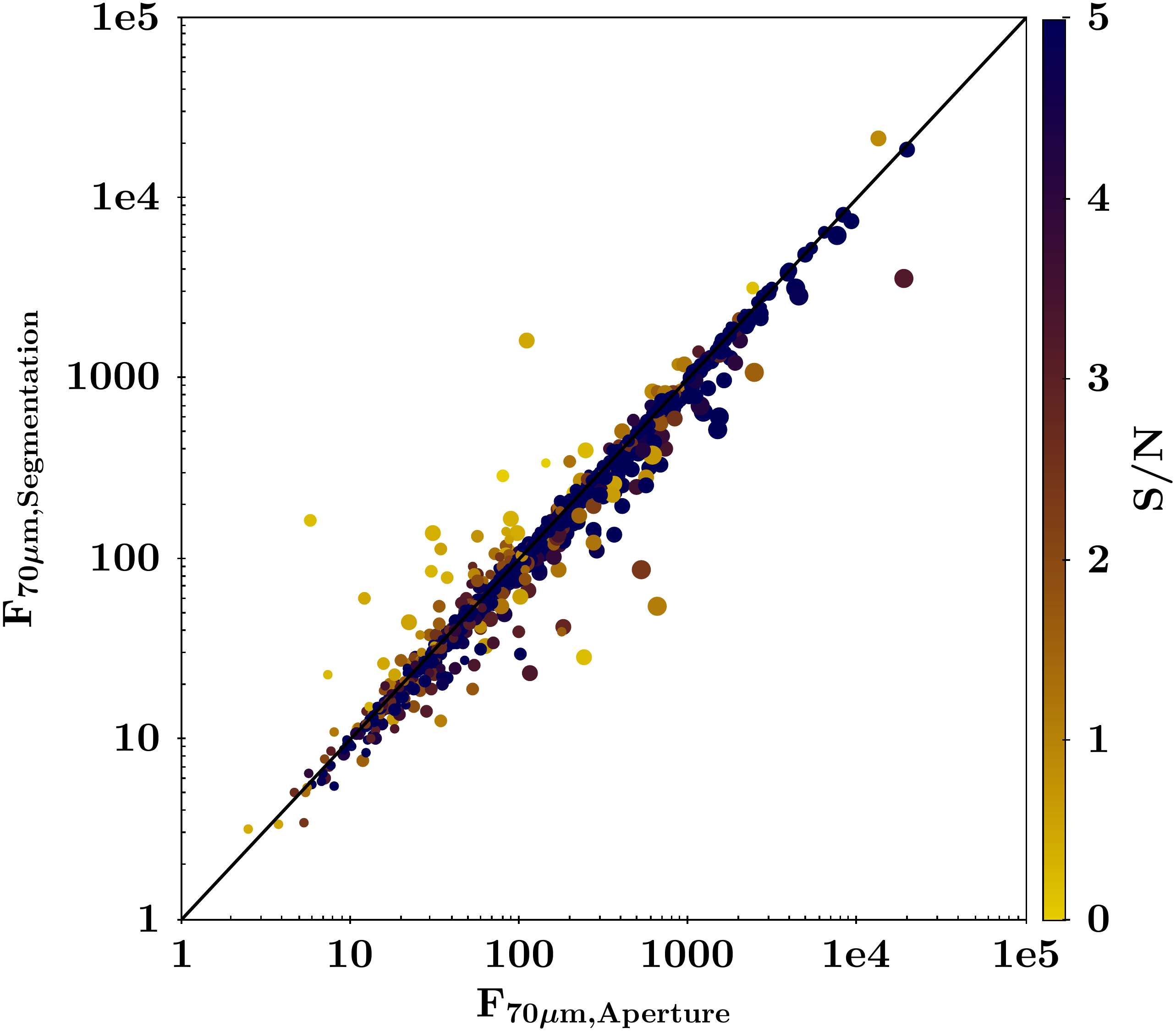}
\includegraphics[width=0.47\textwidth]{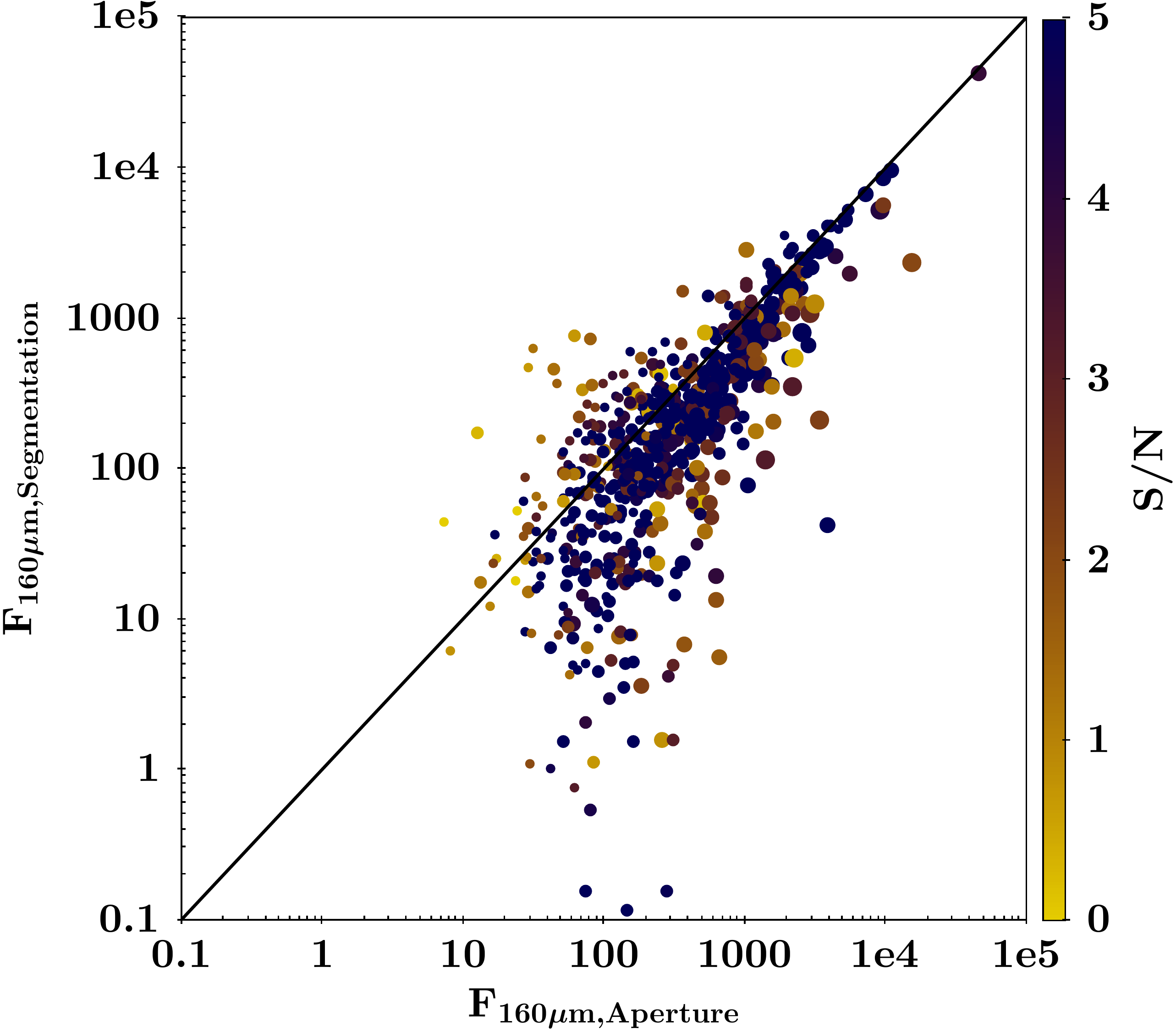}
 \caption{Comparison of flux estimates using the aperture photometry against the segmentation one. Dots size is proportional to the logarithm of the bubble radius ($R_{cat}$) and the color scales based on the S/N of the relative aperture flux estimation. } \label{dist_flux_fig}
\end{figure*}

\begin{figure*}
\includegraphics[width=0.485\textwidth]{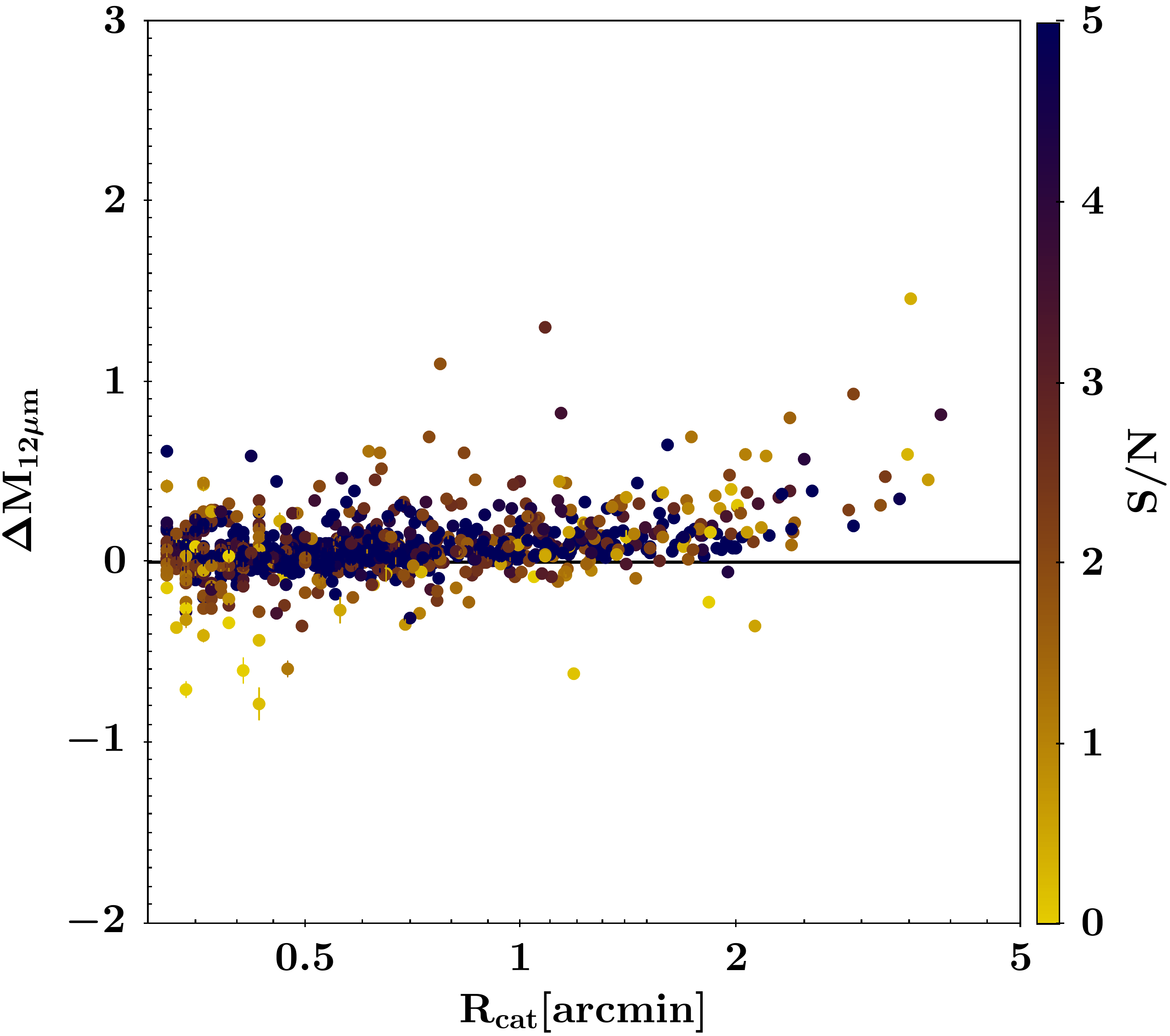}
\includegraphics[width=0.485\textwidth]{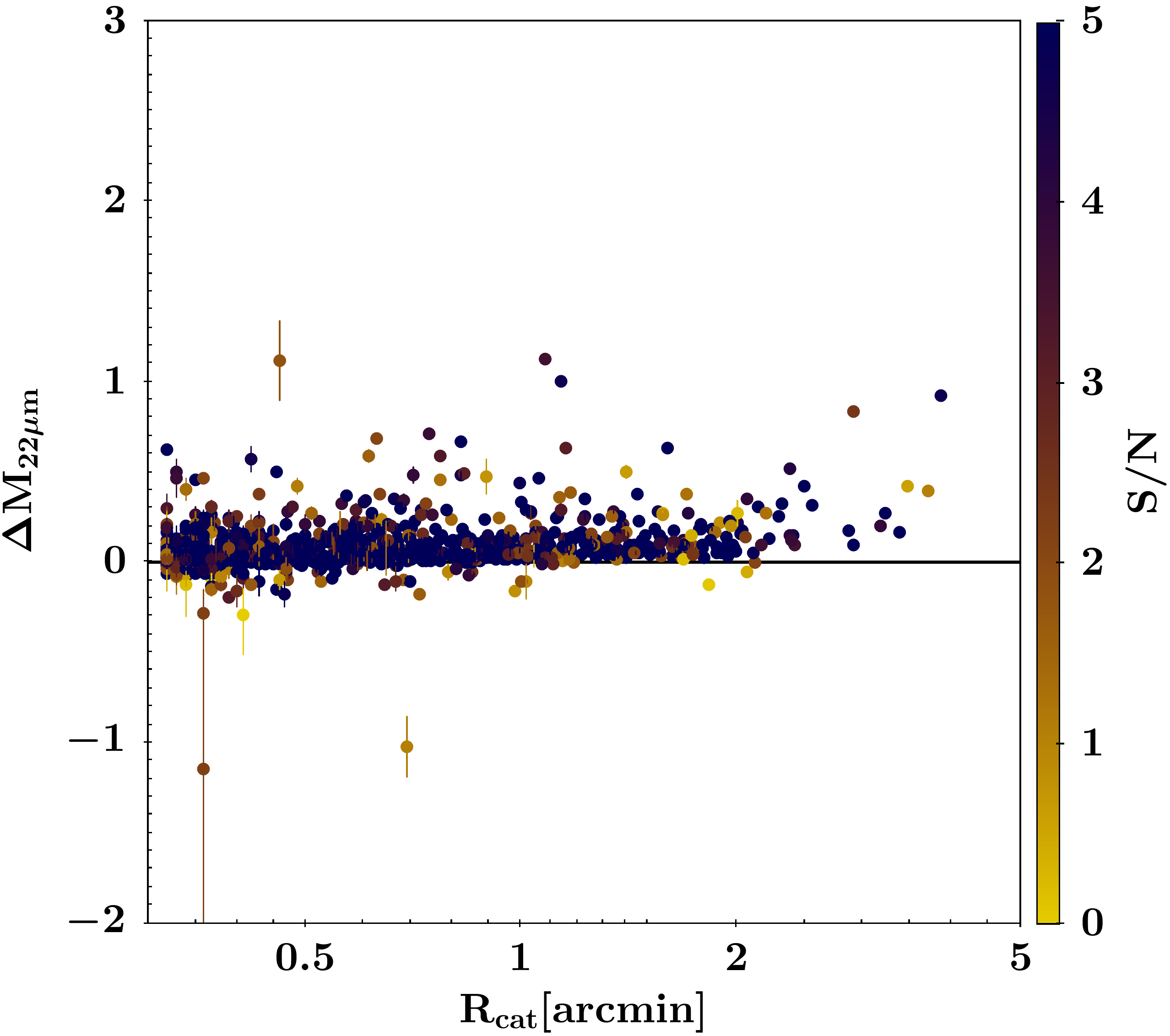}
\includegraphics[width=0.485\textwidth]{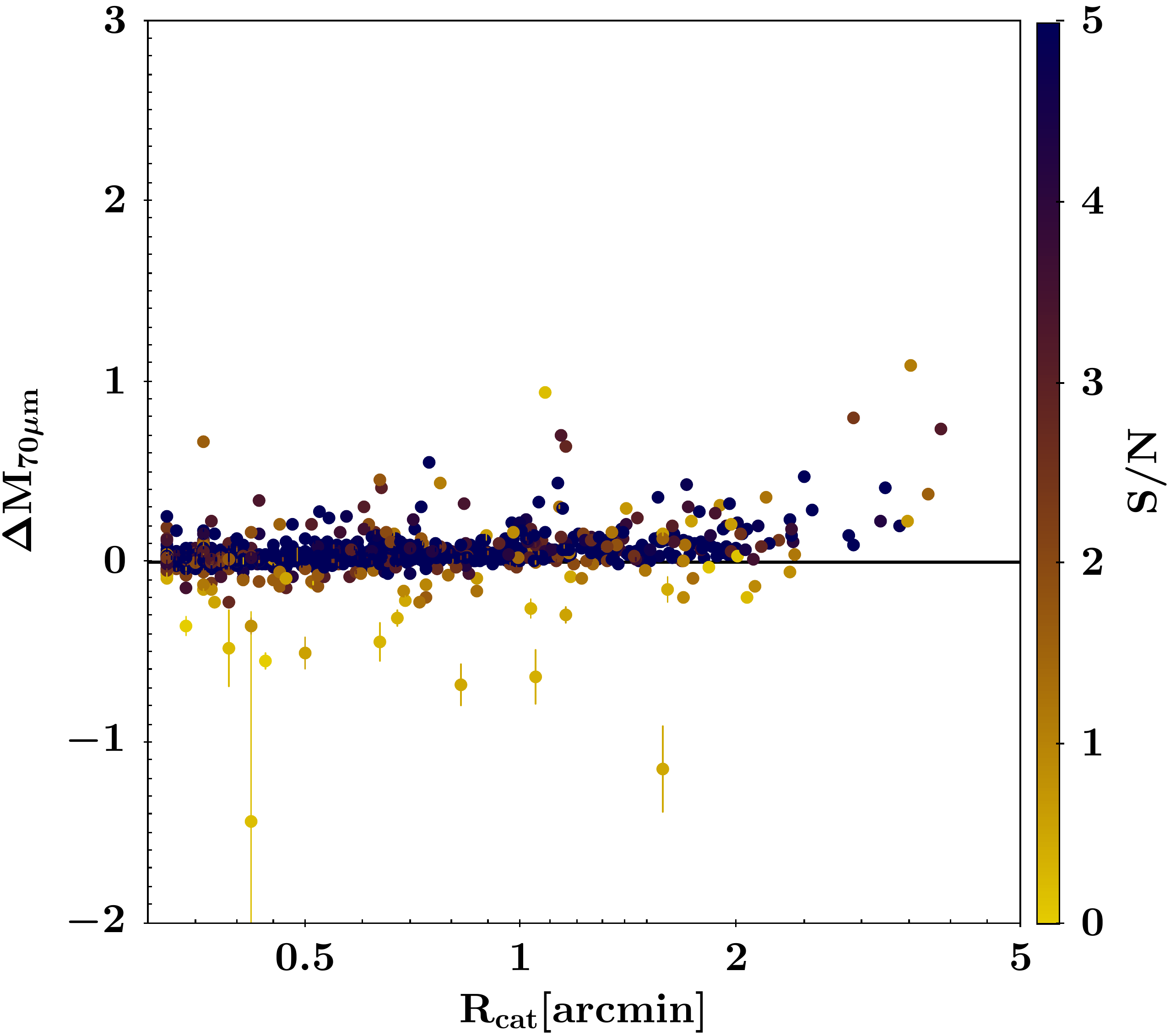}
\includegraphics[width=0.485\textwidth]{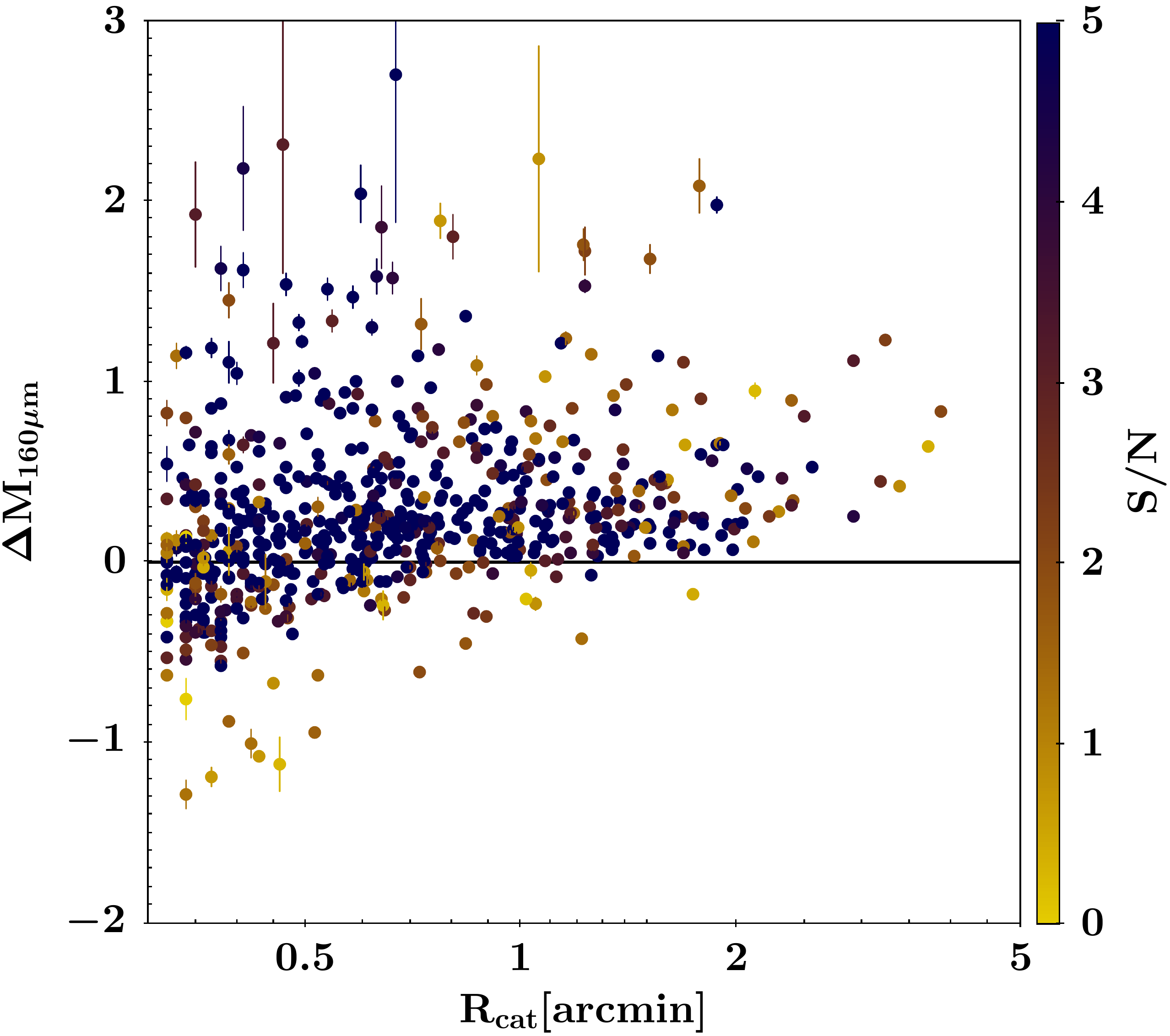}
 \caption{Distribution of flux differences between aperture and segmentation photometric method as function of the angular extension of the bubble. Radii are given in arcminutes.The color scales based on the S/N of the relative aperture flux estimation. } \label{distr_radius}
\end{figure*}

\begin{figure*}
\includegraphics[width=0.48\textwidth]{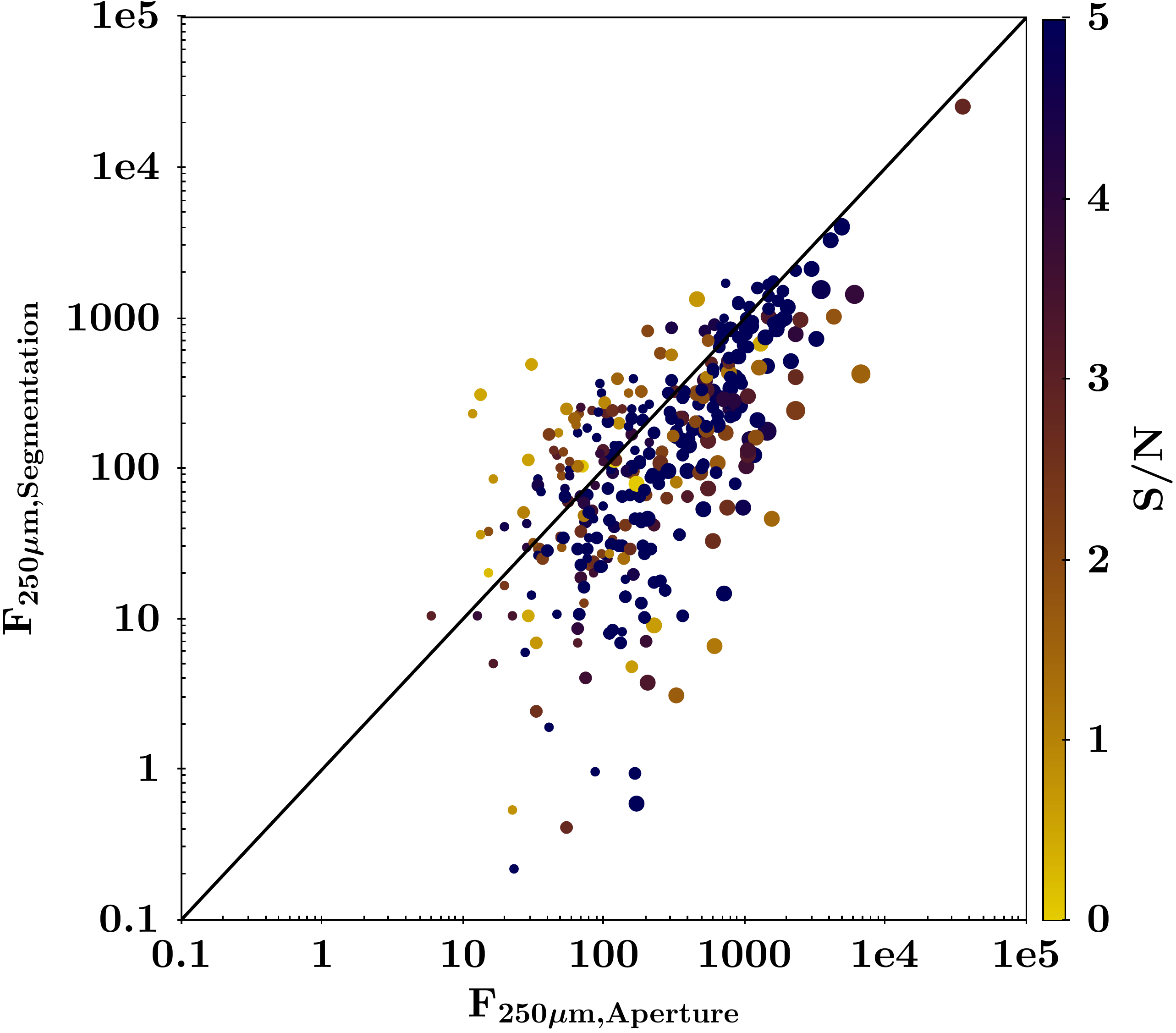}
\includegraphics[width=0.475\textwidth]{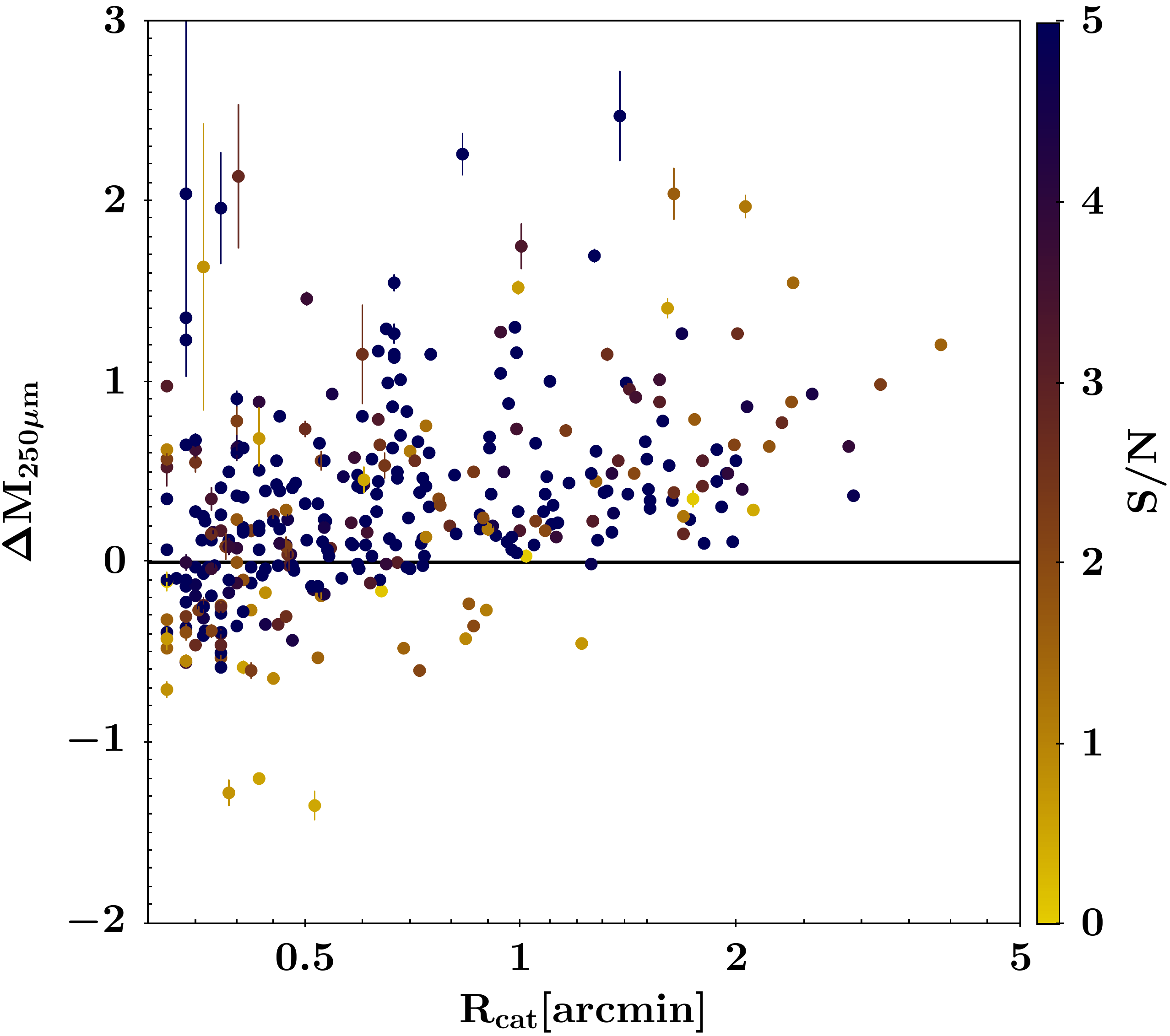}
\includegraphics[width=0.48\textwidth]{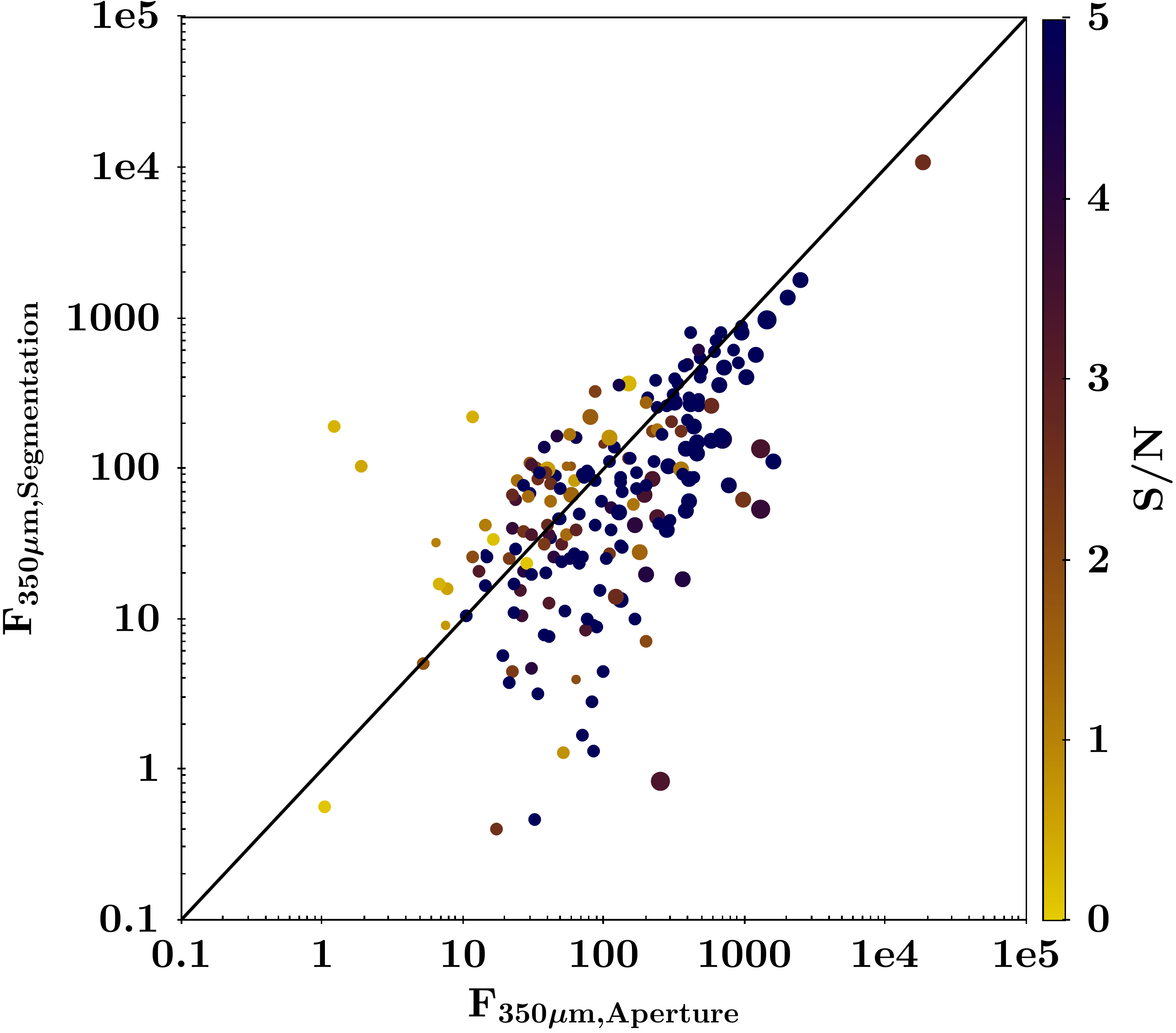}
\includegraphics[width=0.475\textwidth]{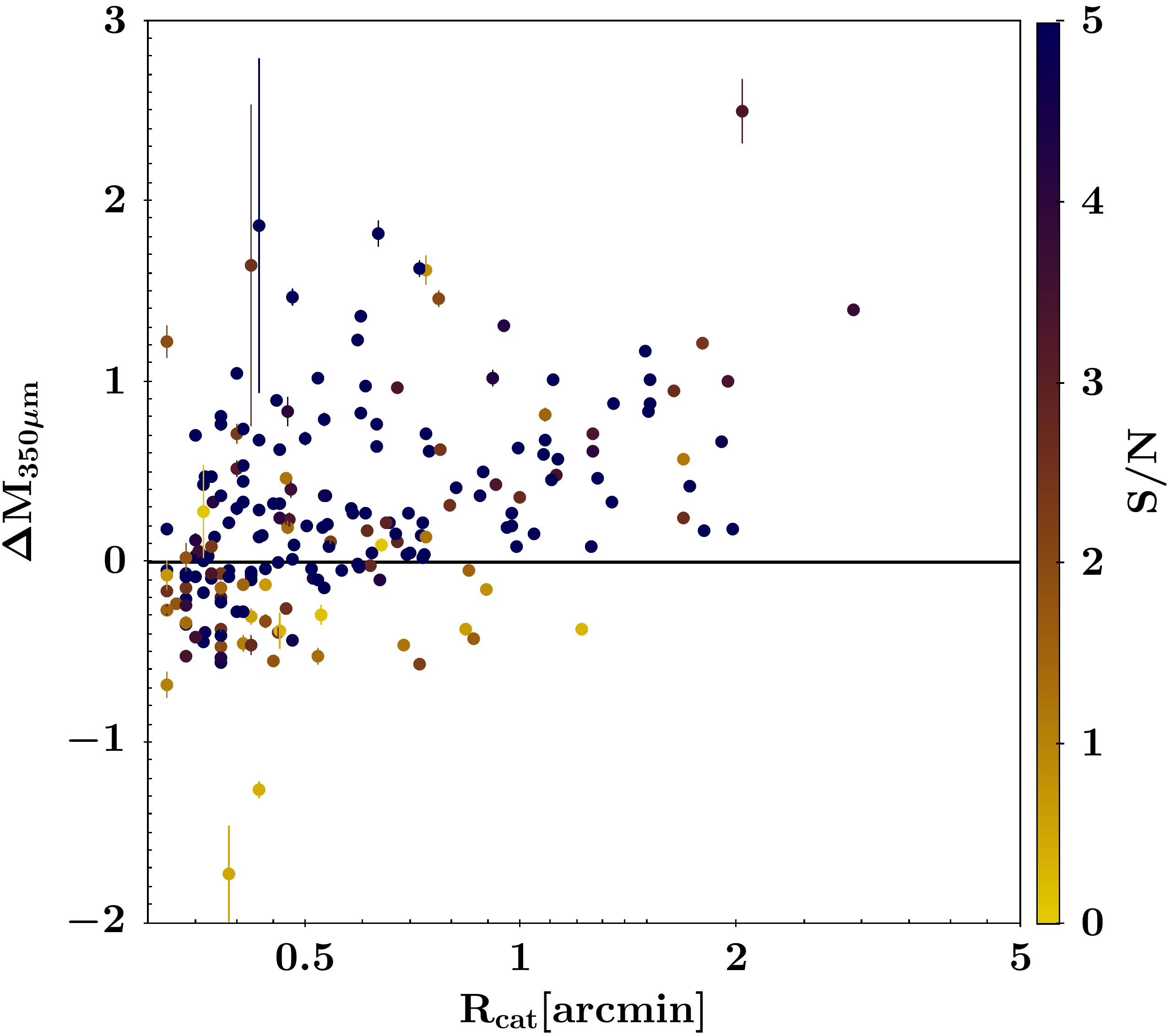}
\includegraphics[width=0.48\textwidth]{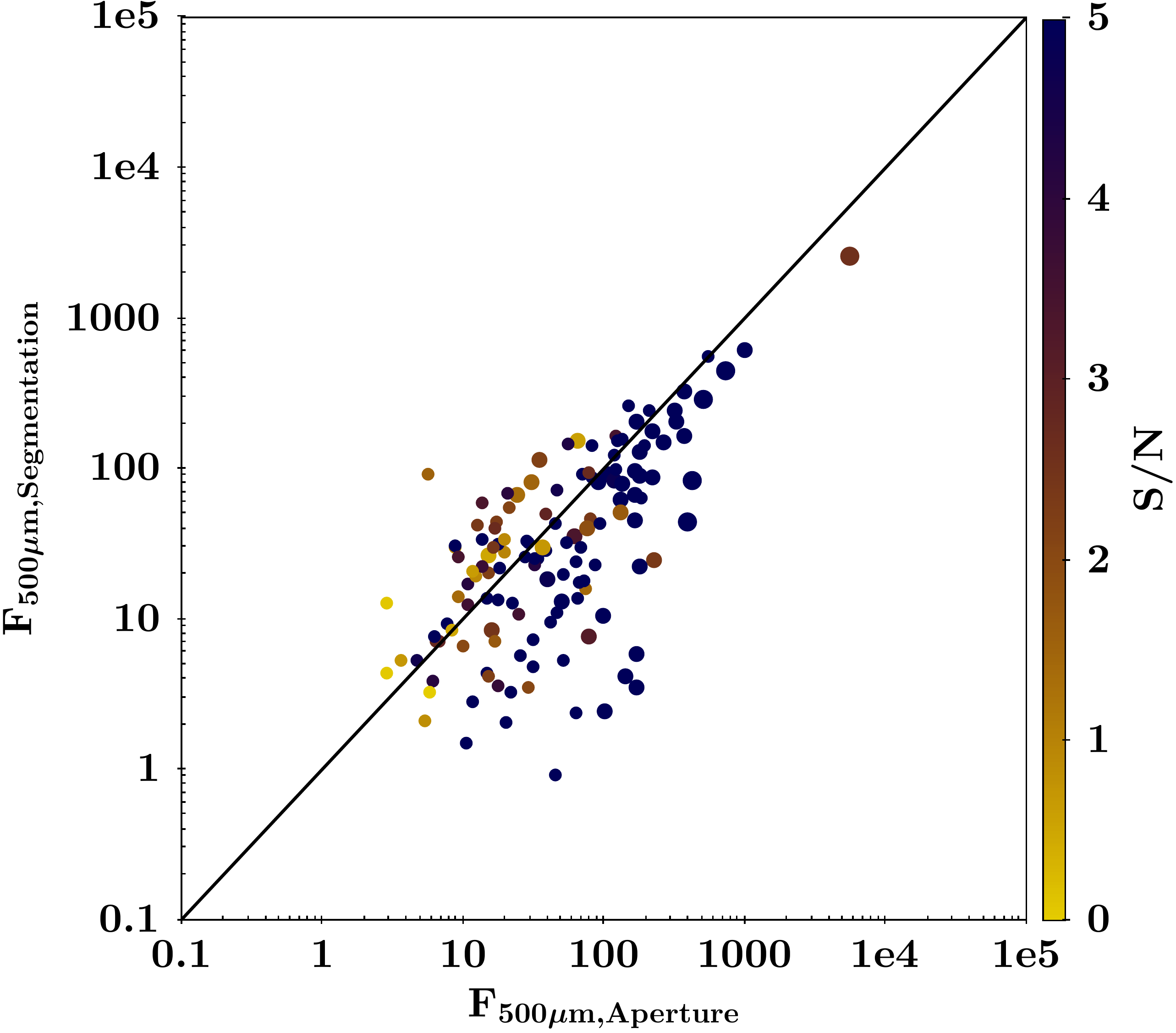}
\includegraphics[width=0.475\textwidth]{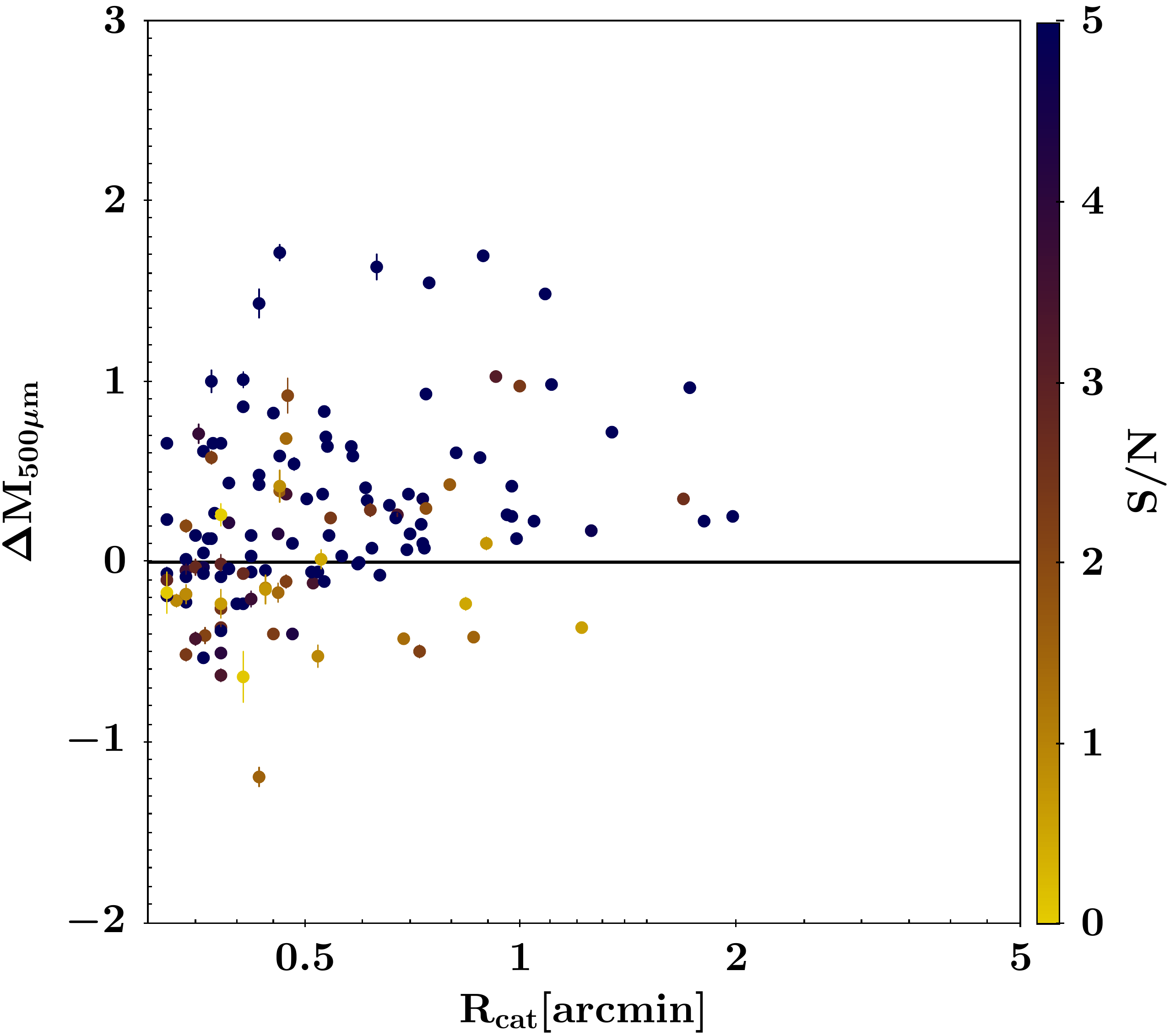}
 \caption{{\it Left Column:} Comparison of flux estimates using the aperture photometry against the segmentation one. Dots size is proportional to the logarithm of the bubble radius ($R_{cat}$). {\it Right Column:} Distribution of flux differences between aperture and segmentation photometric method as function of the angular extension of the bubble. Radii are given in arcminutes. For all the plots, color scales based on the S/N of the relative aperture flux estimation.} \label{dist_red}
\end{figure*}

 \section{Discussion} \label{discussion} 
 
\begin{figure}
\includegraphics[width=0.47\textwidth]{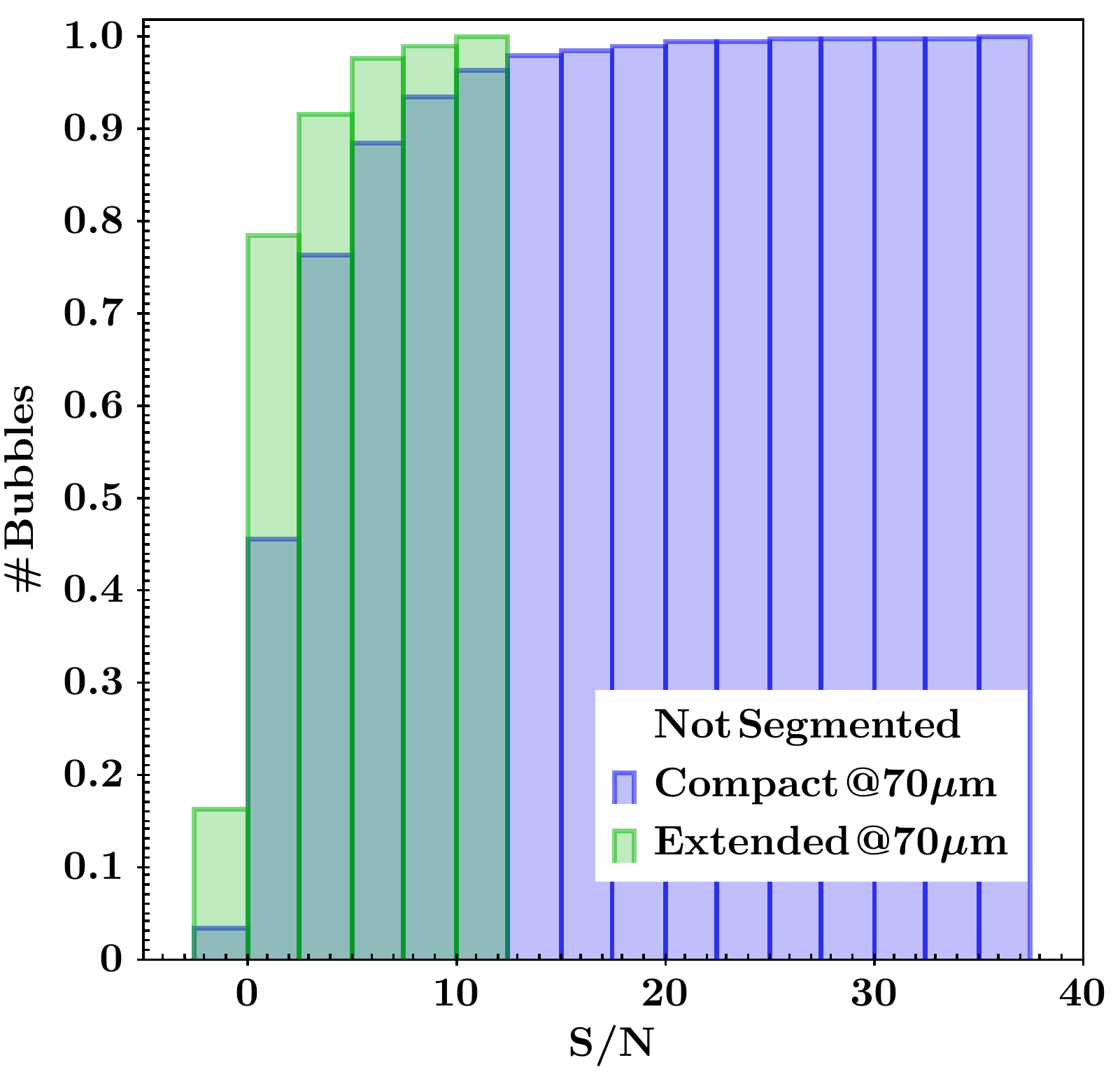}
 \caption{Cumulative distribution of bubbles with failed segmentation at 70\um\, as function of the S/N from aperture flux estimation. } \label{histo}
\end{figure}

\begin{figure}
\includegraphics[width=0.47\textwidth]{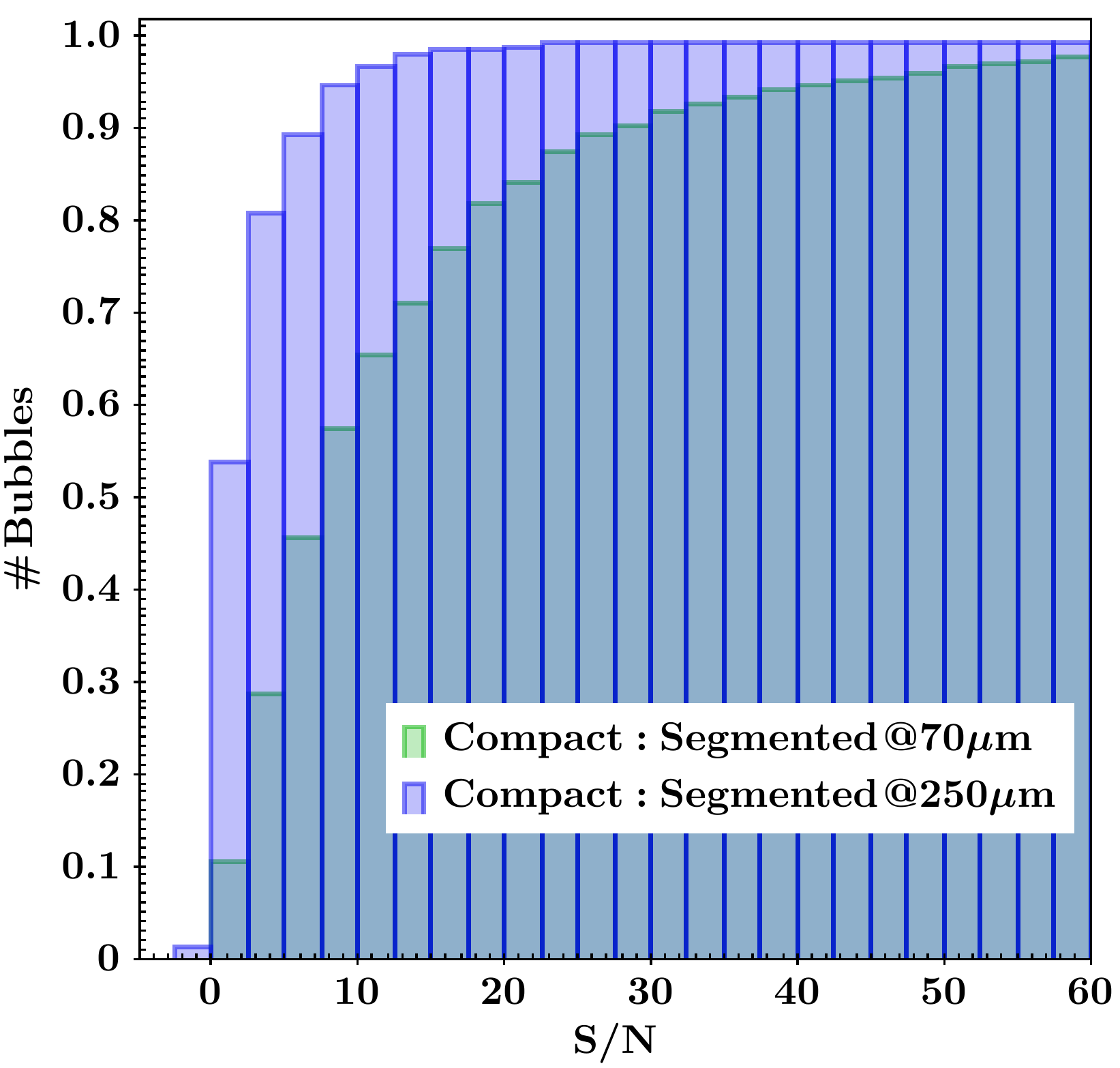}
\includegraphics[width=0.47\textwidth]{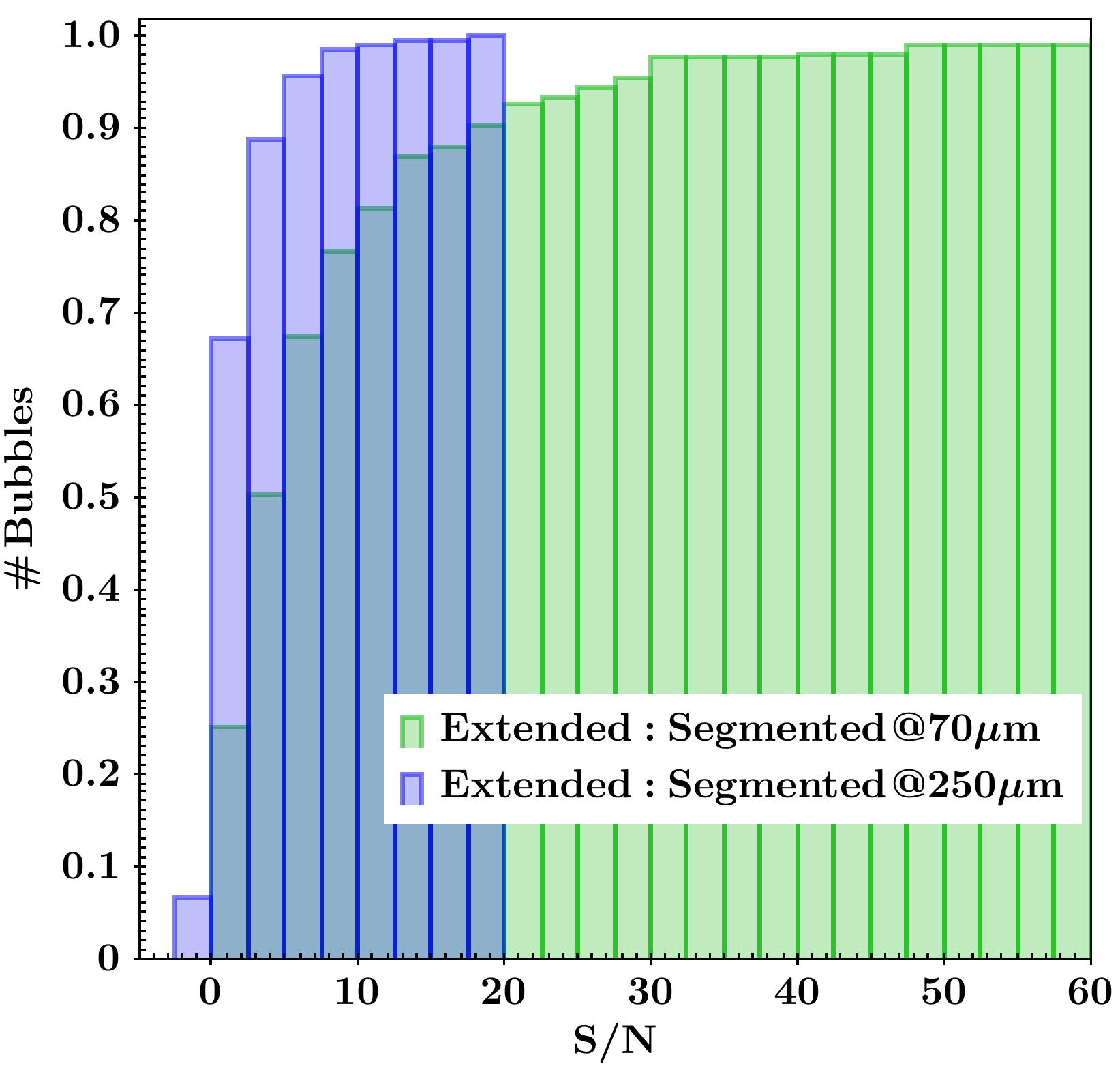}
 \caption{Cumulative distribution of segmented bubbles at 70\um\, and 250\um\, as function of the S/N of the aperture flux estimation.  } \label{histo2}
\end{figure}

Fluxes obtained from the aperture and segmentation method are presented in Figure\,\ref{dist_flux_fig}.
For each bubble, we plotted the aperture photometry flux against the segmentation one, with the size of the dots proportional to the logarithm of the bubbles radius ($R_{cat}$) and the color scaled based on the S/N of the corresponding aperture flux estimation.
The latter efficiently conveys the brightness of a source over the possible complex background.
We defined the ratio  between the two flux measurements for each bubble as        
\begin{equation}\label{eq2}
\Delta M_\lambda = Log F_{\lambda, Ap.} - Log F_{\lambda, Segm.} = Log ({F_{\lambda, Ap.} }/{F_{\lambda, Segm.}})
\end{equation}
and plotted it as a function of the angular dimension of the bubble in Figure\,\ref{distr_radius}.
In Table\,\ref{distr_flux}, we reported the average $\langle\Delta M_\lambda\rangle$ value for each bandpass,
obtained after clipping values more than  3-$\sigma$ away from the average. The total numbers of detected/clipped bubbles sample are also given.\\ 
A very good agreement between aperture and segmentation photometry is visible at 70\um, where $\langle F_{\lambda, Ap.} / F_{\lambda, Segm.}\rangle$ is close to 1, giving a relative difference of around 7\%:
this bandpass is the one that better quantifies the deviation between the two methods, since the segmentation masks are produced using the images taken at this wavelength,
thus the contours best trace the bubble shape.
Consequently, we can assess that at 70\um\, the aperture photometry fluxes are generally slightly larger than the segmentation ones.
 This is what we found also in the comparison of the two methods at the other bandpasses,
having convolved the mask to reproduce instrumental differences from 70\um\, images,
but also assumed a physical similar shape. 
 Higher aperture flux is likely ascribable to the contamination
in the aperture method from the local background flux that cannot be totally removed by subtracting the average level and/or that falls within the aperture but ``outside'' the bubble contours as defined by the segmentation method.
The effects of a complex background could explain why $\Delta M_\lambda$ increases at larger radii ($R_{cat}> 2\arcmin$), as visible in Figure\,\ref{distr_radius}.
This could also explain why bubbles for which $\Delta M_\lambda$ is below zero ($F_{\lambda, Ap.} <F_{\lambda, Segm.}$)  all have a low S/N.
Thus, the fact that low S/N bubbles preferentially leads to $\Delta M_\lambda < 0$ seems to suggest that the segmentation is the better choice, when available, for the flux measurements of extended sources.\\
The good agreement between aperture and segmentation photometry stands out, especially at short wavelengths, with a difference $\left|\Delta M_\lambda\right|<0.1$ for  65\%, 68\% and 81\% of the cases at 12\um, 22\um\, and 70\um, respectively.
The distribution becomes more widely spread at 160\um\, as visible in Figures\,\ref{dist_flux_fig} and \ref{distr_radius} and as indicated by $\langle\Delta M_\lambda\rangle$
in Table\,\ref{distr_flux}. This is  a consequence of a dominant emission from the background that, moving red-ward, increasingly contaminates the flux measurements.
At 160\um, the number of bubbles with $\left|\Delta M_\lambda\right|<0.1$ drops significantly to 19\% and at 250\um\, to 14\%, with average   relative difference that could reach $\sim$50--90\%.\\
As showed in Table\,\ref{bubbles}, we found that the active contour segmentation method failed for about 45\% of the detected bubbles.
The reason of such a fraction of failed segmentation could be attributed to the presence of a possibly complex background, which is expressed by a low S/N of the measurements. \\
 We split the bubbles in {\it extended} ($R_{cat}>60\arcsec$) and {\it compact}   ($R_{cat}\le60\arcsec$)
sources and showed in Figure\,\ref{histo} the relative distribution of bubbles with a failed segmentation as a function of the S/N.
Since no S/N is estimated for non-segmented bubbles, for consistency we used that from aperture photometry instead.
As expected extended bubbles with strong contamination or the presence of a structurally complex background affects the boundary found by the algorithm. 
 Indeed, at  70\um\, the fraction of extended bubbles with a failed segmentation is equal to 55\%, and 92\% of them has a S/N lower than 5 (see Figure\,\ref{histo}), while  
such fraction is $\sim$70\% if we include segmented bubbles. \\
For compact bubbles, such effect seems to affect the distribution less, since they are characterized by a lower fraction of unsegmented bubbles with low S/N than the extended bubbles: 40\% of
compact bubbles have a failed segmentation, and 76\% of them are characterized by a S/N $<$ 5, corresponding to a slightly smaller fraction ($\sim$64\%) compared to the whole low S/N sample. 
This confirms that extended bubbles are more affected by the failures of the segmentation method  surely because of the background contamination.\\
Average $\langle\Delta M_\lambda\rangle$ is always positive at all bandpasses and increases moving to longer wavelengths:  
the relative difference between the two methods obtained from $\Delta M_\lambda$ is less than 15\% for 12\um\, and 22\um\, and goes up to 50--70\% at longer wavelengths with a peak at 250\um\, ($\sim$93\%). 
This effect clearly points to a relevant flux contamination from the background.
This is supported also by the steep drop of the number of bright bubbles (see Table\,\ref{distr_flux}).
 In Figure\,\ref{histo2},  we show the cumulative distributions of segmented compact and extended bubbles at 70\um\, and 250\um\, as function of the S/N of the flux measurements.
The S/N characterizing the bubbles sensibly decreases at longer wavelengths  due to a stronger background emission or/and to a weaker source emission: the fraction of extended bubbles with a S/N $> $5 
goes from  53\%  at 70\um\, down to 21\%  at  250\um, while for compact bubbles the fraction changed from 72\% to 39\%  at 70\um\, and 250\um, respectively.\\
Thus,  giving the  underestimation of the total emitted flux of the segmentation method with respect to the aperture observed in Fig.\,\ref{dist_flux_fig}
and obtained from the analysis of low S/N bubbles, segmentation method demonstrates to be closer to the real flux than the aperture one.
Nevertheless, when using it, the fact that the real shape of the bubble contours  at $\lambda\ne$70\um\,  could differ from that assumed by using the segmentation masks 
should be taken into consideration.
\\

 \begin{table}
 \caption{Average difference between bubbles fluxes measured with aperture and segmentation photometric method. The number of ``clipped'' bubbles over the total number of bubbles are given, which turn out to be ``bright''  after the application of both two methods. } \label{distr_flux}
 \centering
 \begin{tabular}{@{}rcc@{}}   
      \hline
  Band &  $\langle\Delta M_\lambda\rangle$     & 			  Clipped/Bright Bubbles	\\
\hline
 12\um  &  0.06  $\pm$   0.11&    954/1008  \\
22\um  &  0.06  $\pm$ 0.07&       950/1012        \\
70\um   & 0.03  $\pm$ 0.05&     935/1020        \\
160\um  & 0.23 $\pm$ 0.37&     632/663       \\
250\um  &  0.28 $\pm$ 0.47&     336/346        \\
350\um  &    0.25 $\pm$ 0.52&   203/206        \\
500\um   &  0.18 $\pm$ 0.47&   136/138        \\
   \hline
 \end{tabular}
 \end{table}

\section{Conclusions}\label{conclusions}
\begin{figure}
\includegraphics[width=0.475\textwidth]{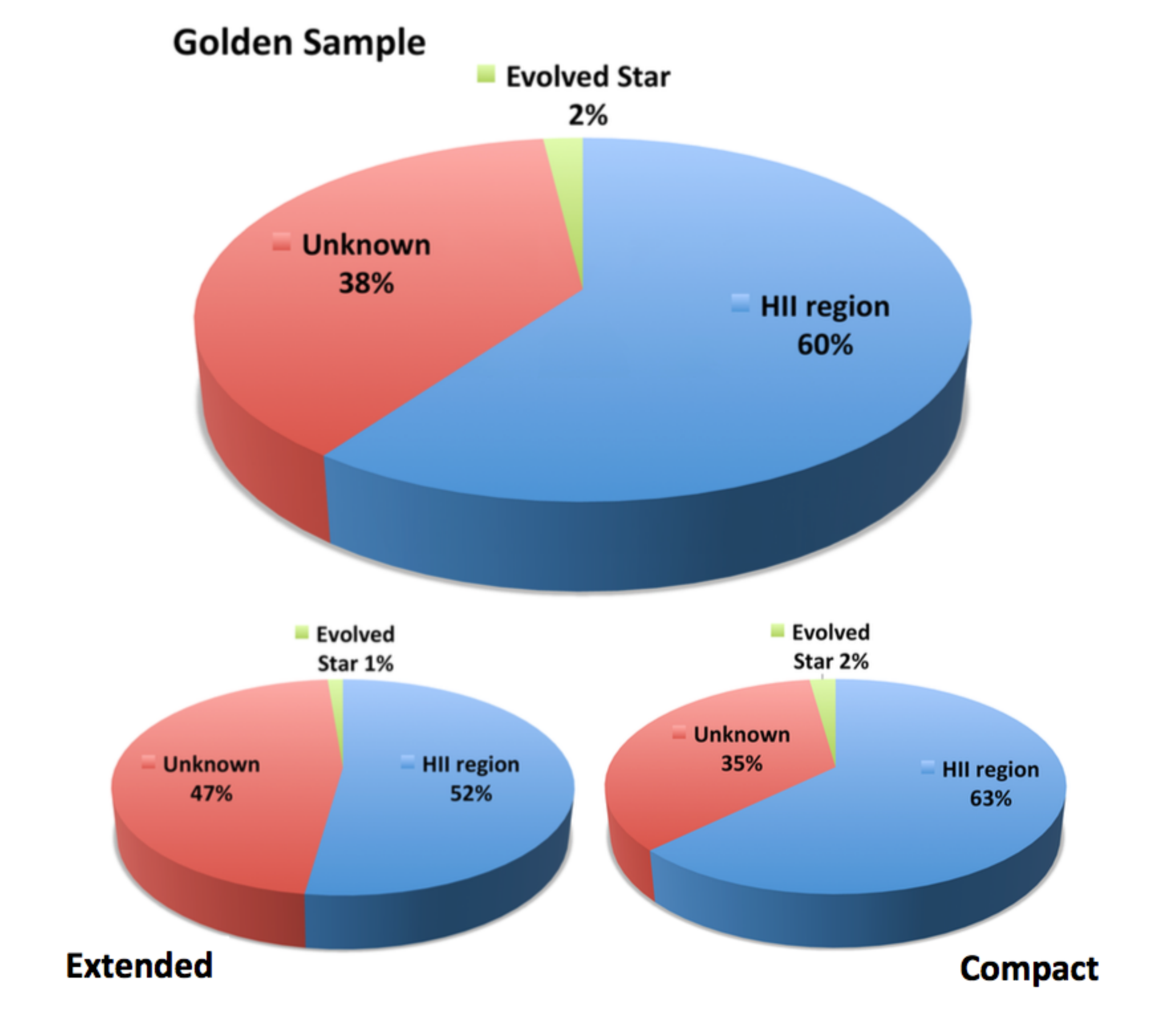}
 \caption{Fraction of classified/unclassified golden sample bubbles from SIMBAD database.  } \label{torta}
\end{figure}

This work, born in the wide framework of the VIALACTEA project\footnote{http://vialactea.iaps.inaf.it}, has been inspired by
the unique opportunity that  \herschel\, telescope offers, thanks to its sensitivity and its large wavelength coverage in the far infrared, to derive 
the physical conditions in Galactic bubbles, whose origins  can radically differ being the yield of different stages of star evolution.
We took advantage of the availability of the wide image dataset collected from the Hi-GAL survey, flanked it with the \wise\, survey data
sampling the emissions at shorter wavelengths, and produced the most extensive catalogue of IR fluxes of extended sources.
Thus, in this work, we presented the fluxes of a golden sample of 1814 Galactic bubbles taken from \citet{Simpson}, acquired
at 12\um, 22\um, 70\um, 160\um, 250\um, 350\um\,  and 500\um\, bandpasses.\\
We used two approaches for the flux estimation: a classical aperture photometry  and a more innovative method, based on the use of segmentation masks,
produced by an image analysis algorithm, called active contour, which defines the boundaries of the bubbles (see Appendix\,\ref{segmentation} for a brief explanation of the method).
In both methods, we used the bubbles dimension provided by \citet{Simpson}, to define a circular aperture region centered on the bubble centroid where we estimated the source fluxes and an annular region around it for the local average background level definition.\\
Fluxes obtained with both aperture and segmentation photometry were checked comparing them with those of a more limited sample of \hiirs\, and PNe from A12, obtained with an interactive method. We found a very good agreement, especially with the aperture method results. 
On the other hand, segmentation photometry seems to work better at short wavelengths but fails over compact objects, for which segmentation algorithm shows to have a high failure rate in producing bubble masks.\\
Finally we compared fluxes of the golden sample bubbles obtained with the two methods, finding a very good agreement especially at the shorter wavelengths 
(average difference does not exceed 15\%). Generally, aperture photometry fluxes turn out to be larger than the segmentation ones, possibly a consequence of a contaminating complex background, whose subtraction can be a tricky task. Indeed such effect gets stronger  at long wavelengths ($>$160\um) where background dust emission increases. \\ 
With this work we offer for the first time a wide catalogue of bubble IR fluxes, produced using fully automated methods.
This kind of approach, together with automated algorithms using for instance data mining capabilities for e.g. the source extraction of extended sources or the automated definition of the source contours (see e.g. Carey et al, in prep; \citealt{Riggi}), is a necessary choice in the astrophysical data analysis considering the new generation instruments (e.g. LSST in the optical, JWST in the IR and SKA at the radio frequencies), which will survey wide sky regions providing a gigantic amount of data. \\
We checked, querying the SIMBAD astronomical database \citep{Simbad}\footnote{We noticed that \hiirs\, identified by \citet{Anderson14}, as well as \citet{Anderson11} and \citet{Paladini03}, were not included in the SIMBAD database. Thus, we additionally check on such catalogues for our statistics.}, if the bubbles in the golden sample have been identified with a specific star evolution event. 
We considered all the objects in a circular area of radius $R_{cat}$ centered on each bubble centroid coordinates and selected the one at the minimum distance as the literature object associated with the bubble (assuming negligible the probability of false matches coming from perspective coincidences).
Finally, we split the objects in {\it \hiirs,\, Evolved Stars} (which includes LBV stars, AGB and post-AGB stars, SNRs, PNe, etc.) and {\it Unknown}.
We noticed that the fraction of bubbles that are classified as \hiirs\, is the highest one (60\%), against the very low 2\% of the evolved stars (see Figure\,\ref{torta}).
Fractions do not change significantly if we split bubbles in {\it extended} and {\it compact}, as done in the previous Section.
In any case, a large fraction (38\%) remains unclassified.      
Such finding together with the future perspective of large amount of available data, strengthens the need for an automated method for the bubbles classification, possibly based on their Spectral Energy Distribution and/or on their morphology at different wavelength. This issue exceeds the purpose  of this work, but it will be matter of discussion of a forthcoming paper.

\section*{Acknowledgments}
The authors acknowledgethe anonymous referee for the very useful comments that helped in improving the paper.
This work is part of the VIALACTEA Project, a Collaborative Project under Framework Programme 7
of the European Union, funded under Contract \# 607380 that is hereby acknowledged.
F.B. acknowledges support from the VIALACTEA Project.
\herschel is an ESA space observatory with science instruments provided by
European-led Principal Investigator consortia and with important participation
from NASA.
PACS has been developed by a consortium of institutes led by MPE
(Germany) and including UVIE (Austria); KUL, CSL, IMEC (Belgium); CEA,
OAMP (France); MPIA (Germany); IAPS, OAP/OAT, OAA/CAISMI, LENS,
SISSA (Italy); IAC (Spain). This development has been supported by the funding
agencies BMVIT (Austria), ESA-PRODEX (Belgium), CEA/CNES (France),
DLR (Germany), ASI (Italy), and CICYT/MCYT (Spain).
SPIRE has been developed by a consortium of institutes led by Cardiff
Univ. (UK) and including Univ. Lethbridge (Canada); NAOC (China); CEA,
LAM (France); IAPS, Univ. Padua (Italy); IAC (Spain); Stockholm Observatory
(Sweden); Imperial College London, RAL, UCL-MSSL, UKATC, Univ. Sussex
(UK); Caltech, JPL, NHSC, Univ. colourado (USA). This development has been
supported by national funding agencies: CSA (Canada); NAOC (China); CEA,
CNES, CNRS (France); ASI (Italy); MCINN (Spain); Stockholm Observatory
(Sweden); STFC (UK); and NASA (USA).\\
This publication makes use of data products from the Wide-field Infrared Survey Explorer, which is a joint project of the University of California, Los Angeles, and the Jet Propulsion Laboratory/California Institute of Technology, funded by the National Aeronautics and Space Administration.
This research has made use of the SIMBAD database, operated at CDS, Strasbourg, France.
This research made use of Montage. It is funded by the National Science Foundation under Grant Number ACI-1440620, and was previously funded by the National Aeronautics and Space Administration's Earth Science Technology Office, Computation Technologies Project, under Cooperative Agreement Number NCC5-626 between NASA and the California Institute of Technology.

\appendix
\section{The Active Contours Method for Segmentation }\label{segmentation}

Active Contours are a family of popular curve deformation techniques that are often applied within computer vision for the unsupervised segmentation of image objects. Their simple mode of operation has allowed them to be applied to a variety of different problems (\citealt{Akram}; \citealt{Lankton};  \citealt{Yilmaz}). They are especially useful in instances where the use of supervised machine learning approaches is implausible. This is because labelled data is expensive to generate and the robustness of these techniques makes them good candidates for situations where there is a lack of expert annotations. However, the drawback of most contouring methods is that they come with the caveat that parameters have to be substantially adjusted before segmentations will meet human expectations. 
For example, a well-established, and parameter heavy, Active Contour model is the Localised Variant (\citealt{Lankton}; \citealt{Wang09}; \citealt{Yang12}) which evaluates contour deformation functions within the bounds of predefined kernels. The size of the kernels largely dictate the end segmentation result and since they have to be defined a priori, its use in real world problems is often limited. 
Therefore, this paper addresses the above issue with a novel generic adaptive kernel selection scheme that also makes use of the sign of Magnetostatic forces \citep{Xie}. The use of signed electrostatic information enables textured foreground regions to be delineated from low gradient background areas. However, as astronomy images are composed of complex objects of varying intensity, the segmentation results of Magnetostatic Active Contours cannot be relied upon and more appropriate results are achieved when this technique is aided by local statistical information.  
The main steps adopted in this paper are reported in the following text, while a more detailed description of the method is given in Carey et al. (2017, in prep.).

\subsection{Preprocessing}
\begin{figure*}
\includegraphics[width=1.\textwidth]{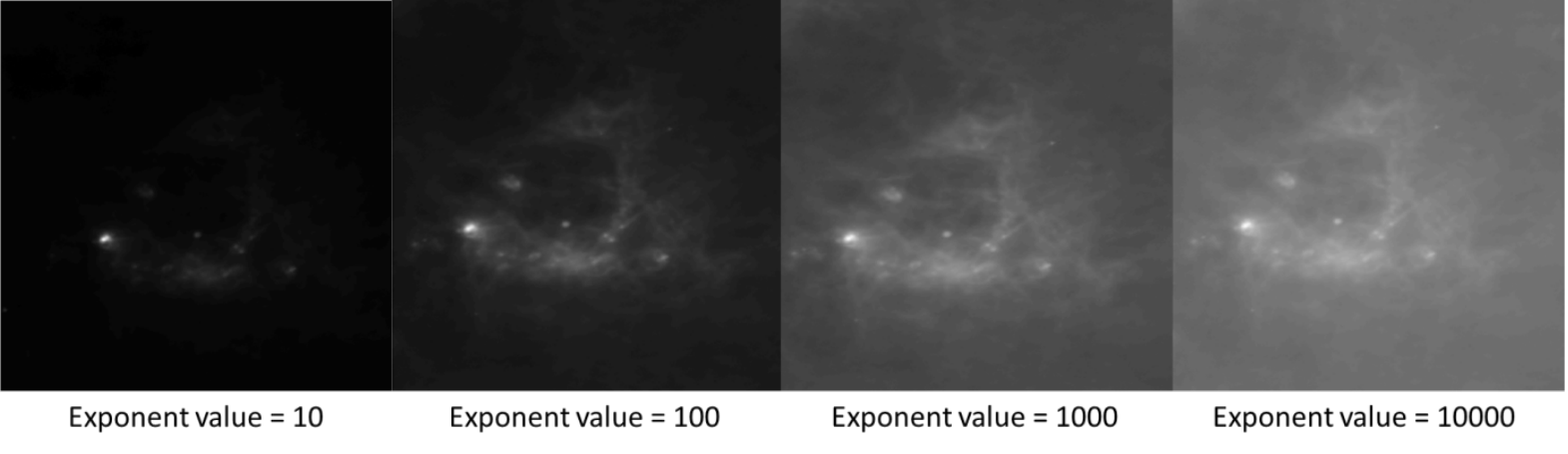}
\caption{ The result of using a range of different exponent values. The gradient magnitude of these are correlated against those of the log of the magnitude of the original data so that an appropriate exponent values can be found. \label{duane3}}
\end{figure*}

The above approaches work best when the image data under consideration is made more amenable to segmentation. In this work, preprocessing amounted to:  adjusting the dynamic range of the original image data by selecting a log transformation       coefficient which maximised the correlation between the transformed and the log of the original data's numerical gradients;  locally maximising the contrast (defined as the difference in intensity between local pixels)  of the transformed data; data  smoothing and compression via a discrete wavelet transformation and appropriately initialising the active contour algorithm. \\
\noindent{\bf Automatic Log Exponent Selection}--  Log transformation is a standard tool used in astronomy and results in the compression of the dynamic range of the data. The skewness of the original image, where there are only a few pixels with high intensity values, means that if active contours were applied to this data then it will only fit around very bright objects. Therefore, the difference in intensity between high  and low valued pixels needs to be decreased and this is accomplished by Log transformations. When too high exponent values are used in this process, more pixels will reach the maximum transformed intensity value of 255. This will make the boundaries of the bubble features more difficult to locate as is evident from Figure\,\ref{duane3} where the definition of the edges of the image object has varying degrees of ambiguity over the range of exponent values used.
Notice that when using a value of 10,000, the bright source in the bottom left of the image dominates the local region and the true edge of the object is lost. Therefore, similar problems as using the original image data will persist. The same is true when using exponent values that are too low, where differences in pixel intensity values with respect to the images background are lost. The edges of the transformed data will be weak when definition is lost and so correlating the transformed images gradient with that of the log of the gradient magnitude of the original will mean that an exponent value is selected that preserves as much authentic boundary information as possible.

\begin{figure}
\includegraphics[width=0.45\textwidth]{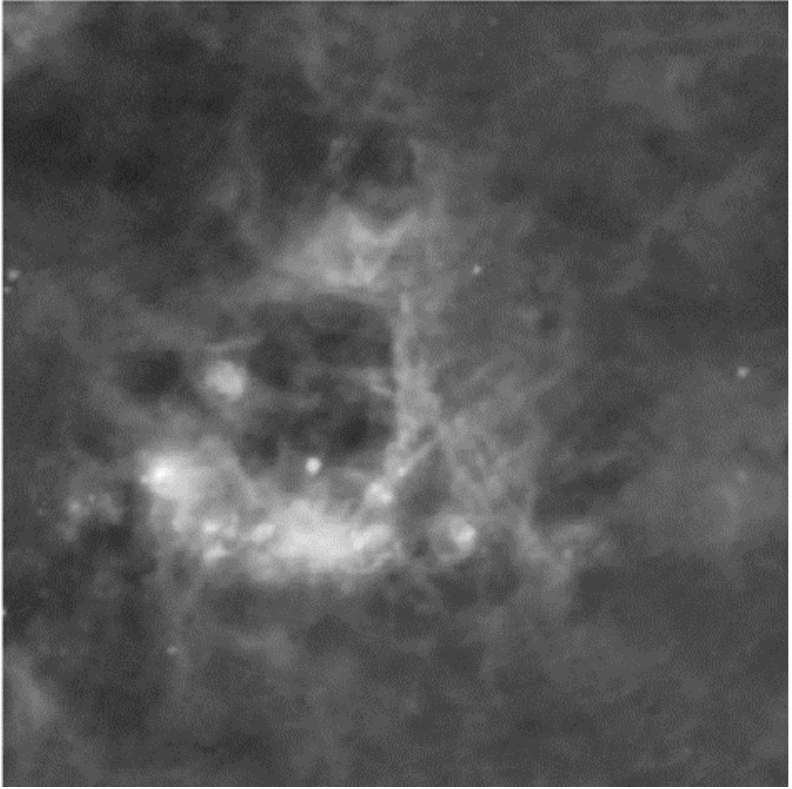}
\caption{ The result of using the 1000 exponent transformed image of Figure\,\ref{duane3} with Localised Contrast Enhancement with a tile size of 40 by 40 pixels. \label{duane4}}
\end{figure}

\begin{figure}
\includegraphics[width=0.5\textwidth]{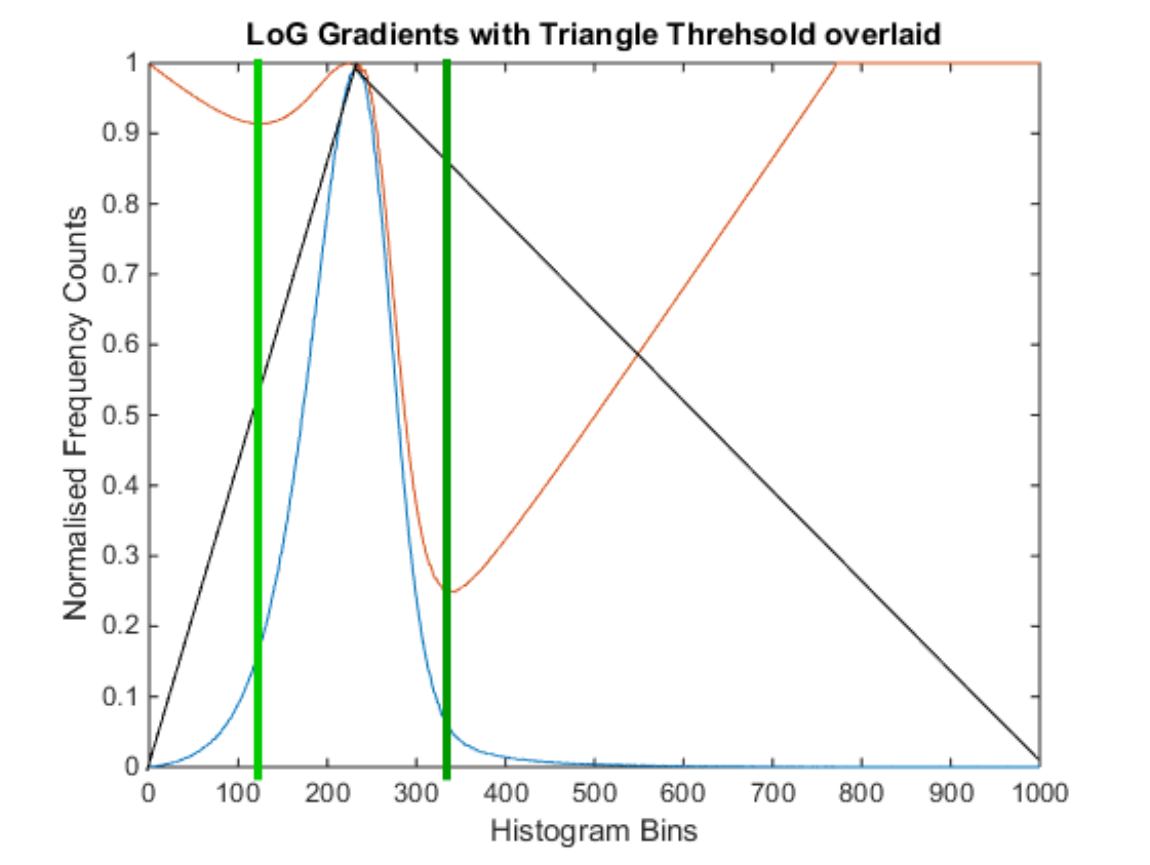}
\caption{ A histogram of the original imageÕs Log gradients. The blue line is the histogram, the black lines are the triangles formed by the triangle thresholding, the red lines are the distances of the histogramÕs bin counts away from their nearest triangle hypotenuse and the green line are the selected threshold points.   Please note that the red lines contained in the chart do not relate to the y axis. Depending on the distance calculation used - the calculated "distance" can either be negative or positive. In this case, the negative output was used so the largest negative value was found for the selected thresholds.
\label{duane5}}
\end{figure}

\noindent{\bf Local Adaptive Contrast Histogram Equalisation}--
The selection of an appropriate log transformation preserves as much boundary information as possible but the definition of the objects boundaries and their contrast with respect to the background can still be poor (due to the inherent skewness of the original data). Therefore, a tiling process, known as Locally Adaptive Contrast Histogram Equalization (LACHE), was used to adjust the local pixel intensity histograms of patches of the original image   \citep{LACHE}: in this paper a local tile of 40 by 40 pixels was used. The adjustments result in the stretching of the local histograms so that very weak pixels become more apparent. This greatly improves the definition of the bubble features as is apparent when Figure\,\ref{duane4} is compared to those present in Figure\,\ref{duane3}. \\
\noindent{\bf Discrete Wavelet Transformation}--The localised enhancement of the image features helps in the segmentation process but the large variations introduced by the above process needs to be corrected for. In this instance, this was facilitated by a Discrete Wavelet Transformation (DWT). 
 As LACHE is a local technique, errors in enhancement can happen at the edges of the local tiles this algorithm uses. This can effect where the boundaries of bubble object reside and so this essential smoothing process enables the active contours to better fit the features of interest. DWT not only smooths out the data but also reduces its size. \\
\noindent{\bf Initialisation}-- The starting position of the active contour, from which it will grow, also has a large impact on the end segmentation result. In this paperÕs approach, the incorporation of Magnetostatic forces relieved this difficulty to an extent, but the contours still needed to be started fairly close to object of interest.
For example, every pixel in an image will have gradient and these will not necessarily correspond to image objects of interest. Therefore, the simple use of thresholding to provide active contour intialisation points will mean that contours could grow around undesired structures within the data. To avoid this, the triangle thresholding algorithm \citep{Rogers} was used with the log of the gradient magnitude of the original image data. Its mode of operation is demonstrated in Figure\,\ref{duane5}. In this technique, a histogram is formed from the log gradient magnitude image.  Its maximum peak, the first and last non empty histogram bins, are  found so that two triangles can be formed between its two extrema and the largest concavity (the largest distance) found for each triangle allows an objective threshold to be found for automatic active contour initialisation.
 Since two thresholds are found, the highest is selected for this segmentation pipeline assuming that celestial objects are usually brighter than their background, as usually observed.

\subsection{Adaptive Kernel Selection}
\begin{figure}
\includegraphics[width=0.45\textwidth]{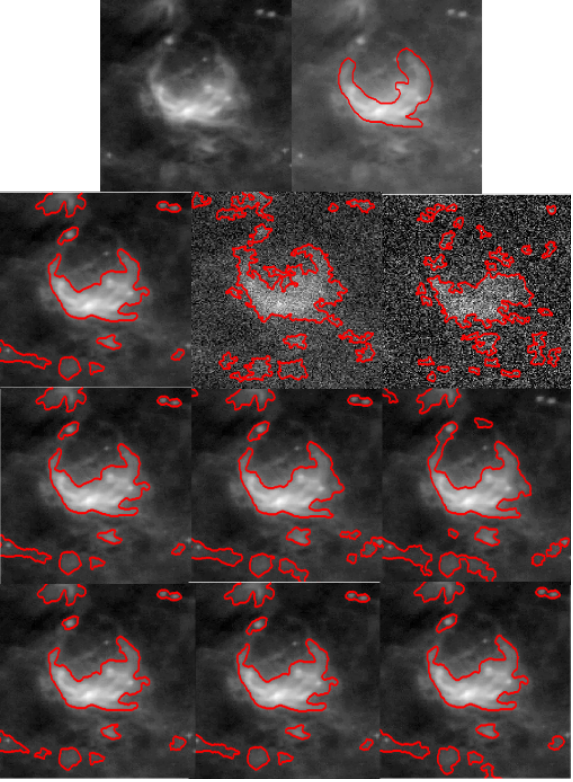}
\caption{The effect of noise and parameters on segmentation quality. Top row) The original data and an astronomer ground truth in red. Second row) The effect of Gaussian noise of zero mean and 0, 0.01 and 0.03 variance on the segmentation. Third row) The effect of different $\tau$ values (0.5, 1 and 2) with a fixed initial kernel size. Fourth row) The effect of different initial kernel sizes (10 by 10 pixels, 20 by 20 pixels and 40 by 40 pixels).  \label{duane1}}
\end{figure}

Once the data was made more amenable to segmentation, active contours can be used for segmentation. The approach taken in this paper relied upon the use of gradient and local statistical information. Gradient information was incorporated via Magnetostatic forces and local statistical information was acquired by the use of an adaptive kernel selection scheme. In traditional approaches, the size of the local area used in the collection of image statistics is predetermined but this is unlikely to reflect the changing content of an image over which the contour evolves. To reflect this changing texture, an initial kernel size, set to be 20 by 20 pixels in this paper, is used to compute the local parametric Bhattacharyya distance between the inside and outside regions of an evolving contour
which is defined as:

\[  BD = exp\Bigg(\!\!-\frac{1}{4}ln\bigg(\!\frac{1}{4}\bigg(\!\frac{\sigma_{in}^2}{\sigma_{out}^2}+\frac{\sigma_{out}^2}{\sigma_{in}^2}+2   \bigg)\!\bigg) +  \]
\begin{equation}\label{BCdistance}
+\frac{1}{4}\bigg( \frac{(\mu_{in}\!-\!\mu_{out})^2}{\sigma_{in}^2\!+\!\sigma_{out}^2}\bigg)\!\Bigg)
\end{equation}

where $\mu$ and $\sigma$ are the respective mean and standard deviations of the intensities on the inside and outside of a local region around the evolving contour.  This gives a normalised measure of how similar the inside and outside regions of the evolving curve are,  allowing large kernels to be selected for regions of homogeneity and small sizes for areas of texture. 
The weighting of this with a user defined parameter, $\tau$ (set to be 0.5 throughout), allowed kernel sizes to be selected in a single pass approach by multiplying Equation\, \ref{BCdistance} with a maximum desired kernel size. 
Once appropriate kernel sizes have been selected, any localised regional evolution 
function can be used. In this instance, the following function was used to guide contours towards salient objects:
\begin{equation}
F(x) = \frac{\big(I(x)-\mu_{in}\big)^2}{2\sigma_{in}^2}-\frac{\big(I(x)-\mu_{out}\big)^2}{2\sigma_{out}^2}+log\bigg( \frac{\sigma_{out}}{\sigma_{in}}\bigg)
\end{equation}
where $F$ is the active contour energy, $I$ is the image of interest, $x$ is a point along the evolving contour and the rest of the nomenclature is the same as Eq.\,\ref{BCdistance}. \\
Experiments have suggested that the choice of the weighting parameter, as well as the quality of the underlying data and the initial kernel size used to probe texture has little effect on the end segmentation result, e.g. Figure\,\ref{duane1}. We calculated the Dice values, which are a measure of how well automatically generated segmentations overlap with ground truths. For the results within Figure\,\ref{duane1}, Dice values vary within a fairly decent range of 0.55 (in the extreme noise case) to 0.84.

\subsection{Magnetostatic Forces}
\begin{figure}
\includegraphics[width=0.45\textwidth]{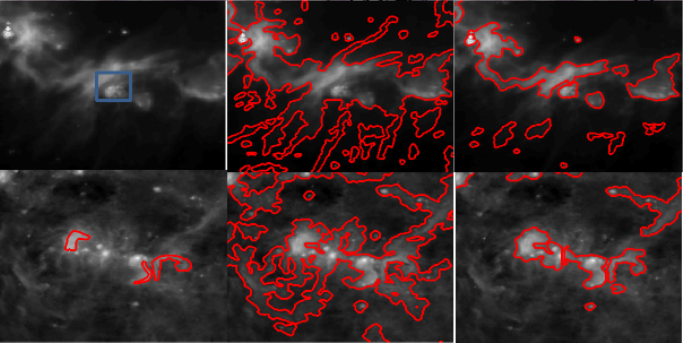}
\caption{ The improved results achieved when adaptive local active contours are used with Magnetostatic forces. The first column is the original data with examples of objects of interest highlighted by blue boxes/red outlines. The middle column is the Magnetostatic forces by themselves and the last column is the result of the adaptive localised contour.  \label{duane2}}
\end{figure}

The sole reliance on either localised or Magnetostatic forces to bring about acceptable segmentations is limiting and this is evident from Figure\,\ref{duane2}. In this figure, a comparison is made against the use of just Magnetostatic information for segmentation against its combination with local image statistics. It can be seen that this complementary methodology can aid the image segmentation process. It is also preferable to just using local information as the whole process would then be reliant upon the statistical approximation being made. 

Magnetostatic forces are derived via a magnetic density flux coefficient matrix, {\bf B}, which is given by:
\begin{center}
\begin{eqnarray}
e1=g \cdot (-\nabla y \otimes)\frac{1}{r^2} \qquad ; \qquad e2=g \cdot \nabla x \otimes \frac{1}{r^2} \nonumber \\
{\bf B}=\nabla e2 - \nabla e1
\end{eqnarray}
\end{center}

Where $\otimes$ is the convolution operation performed in the frequency domain, $g$ is the gradient magnitude of the image and $r^2$ is a centred Euclidean distance matrix padded to the maximum length of an image of interest for frequency filtering.
Magnetostatic forces operate by creating a signed map of where regions of gradient reside in an image and the simple thresholding of this at zero, or another user defined constant, produces a mask which can constrain the growth of the contour around areas of high gradient. This negates the use of background areas in the statistical analysis step of the segmentation. 

The refinement of the segmentation with Magnetostatic field diffusion and local image statistics has
enabled an image segmentation pipeline  that requires little human intervention.
Intervention in this paper, took the form of experimenting with different active contour energies and testing different kernel selection schemes. There is unlikely to ever be one active contour method that would be suitable for all astronomy data and an investigation of this sort should always be conducted before any attempt at the use of active contours is made.
However, currently the segmentation just provides outlines of astronomical objects and makes no
discrimination between the features it finds. Therefore, in the future, approaches will be developed
that will allow machine learning techniques to be utilised in the identification of objects of interest
(bubbles for example).

\section{Photometry Tables}\label{photometrytables}


\begin{table*}
 \centering
 \begin{minipage}{140mm}
\caption{Golden Sample Aperture Photometry at 12\um\, and 22\um. }\label{wiseap}
\begin{tabular}{@{}lrrrrrrrrrr@{}}
\hline
&&&&\multicolumn{3}{c}{12\um}&&\multicolumn{3}{c}{22\um} \\
\cline{5-7}
\cline{9-11}
Bubble ID & Long\footnote{Bubble Galactic Coordinates given by \citet{Simpson}.} & Lat$^a$ & $R_{cat}$\footnote{$R_{cat}$ is taken from \citet{Simpson} catalogue and corresponds to the effective radius for the small bubbles or to half the outer diameter in case of large ones.}&Flux& err&S/N& &Flux & err&S/N \\
&(deg)&(deg)&(arcsec)&(Jy)&(Jy)&&&(Jy)&(Jy)&\\
\hline
 MWP1G358760-007700S & 358.76   & -0.77    & 21.6    & 0.72   & 0.01 & 3.1  & & 2.01    & 0.06 & 4.6   \\
  MWP1G358770+001090  & 358.77   & 0.109    & 20.2    & 0.25   & 0.01 & 5.3  & & 1.00    & 0.06 & 4.4   \\
  MWP1G358840-007400S & 358.84   & -0.74    & 27.0    & 2.27   & 0.01 & 5.3  & & 6.66    & 0.07 & 22.1  \\
  MWP1G358881+000576  & 358.881  & 0.058    & 40.0    & 18.02  & 0.02 & 33.5 & & 52.88   & 0.13 & 27.2  \\
  MWP1G358890+000800S & 358.89   & 0.08     & 20.4    & 1.15   & 0.01 & 8.5  & & 2.81    & 0.06 & 9.2   \\
  MWP1G358950-000200S & 358.95   & -0.02    & 25.2    & 1.31   & 0.01 & 4.1  & & 3.44    & 0.10 & 1.7   \\
  MWP1G359275-000403  & 359.275  & -0.04    & 23.5    & 1.99   & 0.03 & 1.9  & & 12.69   & 0.13 & 2.8   \\
  MWP1G359282-008955  & 359.282  & -0.895   & 83.0    & 7.78   & 0.06 & 0.4  & & 7.33    & 0.21 & 0.3   \\
  MWP1G359300+002883  & 359.3    & 0.288    & 195.7   & 55.49  & 0.14 & 0.3  & & 162.05  & 0.56 & 0.3   \\
  MWP1G359350-004141  & 359.35   & -0.414   & 49.5    & 2.68   & 0.03 & 1.9  & & 34.60   & 0.14 & 7.3   \\
  MWP1G359411+000363  & 359.411  & 0.036    & 42.6    & 10.31  & 0.07 & 1.9  & & 60.63   & 0.19 & 5.4   \\
  MWP1G359420+000200S & 359.42   & 0.02     & 19.2    & 1.76   & 0.03 & 0.4  & & 12.47   & 0.15 & 1.4   \\
  MWP1G359450-000200S & 359.45   & -0.02    & 24.0    & 0.23   & 0.02 & 0.2  & & 7.63    & 0.12 & 1.5   \\
  MWP1G359514+002727  & 359.514  & 0.273    & 166.0   & 16.87  & 0.12 & 0.2  & & 198.83  & 0.51 & 1.8   \\
  MWP1G359569-004772  & 359.569  & -0.477   & 199.0   & --     & --   & --   & & --      & --   & --    \\
  MWP1G359740-005900S & 359.74   & -0.59    & 22.8    & 1.62   & 0.01 & 6.8  & & 7.16    & 0.06 & 34.8  \\  
  ...&&&&&&&&&&\\
\hline
\end{tabular}
\end{minipage}
\end{table*}


\begin{table*}
 \centering
 \begin{minipage}{140mm}
\caption{Golden Sample Segmentation Photometry at 12\um\, and 22\um.} \label{wiseseg}
\begin{tabular}{@{}lrrrrrrrrrr@{}}
\hline
&&&&\multicolumn{3}{c}{12\um}&&\multicolumn{3}{c}{22\um} \\
\cline{5-7}
\cline{9-11}
Bubble ID & Long\footnote{Bubble Galactic Coordinates given by \citet{Simpson}.} & Lat$^a$ & $R_{cat}$\footnote{$R_{cat}$ is taken from \citet{Simpson} catalogue and corresponds to the effective radius for the small bubbles or to half the outer diameter in case of large ones.}&Flux& err&S/N& &Flux & err&S/N \\
&(deg)&(deg)&(arcsec)&(Jy)&(Jy)&&&(Jy)&(Jy)&\\
\hline
  MWP1G358760-007700S & 358.76   & -0.77    & 21.6    & 0.85   & 0.01 & 4.0  & & 2.25    & 0.06 & 5.7  \\
  MWP1G358770+001090  & 358.77   & 0.109    & 20.2    & --     & --   & --   & & --      & --   & --   \\
  MWP1G358840-007400S & 358.84   & -0.74    & 27.0    & 2.41   & 0.01 & 6.3  & & 6.62    & 0.06 & 21.5 \\
  MWP1G358881+000576  & 358.881  & 0.058    & 40.0    & 17.10  & 0.02 & 23.9 & & 49.11   & 0.10 & 23.0 \\
  MWP1G358890+000800S & 358.89   & 0.08     & 20.4    & 0.91   & 0.01 & 4.4  & & 1.80    & 0.04 & 4.0  \\
  MWP1G358950-000200S & 358.95   & -0.02    & 25.2    & 1.27   & 0.01 & 3.2  & & 4.60    & 0.07 & 3.0  \\
  MWP1G359275-000403  & 359.275  & -0.04    & 23.5    & --     & --   & --   & & --      & --   & --   \\
  MWP1G359282-008955  & 359.282  & -0.895   & 83.0    & --     & --   & --   & & --      & --   & --   \\
  MWP1G359300+002883  & 359.3    & 0.288    & 195.7   & --     & --   & --   & & --      & --   & --   \\
  MWP1G359350-004141  & 359.35   & -0.414   & 49.5    & 2.34   & 0.01 & 1.5  & & 7.57    & 0.06 & 3.3  \\
  MWP1G359411+000363  & 359.411  & 0.036    & 42.6    & --     & --   & --   & & --      & --   & --   \\
  MWP1G359420+000200S & 359.42   & 0.02     & 19.2    & --     & --   & --   & & --      & --   & --   \\
  MWP1G359450-000200S & 359.45   & -0.02    & 24.0    & --     & --   & --   & & --      & --   & --   \\
  MWP1G359514+002727  & 359.514  & 0.273    & 166.0   & --     & --   & --   & & --      & --   & --   \\
  MWP1G359569-004772  & 359.569  & -0.477   & 199.0   & --     & --   & --   & & --      & --   & --   \\
  MWP1G359740-005900S & 359.74   & -0.59    & 22.8    & 1.48   & 0.01 & 4.3  & & 6.78    & 0.05 & 23.9 \\
...&&&&&&&&&\\
\hline
\end{tabular}
\end{minipage}
\end{table*}

\begin{landscape}
\begin{table}
 \begin{minipage}{240mm}
\caption{Golden Sample Aperture Photometry at 70\um, 160\um, 250\um, 350\um\, and 500\um.} \label{hgap}
\begin{tabular}{@{}lrrrrrrrrrrrrrrrrrrrrrr@{}}
\hline
&&&&\multicolumn{3}{c}{70\um}&&\multicolumn{3}{c}{160\um} &&\multicolumn{3}{c}{250\um}&&\multicolumn{3}{c}{350\um}&&\multicolumn{3}{c}{500\um}\\
\cline{5-7}
\cline{9-11}
\cline{13-15}
\cline{17-19}
\cline{21-23}
Bubble ID & Long\footnote{Bubble Galactic Coordinates given by \citet{Simpson}.} & Lat$^a$ & $R_{cat}$\footnote{$R_{cat}$ is taken from \citet{Simpson} catalogue and corresponds to the effective radius for the small bubbles or to half the outer diameter in case of large ones.}&Flux& err&S/N& &Flux & err&S/N&&Flux & err&S/N&&Flux & err&S/N&&Flux & err&S/N \\
&(deg)&(deg)&(arcsec)&(Jy)&(Jy)&&&(Jy)&(Jy)&&&(Jy)&(Jy)&&&(Jy)&(Jy)&&&(Jy)&(Jy)&\\
\hline
  MWP1G358760-007700S & 358.76   & -0.77    & 21.6    & 47.20    & 0.68   & 6.3   & & 128.25     & 2.22     & 9.8     & & 89.05    & 1.74   & 7.3   & & 49.06    & 1.13   & 7.7   & & 18.37   & 0.62  & 7.1  \\
  MWP1G358770+001090  & 358.77   & 0.109    & 20.2    & 23.39    & 1.15   & 3.5   & & 31.23      & 6.15     & 0.7     & & 17.52    & 4.58   & 0.5   & & 4.65     & 2.57   & 0.6   & & 4.03    & 1.16  & 0.7  \\
  MWP1G358840-007400S & 358.84   & -0.74    & 27.0    & 166.08   & 0.99   & 20.5  & & 289.98     & 3.48     & 11.1    & & 179.74   & 2.82   & 8.7   & & 87.14    & 1.86   & 8.1   & & 31.63   & 0.96  & 9.0  \\
  MWP1G358881+000576  & 358.881  & 0.058    & 40.0    & 1103.74  & 6.95   & 11.4  & & 789.43     & 12.56    & 5.9     & & 372.57   & 7.56   & 4.5   & & 170.20   & 4.10   & 4.5   & & 59.86   & 1.91  & 4.5  \\
  MWP1G358890+000800S & 358.89   & 0.08     & 20.4    & 98.44    & 1.88   & 6.7   & & 148.42     & 6.62     & 3.2     & & 84.45    & 5.64   & 2.1   & & 29.93    & 4.48   & 1.4   & & 13.25   & 2.67  & 0.9  \\
  MWP1G358950-000200S & 358.95   & -0.02    & 25.2    & 57.45    & 5.42   & 0.8   & & 44.45      & 7.86     & 1.3     & & 41.75    & 4.26   & 2.3   & & 22.82    & 2.45   & 2.8   & & 13.90   & 1.23  & 3.7  \\
  MWP1G359275-000403  & 359.275  & -0.04    & 23.5    & 243.59   & 5.40   & 3.2   & & 155.56     & 14.64    & 0.6     & & 59.48    & 11.92  & 0.4   & & 29.32    & 7.17   & 0.4   & & 16.79   & 3.54  & 0.4  \\
  MWP1G359282-008955  & 359.282  & -0.895   & 83.0    & 181.10   & 4.89   & 0.1   & & 561.47     & 15.14    & 0.3     & & 292.88   & 9.32   & 0.3   & & 116.16   & 4.97   & 0.3   & & 40.35   & 2.21  & 0.4  \\
  MWP1G359300+002883  & 359.3    & 0.288    & 195.7   & 2875.17  & 22.17  & 0.2   & & 7833.53    & 51.46    & 0.9     & & 4420.15  & 30.40  & 1.4   & & 1995.04  & 16.34  & 1.6   & & 660.95  & 7.46  & 1.6  \\
  MWP1G359350-004141  & 359.35   & -0.414   & 49.5    & 12.27    & 3.31   & 0.5   & & 70.59      & 8.79     & 0.9     & & 84.00    & 5.12   & 1.4   & & 52.52    & 2.73   & 2.0   & & 21.17   & 1.32  & 2.3  \\
  MWP1G359411+000363  & 359.411  & 0.036    & 42.6    & 1262.36  & 9.62   & 3.6   & & 973.32     & 15.11    & 2.0     & & 256.28   & 9.06   & 0.3   & & 59.42    & 5.07   & -0.6  & & 11.02   & 2.50  & -0.7 \\
  MWP1G359420+000200S & 359.42   & 0.02     & 19.2    & 411.34   & 6.92   & 2.8   & & 345.95     & 13.36    & 2.2     & & 88.14    & 10.27  & 0.9   & & 41.18    & 5.30   & 1.1   & & 20.67   & 2.27  & 1.4  \\
  MWP1G359450-000200S & 359.45   & -0.02    & 24.0    & ***      & ***    & ***   & & -267.39    & 15.42    & -1.4    & & ***      & ***    & ***   & & ***      & ***    & ***   & & ***     & ***   & ***  \\
  MWP1G359514+002727  & 359.514  & 0.273    & 166.0   & 3960.82  & 12.21  & 1.9   & & 3574.36    & 46.98    & 0.4     & & 564.90   & 36.44  & -0.1  & & 116.42   & 20.75  & -0.2  & & 45.81   & 9.36  & -0.2 \\
  MWP1G359569-004772  & 359.569  & -0.477   & 199.0   & 8392.94  & 18.38  & 1.6   & & 8238.16    & 72.53    & 0.4     & & 1229.62  & 55.34  & -0.0  & & 242.08   & 30.33  & -0.1  & & 42.60   & 13.86 & -0.2 \\
  MWP1G359740-005900S & 359.74   & -0.59    & 22.8    & 111.24   & 0.97   & 9.6   & & 104.36     & 1.88     & 10.7    & & --       & --     & --    & & --       & --     & --    & & --      & --    & --   \\
...&&&&&&&&&&&&&&&\\
\hline
\end{tabular}
\end{minipage}
\end{table}
\end{landscape}

\begin{landscape}
\begin{table}
 \centering
 \begin{minipage}{240mm}
\caption{Golden Sample Segmentation Photometry at 70\um, 160\um, 250\um, 350\um\, and 500\um.}\label{hgseg}
\begin{tabular}{@{}lrrrrrrrrrrrrrrrrrrrrrr@{}}
\hline
&&&&\multicolumn{3}{c}{70\um}&&\multicolumn{3}{c}{160\um} &&\multicolumn{3}{c}{250\um}&&\multicolumn{3}{c}{350\um}&&\multicolumn{3}{c}{500\um}\\
\cline{5-7}
\cline{9-11}
\cline{13-15}
\cline{17-19}
\cline{21-23}
Bubble ID & Long\footnote{Bubble Galactic Coordinates given by \citet{Simpson}.} & Lat$^a$ & $R_{cat}$\footnote{$R_{cat}$ is taken from \citet{Simpson} catalogue and corresponds to the effective radius for the small bubbles or to half the outer diameter in case of large ones.}&Flux& err&S/N& &Flux & err&S/N&&Flux & err&S/N&&Flux & err&S/N&&Flux & err&S/N \\
&(deg)&(deg)&(arcsec)&(Jy)&(Jy)&&&(Jy)&(Jy)&&&(Jy)&(Jy)&&&(Jy)&(Jy)&&&(Jy)&(Jy)&\\
\hline
  MWP1G358760-007700S & 358.76   & -0.77    & 21.6    & 47.15    & 0.22  & 10.0  & & 235.62   & 3.00   & 1.9    & & 158.04   & 1.91  & 2.0     & & 72.29    & 0.86  & 2.3   & & 21.47   & 0.32  & 2.2   \\
  MWP1G358770+001090  & 358.77   & 0.109    & 20.2    & --       & --    & --    & & --       & --     & --     & & --       & --    & --      & & --       & --    & --    & & --      & --    & --    \\
  MWP1G358840-007400S & 358.84   & -0.74    & 27.0    & 165.75   & 0.41  & 19.8  & & 280.10   & 2.80   & 3.3    & & 107.62   & 0.63  & 9.3     & & 40.84    & 0.32  & 9.2   & & 4.80    & 0.12  & 7.8   \\
  MWP1G358881+000576  & 358.881  & 0.058    & 40.0    & 1102.47  & 2.34  & 12.1  & & 527.76   & 8.96   & 2.1    & & ***      & ***   & ***     & & ***      & ***   & ***   & & ***     & ***   & ***   \\
  MWP1G358890+000800S & 358.89   & 0.08     & 20.4    & 85.73    & 0.63  & 5.6   & & ***      & ***    & ***    & & ***      & ***   & ***     & & ***      & ***   & ***   & & ***     & ***   & ***   \\
  MWP1G358950-000200S & 358.95   & -0.02    & 25.2    & 131.68   & 1.00  & 4.1   & & 452.10   & 9.99   & 1.4    & & 168.01   & 5.60  & 1.5     & & 67.34    & 2.39  & 1.5   & & 22.26   & 0.81  & 1.6   \\
  MWP1G359275-000403  & 359.275  & -0.04    & 23.5    & --       & --    & --    & & --       & --     & --     & & --       & --    & --      & & --       & --    & --    & & --      & --    & --    \\
  MWP1G359282-008955  & 359.282  & -0.895   & 83.0    & --       & --    & --    & & --       & --     & --     & & --       & --    & --      & & --       & --    & --    & & --      & --    & --    \\
  MWP1G359300+002883  & 359.3    & 0.288    & 195.7   & --       & --    & --    & & --       & --     & --     & & --       & --    & --      & & --       & --    & --    & & --      & --    & --    \\
  MWP1G359350-004141  & 359.35   & -0.414   & 49.5    & 59.75    & 0.53  & 0.3   & & ***      & ***    & ***    & & ***      & ***   & ***     & & ***      & ***   & ***   & & ***     & ***   & ***   \\
  MWP1G359411+000363  & 359.411  & 0.036    & 42.6    & --       & --    & --    & & --       & --     & --     & & --       & --    & --      & & --       & --    & --    & & --      & --    & --    \\
  MWP1G359420+000200S & 359.42   & 0.02     & 19.2    & --       & --    & --    & & --       & --     & --     & & --       & --    & --      & & --       & --    & --    & & --      & --    & --    \\
  MWP1G359450-000200S & 359.45   & -0.02    & 24.0    & --       & --    & --    & & --       & --     & --     & & --       & --    & --      & & --       & --    & --    & & --      & --    & --    \\
  MWP1G359514+002727  & 359.514  & 0.273    & 166.0   & --       & --    & --    & & --       & --     & --     & & --       & --    & --      & & --       & --    & --    & & --      & --    & --    \\
  MWP1G359569-004772  & 359.569  & -0.477   & 199.0   & --       & --    & --    & & --       & --     & --     & & --       & --    & --      & & --       & --    & --    & & --      & --    & --    \\
  MWP1G359740-005900S & 359.74   & -0.59    & 22.8    & 109.41   & 0.37  & 11.5  & & 14.05    & 0.52   & 8.8    & & --       & --    & --      & & --       & --    & --    & & --      & --    & --    \\
...&&&&&&&&&&&&&&&\\
\hline
\end{tabular}
\end{minipage}
\end{table}
\end{landscape}


\end{document}